\documentclass[11pt, oneside]{article}   	

\usepackage{jheppub}

\usepackage{amsmath}
\usepackage{amssymb}
\usepackage{amsfonts}
\usepackage{amsthm}
\usepackage{amsbsy}
\usepackage{array}
\usepackage[linesnumbered,lined,boxed,commentsnumbered]{algorithm2e}

\usepackage[OT2,OT1]{fontenc}
\newcommand\cyr
{
\renewcommand\rmdefault{wncyr}
\renewcommand\sfdefault{wncyss}
\renewcommand\encodingdefault{OT2}
\normalfont
\selectfont
}
\DeclareTextFontCommand{\textcyr}{\cyr}

\usepackage{mathtools}
\usepackage[usenames,dvipsnames]{xcolor}

\usepackage{subcaption}

\usepackage{dsdshorthand}

\usepackage[vcentermath]{youngtab}

\usepackage{tikz}
\usetikzlibrary{trees}
\usetikzlibrary{decorations.pathmorphing}
\usetikzlibrary{decorations.markings}
\usetikzlibrary{shapes.misc}
\usepackage{tikzit}

\tikzstyle{zero}=[fill=white, draw=black, shape=circle, inner sep=0pt, minimum size=4pt]
\tikzstyle{zeropole}=[fill={rgb,255: red,213; green,0; blue,3}, draw=black, shape=circle, inner sep=0pt, minimum size=4pt]

\tikzstyle{new edge style 0}=[->]

\tikzset{
  threept/.style={
    circle,
    draw,
    inner sep=2pt,
  },
  twopt/.style={
    circle,
    draw,
    fill=black,
    inner sep=1pt,
    minimum size=1pt
  },
  cross/.style={
    cross out,
    draw=black, 
    minimum size=7pt, 
    inner sep=0pt,
    outer sep=0pt
  },
  scalar/.style={
    thick,
    dashed,
    postaction={
      decorate,
      decoration={
        markings,
        mark=at position 0.5 with {\arrow{>}}
      }
    }
  },
  spinning/.style={
    thick,
    postaction={
      decorate,
      decoration={
        markings,
        mark=at position 0.5 with {\arrow{>}}
      }
    }
  },
  spinning no arrow/.style={
    thick,
  },
  finite with arrow/.style={
    decoration={
      snake,
      amplitude=1pt,
      segment length=6pt,
      post length=2pt
    },
    decorate,
    thick,->
  },
  finite/.style={
    decoration={
      snake,
      amplitude=1pt,
      segment length=6pt,
    },
    decorate,
    thick
  }
}

\makeatletter
\def\@fpheader{\ }
\makeatother

\title{Recursion relation for general 3d blocks}
\author{Rajeev S. Erramilli$\,^a$, Luca V. Iliesiu$\,^b$ and Petr Kravchuk$\,^c$}

\affiliation{${}^a$Department of Physics, Yale University, New Haven, CT 06520, USA \\
${}^b$Joseph Henry Laboratories, Princeton University, Princeton, NJ 08544, USA \\
${}^c$School of Natural Sciences, Institute for Advanced Study, Princeton, NJ 08540, USA}

\abstract{
We derive closed-form expressions for all ingredients of the Zamolodchikov-like recursion relation for general spinning conformal blocks in 3-dimensional conformal field theory. This result opens a path to efficient automatic generation of conformal block tables, which has immediate applications in numerical conformal bootstrap program. Our derivation is based on an understanding of null states and conformally-invariant differential operators in momentum space, combined with a careful choice of the relevant tensor structures bases. This derivation generalizes straightforwardly to higher spacetime dimensions $d$, provided the relevant Clebsch-Gordan coefficients of $\mathrm{Spin}(d)$ are known.
}
\date{}
\preprint{PUPT-2593}

\begin{document}

\maketitle

\section{Introduction}

In the past decade, the conformal bootstrap \cite{Ferrara:1973yt,Polyakov:1974gs,Mack:1975jr} has resurfaced as a powerful tool by numerically constraining the algebra of local operators in conformal field theories (CFTs) in $d >2$ space-time dimensions \cite{Rattazzi:2008pe} (for a recent review see \cite{Poland:2018epd}). Most of the numerical bootstrap studies to date focused on crossing equations for four-point functions of scalar operators, which turned out to be sufficient to significantly constrain and, to some extent, numerically solve interesting conformal field theories. However, in order to place further constraints on the space of unitary CFTs, the analysis of four-point functions that include spinning operators has also proven fruitful. Beyond providing universal constraints on the CFTs that include fermionic operators \cite{Iliesiu:2015qra, Iliesiu:2017nrv, Karateev:2019pvw}, a conserved current \cite{Dymarsky:2017xzb}, or a stress-energy tensor \cite{Dymarsky:2017yzx}, the analysis of four-point functions of spinning-operators has provided access to specific theories which have proven hard to study by only analyzing scalar correlators.  For instance, by constraining the four-point function of spin-$1/2$ operators in $d=3$, strong bounds were placed on the spectrum of operators in critical Gross-Neveu-Yukawa models \cite{Iliesiu:2015qra, Iliesiu:2017nrv} or in the $\cN=1$ supersymmetric Ising model \cite{Iliesiu:2015qra}. 

Nevertheless, to date there have been relatively few bootstrap studies of spinning four-point functions --- the above list exhausts them all --- and we believe that to a large extent this is due to the technical difficulties with setting up the numerical bootstrap. In the case of scalar four-point functions there are numerous tools designed to automate all steps of this process \cite{Simmons-Duffin:2015qma, Behan:2016dtz, Nakayama:2016jhq, Paulos:2014vya, Go:2019lke}. For spinning correlators only \cite{Cuomo:2017wme} is available which facilitates, but does not automate, some of these steps.
A complete automation in the spinning case is mainly hindered by the absence of a critical ingredient: a tool for computing conformal blocks for external operators with any spin. Specifically, one would like a computer program which, given the parameters of a conformal block, would automatically produce an approximation of the block suitable for use in numerical bootstrap algorithms.

Conformal blocks capture the contribution of a primary operator and its descendants to an operator product expansion of a four-point function.  
Scalar conformal blocks are parametrized by only a few numbers:  the scaling dimensions of external operators and the scaling dimension and spin of the intermediate operator. Once these values are provided, a numerical approximation to the relevant scalar block can be computed algorithmically~\cite{DO1,DO2,DO3,Hogervorst:2013kva,Kos:2014bka,Hogervorst:2013sma}, and indeed efficient computer implementations exist~\cite{WalterScalar,Behan:2016dtz, Nakayama:2016jhq, Paulos:2014vya}.

The situation with general spinning blocks is more complicated. In addition to the parameters that go into specifying a scalar block, the spinning blocks depend on additional information about three- and four-point tensor structures (see section~\ref{sec:definition-conf-blocks}). While these tensor structures can be classified and constructed in general cases~\cite{Kravchuk:2016qvl}, most of the methods for computing spinning conformal blocks~\cite{Costa:2011mg,Costa:2011dw,SimmonsDuffin:2012uy,Iliesiu:2015qra,Iliesiu:2015akf,Penedones:2015aga,Costa:2016hju,Costa:2016xah,Karateev:2017jgd} depend on ad hoc constructions of these structures and often on arbitrary intermediate choices. Furthermore, many of them require a non-trivial amount of algebraic manipulation. These are the reasons why it is difficult to implement a tool for computing a general spinning block.\footnote{This is not to say that the results of~\cite{Costa:2011mg,Costa:2011dw,SimmonsDuffin:2012uy,Iliesiu:2015qra,Iliesiu:2015akf,Penedones:2015aga,Costa:2016hju,Costa:2016xah,Karateev:2017jgd} are in any way incomplete or impractical. On the contrary, it was shown in these works how any spinning conformal block in any number $d$ of spacetime dimensions can be computed, and these results were successfully used in $d=3$ and $d=4$ in~\cite{Iliesiu:2015qra,Iliesiu:2017nrv,Dymarsky:2017xzb,Dymarsky:2017yzx,Karateev:2019pvw}. We are merely pointing out that these methods require calculations that, even though can be carried out in each particular case, are difficult to implement in full generality.} An attempt to resolve these issues was made in~\cite{Kravchuk:2017dzd}, which proposed a purely numerical algorithm which uses a uniform construction of tensor structures. However, this algorithm converges rather slowly, and the workarounds suggested in~\cite{Kravchuk:2017dzd} proved to be hard to implement~\cite{Erramilli:toAppear}.

The most attractive candidate for an algorithm that is both general and efficient seems to be the spinning version~\cite{Penedones:2015aga} of a residue recursion relation~\cite{Zamolodchikov:1987,Kos:2013tga,Kos:2014bka} which we will describe in detail shortly.\footnote{The idea of writing such a recursion relation was initially developed by Zamolodchikov for $2$d Virasoro blocks in~\cite{Zamolodchikov:1987} and was later generalized to any dimension for conformal blocks of external scalar operators by Kos, Poland, and Simmons-Duffin in \cite{Kos:2013tga,Kos:2014bka}, and finally generalized to spinning blocks by Penedones, Trevisani, and Yamazaki in~\cite{Penedones:2015aga}.} This recursion relation is currently the tool of choice for computing numerical approximations to scalar blocks: they produce a quickly-convergent expansion in a form that is perfectly suited for numerical bootstrap algorithms based on semidefinite programming~\cite{Poland:2011ey,Kos:2013tga,Kos:2014bka,Simmons-Duffin:2015qma}. While the theory behind the spinning version of the residue recursion relations was explained in~\cite{Penedones:2015aga}, it remained an open problem to find closed-form expressions for the various coefficients appearing in these relations.

In this paper, we solve this problem in $d=3$. Specifically, we provide closed-form expressions for all the ingredients appearing in Zamolodchikov recursion relations for general spinning conformal blocks in a form that is well-suited for numerical applications. While we only do so explicitly in $d=3$, our results generalize straightforwardly to higher dimensions.

To make the discussion more concrete, let us review the structure of residue recursion relations for spinning external operators. The conformal blocks are functions of the standard cross-ratios $z$ and $\bar z$
\be
	g^{a,b}_{\De,j,I}(z,\bar z),
\ee
which are labeled by a pair of three-point tensor structures $a$ and $b$, a four-point tensor structure $I$, and by scaling dimension $\De$ and spin $j$ of the exchanged operator $\cO$.\footnote{Additionally, they implicitly depend on the quantum numbers of the four external operators (for our precise definition of conformal blocks, see section~\ref{sec:definition-conf-blocks}).} Conformal blocks are known to be meromorphic functions of $\De$, with only simple poles in odd $d$~\cite{Penedones:2015aga}. As we review in section~\ref{sec:correlators-mom-space}, these poles arise due to the appearance, at special values of $\De=\De_\bi$, of null descendants in the conformal representation of $\cO$. Null descendants themselves form a conformal representation, and so the representation of $\cO$ ceases to be irreducible at $\De=\De_\bi$. Therefore, the possible positions $\De_\bi$ of the poles can be classified using the representation theory of the conformal group~\cite{Penedones:2015aga}. Near each pole the conformal block behaves as~\cite{Zamolodchikov:1987,Kos:2013tga,Kos:2014bka,Penedones:2015aga}
\be
		g^{a,b}_{\De,j,I}(z,\bar z)\sim \frac{1}{\De-\De_\bi}(\cL_\bi)^a_{a'}(\cR_\bi)^{b}_{b'}	g^{a',b'}_{\De'_\bi,j'_\bi,I}(z,\bar z),
\ee
where $\cL_\bi$ and $\cR_\bi$ are some constant matrices, which we call ``residue matrices,'' and $\De'_\bi,j'_\bi$ are determined by representation theory and depend on $j$ and $\bi$, the latter being a multi-index labeling the poles. Summation over repeated tensor structure indices is understood. The conformal blocks have an essential singularity at $\De\to \oo$ of the form
\be
	g^{a,b}_{\De,j,I}(z,\bar z)=(a_0 r)^\De h^{a,b}_{\oo,j,I}(z,\bar z)\p{1+O(\De^{-1})},
\ee
where $a_0$ is a convention-dependent constant\footnote{Usually $a_0=1$ or $a_0=4$.} and $h^{a,b}_{\oo,j,I}(z,\bar z)$ is an easy-to-determine function, while $r=\sqrt{\r\bar\r}$ with
\be
\label{eq:rho-brho-defintion}
	\r\equiv \frac{z}{(1+\sqrt{1-z})^2},\qquad
	\bar\r\equiv \frac{\bar z}{(1+\sqrt{1-\bar z})^2}.
\ee
If we define
\be
h^{a,b}_{\De,j,I}(z,\bar z)\equiv(a_0 r)^{-\De}g^{a,b}_{\De,j,I}(z,\bar z),
\ee
we find that $h^{a,b}_{\De,j,I}(z,\bar z)$ is a meromorphic function of $\De$ that is analytic at $\De=\oo$, and we can therefore write
\be
\label{eq:recursionintro}
	h^{a,b}_{\De,j,I}(z,\bar z)=h^{a,b}_{\oo,j,I}(z,\bar z)+\sum_{\bi}\frac{(a_0r)^{n_\bi}}{\De-\De_\bi}(\cL_\bi)^a_{a'}(\cR_\bi)^{b}_{b'}h^{a',b'}_{\De'_\bi,j'_\bi,I}(z,\bar z),
\ee
where $n_{\bi}\equiv \De'_\bi-\De_\bi\geq 1$ are positive integers. 

The formula~\eqref{eq:recursionintro} is what we will refer to as the residue recursion relation. It is useful for calculation of the conformal blocks for the following reason: given $h^{a,b}_{\De,j,I}(z,\bar z)$ in $r$-expansion up to order $r^n$,~\eqref{eq:recursionintro} allows us to obtain $h^{a,b}_{\De,j,I}(z,\bar z)$ at order $r^{n+1}$. One can then determine the conformal blocks by starting with $h^{a,b}_{\De,j,I}(z,\bar z)=0+O(r^0)$ and repeatedly applying~\eqref{eq:recursionintro}.

In order to use~\eqref{eq:recursionintro}, one needs to know the classification of poles $\De_\bi$ and the expressions for the residue matrices $\cL_\bi$, $\cR_\bi$, and the functions $h^{a,b}_{\oo,j,I}(z,\bar z)$.  A complete classification of poles~$\De_\bi$ was given in~\cite{Penedones:2015aga}. However, the residue matrices $\cL_\bi$ and $\cR_\bi$ and the function $h^{a,b}_{\oo,j,I}(z,\bar z)$ are unknown in general, but can be obtained on a case-by-case basis by combining the results of~\cite{Penedones:2015aga,Dymarsky:2017xzb,Karateev:2017jgd}.\footnote{They are known explicitly in the case of scalar~\cite{Kos:2013tga,Kos:2014bka} and in $d=3$ for scalar-fermion blocks~\cite{Iliesiu:2015akf}, for a four-point function including one external vector operator \cite{Penedones:2015aga} or for four vector operators \cite{Dymarsky:2017xzb}.}  In this paper we derive closed-form expressions for these objects in $d=3$ and point out how this can be generalized to higher dimensions. This improves on previous results by eliminating the need for any case-by-case calculations, thus opening a way for a general computer implementation.

The remainder of this paper is organized as follows. In section~\ref{sec:correlators-mom-space} we give a simple derivation of the residue recursion relation using a momentum-space representation of conformal blocks. This allows us to re-derive and somewhat refine the classification~\cite{Penedones:2015aga} of poles $\De_\bi$. In particular, we clarify some questions related to fermionic operators, and strengthen selection rules for some families of poles. In section~\ref{sec:residue-matrices} we derive a closed-form expression for the residue matrices $\cL_\bi$ and $\cR_\bi$ in a general $d=3$ conformal block. Our derivation is in essence using the prescription of~\cite{Penedones:2015aga}, simplified dramatically by judicious use of momentum space following~\cite{Karateev:2018oml}. In section~\ref{sec:leading-term-h} we give a closed-form expression for $h^{a,b}_{\oo,j,I}(z,\bar z)$ in a general $d=3$ conformal block. This result relies on using a particularly simple basis of four-point tensor structures introduced in~\cite{Kravchuk:2016qvl} and on the determination of the leading terms of conformal blocks in~\cite{Kravchuk:2017dzd}. Combined, the results of sections~\ref{sec:correlators-mom-space}-\ref{sec:leading-term-h} provide in closed form all the ingredients required for the implementation of the residue recursion relation~\eqref{eq:recursionintro} for any 3-dimensional conformal block. For ease of reference, we summarize these results in section~\ref{sec:summary-results}. In section~\ref{sec:examples-and-checks} we give examples and perform some checks of our formulas using techniques of~\cite{Penedones:2015aga,Dymarsky:2017xzb,Karateev:2017jgd}. We conclude in section~\ref{sec:discussion} with a discussion of the generalization to higher $d$ and of other possible future directions. Appendices~\ref{sec:conventions}-\ref{sec:deriving-f} provide details on our conventions and contain calculations omitted in the main text. 

\section{Definition of the conformal blocks}
\label{sec:definition-conf-blocks}

We start by giving the precise definition of conformal blocks for external spinning operators. We denote the external primaries by $\cO_1,\cdots,\cO_4$ and the exchanged primary by $\cO$. We allow these primaries to have arbitrary spin, but we often keep their spin indices implicit. To simplify the explanation, we also imagine that these are primaries in an actual CFT, although the definition can be straightforwardly reformulated without this assumption. Furthermore, we assume that all these primaries are Hermitian, which is always possible thanks to reality properties of $\mathrm{Spin}(2,1)$ representations.

We will denote the scaling dimension and spin of $\cO$ by $\De$ and $j$, respectively. The two-point function of $\cO$ can be written as~\cite{Iliesiu:2015qra}
\be\label{eq:standard2pt}
\<\cO(x_1,s_1)\cO(x_2,s_2)\>_\O=c_\cO \frac{i^{2j} (s_1^\a \g^\mu_{\a\b} s_2^\b x_{12,\mu})^{2j}}{x_{12}^{2(\De+j)}},\qquad (x_{12}^2>0),
\ee
where we are using index-free formalism of~\cite{Iliesiu:2015akf} in which
\be
\label{eq:define-polarizations}
\cO(x,s)\equiv s_{\a_1}\cdots s_{\a_{2j}}\cO^{\a_1\cdots \a_{2j}}(x),
\ee
and which we review in section~\ref{sec:spinning-primary-ops}. Here $c_\cO>0$  is a so far arbitrary normalization coefficient. We have also set $x_{ij}=x_i-x_j$ and specified the value of the time-ordered two-point function at spacelike separation, in which case it agrees also with the Wightman two-point function. Finally, we have used a subscript $\<\cdots\>_\O$ to indicate the physical two-point function, as opposed to some standard conformally-invariant expression (see below).\footnote{ Equivalent expressions for the position-space two-point function can be also given in terms of spinor embedding formalism of~\cite{Iliesiu:2015qra} as
\be
\<\cO(X_1,S_1)\cO(X_2,S_2)\>_\O=c_\cO\frac{\<S_1S_2\>^{2j}}{X_{12}^{\De+j}},
\ee
or, for integer $j$, in terms of vector embedding formalism of~\cite{Costa:2011mg}
\be
\<\cO(X_1,Z_1)\cO(X_2,Z_2)\>_\O=c_\cO\frac{H_{12}^{j}}{X_{12}^{\De+j}},
\ee
where we have assumed that the correspondence between spinor and vector polarizations is as defined in our appendix~\ref{sec:conventions}.} In this paper we will assume that the operators are normalized so that 
\be
\label{eq:b-j-normalization}
c_\cO=(4/a_0)^{\De}b_j,
\ee
where $a_0$ and $b_j$ are positive but otherwise unconstrained. One consequence of this normalization is that the primary state in radial quantization always has finite positive norm, irrespective of the value of $\De$.

The physical three-point functions of the operators $\cO_i,\cO$ can be expanded as
\be\label{eq:physical-structure-3pt-relation}
	\<\cO_1(x_1)\cO_2(x_2)\cO(x)\>_\O&=\sum_a \<\cO_1(x_1)\cO_2(x_2)\cO(x)\>^{(a)}\l^L_{a},\nn\\
	\<\cO_4(x_4)\cO_3(x_3)\cO(x)\>_\O&=\sum_a \<\cO_4(x_4)\cO_3(x_3)\cO(x)\>^{(a)}\l^R_{a},
\ee
for some coefficients $\l^L_{a}$ and $\l^R_{a}$, where $\<\cdots\>^{(a)}$ are some standard three-point tensor structures indexed by $a$, and on the left again we use a subscript $\<\cdots\>_\O$ to stress that this is a physical three-point function rather than a tensor structure. We assume that a basis of standard tensor structures have been chosen for this particular ordering of the operators.\footnote{We describe our choices of three-point tensor structures in section~\ref{sec:tensor-struct}.} We work in Lorentzian signature and for concreteness we assume that the points $x_i,x$ above are spacelike separated, and hence the corellators above can be equivalently treated as time-ordered or Wightman. Note that if some of the operators are fermions, the signs of the three-point functions are sensitive to operator ordering.

Similarly, the four-point function can be expanded in a basis of four-point tensor structures as\footnote{We describe our choices of four-point tensor structures in section~\ref{sec:four-point-tensor-struct}.}
\be
\label{eq:4-pt-function-general-z-and-barz}
	\<\cO_1(x_1)\cO_2(x_2)\cO_3(x_3)\cO_4(x_4)\>_\O=\sum_I\<\cO_1(x_1)\cO_2(x_2)\cO_3(x_3)\cO_4(x_4)\>^{(I)}g_{I}(z,\bar z)\,,
\ee
where the cross-ratios $z$ and $\bar z$ are defined in the canonical way,
\be
\label{eq:z-barz-def}
	z\bar z=\frac{x_{12}^2x_{34}^2}{x_{13}^2x_{24}^2},\quad
	(1-z)(1-\bar z)=\frac{x_{14}^2x_{23}^2}{x_{13}^2x_{24}^2}.
\ee
The same remarks regarding $x_i$ being spacelike separated and about the ordering of the operators apply. Region of spacelike separation includes, in particular, the region of real $z,\bar z$ with $0<z<1,0<\bar z<1$. 

The conformal block $g^{a,b}_{I,\De,j}(z,\bar z)$ is then defined as the contribution of the operator $\cO$ and its descendants to the cross-ratio function $	g_I(z,\bar z)$, appearing in \eqref{eq:4-pt-function-general-z-and-barz}:
\be 
	g_I(z,\bar z)\ni \sum_{a,b}\l^L_a\l^R_b g^{a,\,b}_{\De,\,j,\, I}(z,\bar z)\,.
\ee
We emphasize that the definition of the conformal block depends on
\begin{itemize}
	\item the normalization of the two-point function in~\eqref{eq:standard2pt},
	\item the choice of left ($a$) and right ($b$) three-point tensor structures,
	\item the choice of four-point tensor structure basis,
	\item the scaling dimension $\De$ and spin $j$ of the exchanged operator $\cO$,
	\item the scaling dimensions $\De_i$ and spins $j_i$ of the external operators $\cO_i$; this data of course enters implicitly also into the three- and four-point tensor structures.
\end{itemize}
It is easy to show that dependence of $g^{a,b}_{I,\De,j}(z,\bar z)$ on $\De_i$ is only through the differences $\De_{12}$ and $\De_{43}$, where $\De_{ij}=\De_i-\De_j$, and through a simple overall factor that depends on the choice of four-point structure basis. For example, with our choice this factor is $(z\bar z)^{-\frac{\De_1+\De_2}{2}}$.

\section{Null states and differential operators in momentum space} 
\label{sec:correlators-mom-space}

We start with a simple derivation of the residue recursion relations using momentum space. As explained above, a conformal block is simply the contribution of a primary $\cO$ and its descendants to a four-point function of primary operators $\cO_i$. For simplicity, {let us first focus on the case when $\cO_i$ and $\cO$ are Hermitian scalars}. By the definition of a conformal block we have
\be\label{eq:schematicblock}
	\l^L\l^R\<\cO_1\cO_2\cO_3\cO_4\>^{(1)}g_{\De}(z,\bar z)=\sum_{\Psi\in\cO} {\<\O|\cO_1(x_1)\cO_2(x_2)|\Psi\>\<\Psi|\cO_3(x_3)\cO_4(x_4)|\O\>},
\ee
where we have omitted all labels on $g$ except for $\De$ since they can only take one value for scalar $\cO,\cO_i$. We are summing over an orthonormal basis of descendants $\Psi$ of $\cO$ (including $\cO$ itself), and $|\O\>$ denotes the physical vacuum.

The main problem with using this expression for computing a conformal block is in finding an orthonormal basis of descendants $\Psi$. The key idea of our derivation is to work in Lorentzian signature and study the states of the form (assume for now that $\cO$ is a scalar primary)
\be\label{eq:ourstates}
	|\cO(f)\>\equiv\int d^d x f(x)\cO(x)|\O\>\,,
\ee
where $f(x)$ is a Schwartz test function: it is smooth and decays together with its derivatives faster than any power when $x\to \infty$. States $|\cO(f)\>$ are normalizable and thus are elements of the physical Hilbert space $\cH$. More precisely, these states belong to the subspace $\cH_\cO\subset \cH$, where $\cH_\cO=\mathrm{span}\{\Psi\}$ is the representation of the conformal group generated by $\cO$. Furthermore, $|\cO(f)\>$ are dense in $\cH_\cO$~\cite{Mack:1975je}.

Let us now define
\be
	|\cO(p)\>\equiv \int d^dx e^{ipx}\cO(x)|\O\>\,.
\ee
Since the states $|\cO(f)\>$ are dense in $\cH_\cO$, so are the states
\be
	\int d^dp f(p)|\cO(p)\>\,.
\ee
It is immediate to see that the objects $|\cO(p)\>$ form a $\delta$-function normalizable orthogonal basis in $\cH_\cO$,
\be\label{eq:scalar2pt}
	\<\cO(q)|\cO(p)\>=(2\pi)^d\de^d(p-q)\cB(\De)(-p^2)^{\De-\frac{d}{2}}\theta(p)\,.
\ee
Here $\theta(p)=\theta(-p^2)\theta(p^0)$ restricts $p$ to the forward null-cone, and the coefficient $\cB(\De)$ is given by
\be
	\cB(\De)=\frac{2^{d-2\De}2\pi^{\frac{d+2}{2}}}{\Gamma(\De-\tfrac{d-2}{2})\Gamma(\De)}\,,
\ee
assuming that the Euclidean two-point function in position space is normalized as\footnote{I.e.\ as in \eqref{eq:standard2pt} with $c_{\cO} = 1$.}
\be
	\<\cO(x)\cO(0)\>=\frac{1}{x^{2\De}}\,.
\ee
The form of~\eqref{eq:scalar2pt} is fixed by translation, scale, and Lorentz invariance. In particular, the presence of momentum-conserving $\de$-function, which makes the basis orthogonal, is simply due to translation invariance. Using this basis, we can write
\be
\label{eq:completeness-scalars}
	\sum_{\Psi\in\cO} |\Psi\>\<\Psi|=\cB(\De)^{-1}\int_{p>0}\frac{d^dp}{(2\pi)^d}(-p^2)^{-\De+\frac{d}{2}}|\cO(p)\>\<\cO(p)|\,.
\ee
The contribution of $\cO$ to the four-point function then takes the form (where we once again emphasize that we assume, for now, that $\cO_i$ and $\cO$ are scalars)
\be
	&\sum_{\Psi\in\cO}\<\O|\cO_1(x_1)\cO_2(x_2)|\Psi\>\<\Psi|\cO_3(x_3)\cO_4(x_4)|\O\>\nn\\
	&=\cB(\De)^{-1}\int_{p>0}\frac{d^dp}{(2\pi)^d}(-p^2)^{-\De+\frac{d}{2}}\<\O|\cO_1(x_1)\cO_2(x_2)|\cO(p)\>\<\cO(p)|\cO_3(x_3)\cO_4(x_4)|\O\>\,.
\ee
Therefore, we find the following relation for the conformal block,
\be\label{eq:scalarblockequation}
	&\<\cO_1(x_1)\cO_2(x_2)\cO_3(x_3)\cO_4(x_4)\>^{(1)} g_\De(z,\bar z) \nn\\
	&= \cB(\De)^{-1}\int_{p>0}\frac{d^dp}{(2\pi)^d}(-p^2)^{-\De+\frac{d}{2}}\<0|\cO_1(x_1)\cO_2(x_2)|\cO(p)\>^{(1)}\<\cO(p)|\cO_3(x_3)\cO_4(x_4)|0\>^{(1)},
\ee
where our use of $|0\>$ vacuum instead of $|\O\>$ indicates that we have replaced the physical three-point function by standard conformally-invariant structures (i.e.\ we factored out the OPE coefficients).\footnote{The Wightman tensor structures such as $\<0|\cO_1(x_1)\cO_2(x_2)|\cO(p)\>^{(a)}$ are obtained from the standard tensor structures $\<\cO_1(x_1)\cO_2(x_2)\cO(p)\>^{(a)}$ by using the usual $i\e$ prescriptions. Note that by convention, we specify tensor structures in the ordering as in~\eqref{eq:physical-structure-3pt-relation}, and tensor structures for other orderings are obtained by treating them as physical correlators satisfying the usual microcausality properties.} Recall that there is a unique three-point tensor structure when all three operators are scalars. This representation of conformal block was recently used in~\cite{Gillioz:2018mto,Kologlu:2019bco}.\footnote{The completeness relation~\eqref{eq:completeness-scalars} (as well as its spinning version) was also used in~\cite{Gillioz:2016jnn,Gillioz:2018kwh}. Our treatment of spin will be different from~\cite{Gillioz:2018mto,Gillioz:2016jnn,Gillioz:2018kwh}, and instead similar to~\cite{Kologlu:2019bco,Mack:1975je,Karateev:2018oml}.}

From equation~\eqref{eq:scalarblockequation} it is clear that the conformal block has poles in $\De$ coming from the factor $\cB(\De)^{-1}$ at
\be
	\De=1-k,\quad \De=\frac{d}{2}-k,
\ee
with integer $k\geq 1$. This corresponds to families $\mathrm{I}_1$ and $\mathrm{III}$ in classification of~\cite{Penedones:2015aga}, the only two families of poles present for scalar $\cO$. We can immediately track down the origin of these poles. For example, let us take the family $\mathrm{III}$ with $k=1$, i.e. 
\be
	\De=\De_*=\frac{d-2}{2}.
\ee
The pole appears because $\cB(\De_*)=0$. Recall that $\cB(\De)$ appears in the inner product~\eqref{eq:scalar2pt}. Naively, we might conclude that for $\De=\De_*$ this inner product vanishes. However, we should be more careful in taking the limit. Let us set $\De=\De_*+\e$ and rewrite
\be
	\<\cO(q)|\cO(p)\>\approx (2\pi)^d\de^d(p-q)\cB'(\De_*)\e(-p^2)^{-1+\e}\theta(p)\,.
\ee
We have in the limit $\e\to 0$
\be
	\e x^{-1+\e}\theta(x)\to \de(x).
\ee
In our case we can use this to write
\be
	\e (-p^2)^{-1+\e}\theta(p)=\e (-p^2)^{-1+\e}\theta(-p^2)\theta(p^0)\to \de(p^2)\theta(p^0),
\ee
and so we conclude
\be
	\<\cO(q)|\cO(p)\>\to (2\pi)^d\de^d(p-q)\cB'(\De_*)\de(p^2)\theta(p^0).
\ee
In other words, $|\cO(p)\>$ only has modes on the forward null cone $p^2=0$. In particular, $p^2|\cO(p)\>=0$, and thus
\be
	\ptl^2\cO(x)|\O\>=0\,.
\ee
This implies $\ptl^2\cO(x)=0$ as an operator equation. Such differential equations are a manifestation of null states (states of zero norm) in radial quantization. Above we found an equivalent description that the null states (for this pole) are simply $|\cO(p)\>$ with timelike momenta.\footnote{For higher $k$ in family $\mathrm{III}$ we find that $\<\cO(q)|\cO(p)\>\propto \de^{(k-1)}(p^2)$, the $(k-1)$-th derivative of $\de$-function, so the non-zero modes live on a ``fuzzy'' null cone, with $p^{2k}|\cO(p)\>=0$. Similarly, for family $\mathrm{I}_1$ we find that $\<\cO(q)|\cO(p)\>\propto \ptl^{2(k-1)}\de^{d}(p)$. The modes now live at a ``fuzzy'' point $p=0$, and it is possible to show that $(p_{\mu_1}\cdots p_{\mu_k}-\text{traces})|\cO(p)\>=0$.}

We see that using the momentum space representation~\eqref{eq:scalarblockequation} of conformal blocks, we can directly link the poles in $\De$ to degeneration of the inner product~\eqref{eq:scalar2pt}, and thus to null states and differential operators. As emphasized in~\cite{Penedones:2015aga}, these differential operators are important for determining the residue matrices in the residue recursion relation. It is a bit awkward to explain at this point how this, or even the fact that the residues of the poles are related to other conformal blocks, comes about in our approach. We will provide this explanation after some additional preparations, which are as follows.

Unless otherwise stated, in the rest of this paper we specialize to $d=3$. First, in subsections~\ref{sec:spinning-primary-ops} and~\ref{sec:two-point-function} we generalize the above discussion to non-scalar primaries. Then in subsection~\ref{sec:differential-ops} we describe momentum space version of conformally-invariant differential operators (such as $\ptl^2$ above) that are responsible for the existence of null states. Finally, in subsection~\ref{sec:derivation-of-Zamolodchikov-recursion} we present a momentum-space derivation of the residue recursion relation. 

\subsection{Spinning primary operators}
\label{sec:spinning-primary-ops}

 The Lorentz group in $d=3$ is $\mathrm{Spin}(2,1)\simeq \SL(2, \R)$. All possible spin representations are real and the respective primary operators can be described as Hermitian symmetric tensors
\be
	\cO^{\a_1\cdots \a_{2j}}(x),
\ee
where each index $\a_i$ transforms in the spinor representation of $\mathrm{Spin}(2,1)$, which is the fundamental of $\SL(2, \R)$. Here $j$ is the (half-)integral spin of $\cO$. Our conventions for these representations are similar to those of~\cite{Iliesiu:2015qra} and are laid out in appendix~\ref{sec:conventions}. 

As previously stated, we will often work in index-free formalism by contracting the indices $\a_i$ with a single real polarization spinor $s_{\a_i}$, i.e.
\be
	\cO(x,s)\equiv s_{\a_1}\cdots s_{\a_{2j}} \cO^{\a_1\cdots \a_{2j}}(x).
\ee
Clearly, $\cO(x,s)$ is a homogeneous polynomial in $s_\a$ of degree $2j$, and the original symmetric tensor can be restored from it by repeated differentiation.

We define the Fourier-transformed operators as
\be\label{eq:smomentumoperator}
	\cO(p,s)=\int d^3 x\, e^{ipx} \cO(x,s).
\ee
At every point $p$ in momentum space there is a natural subgroup of the Lorentz group, the little group of $p$. It consists of $\L\in \mathrm{Spin}(2,1)$ which leave $p$ invariant
\be
	\L p = p.
\ee
It will suffice for our purposes to consider only timelike $p$, in which case the little group is isomorphic to $\mathrm{Spin}(2)\simeq U(1)$. Let us choose a canonical momentum $q_0=\hat e_0=(1,0,0)$.\footnote{We will often use the notation $\hat e_\mu$ to denote a unit vector in direction $\mu$, i.e. $x=x^\mu \hat e_\mu$.} The little group of $q_0$ is generated by $M_{12}$, and it is convenient to define components of $\cO$ which are diagonal under $M_{12}$,
\be
	[M_{12},\cO_m(q_0)]=-im\,\cO_m(q_0).
\ee
Specifically, we define $\cO_m(q_0)$ in terms of \(\cO(q_0,s)\) through:
\be\label{eq:standardmomentumoperator}
	\cO(q_0,s)=\sum_{m=-j}^j |j,m\>\cO_m(q_0),
\ee
where $m$ runs through $\{-j,-j+1,\cdots j\}$ and $|j,m\>$ are standard polynomials in $s$ that are diagonal under $M_{12}$ rotations.\footnote{One should view $|j, m\>$ as functions of $s_\a$ and not as states in the Hilbert space $\cH_\cO$. However, one should keep in mind that the functions $|j, m\>$ with different values of $m$ form a vector space which furnishes an irreducible representation of $\SU(2)$. The notation $|j, m\>$ is standard for the elements of this representation, and will turn out to be convenient for our purposes. We hope that this will not cause too much confusion.}  The exact relation between $s$ and $|j, m\>$ is not necessary for now and will be further discussed in section \ref{sec:tensor-struct}.

Proceeding with the usual construction, for any future-directed timelike $p$ we choose a standard Lorentz transformation $\L_p\in\mathrm{Spin}(2,1)$ such that
\be
	\L_p q_0=p,\qquad \L_{q_0}=\mathrm{id},
\ee
and define a basis for spin components of $\cO(p)$ by
\be\label{eq:momentumoperator}
	\cO_m(p)\equiv U(\L_p)\cO_m(q_0)U(\L_p)^{-1},
\ee
where $U(\L)$ is the unitary operator implementing Lorentz transformation $\L$. Note that our definition of $\cO_m(p)$ is ambiguous because $\L_p$ can be modified by multiplying on the right by a $p$-dependent element of the little group of $q_0$. This means that choosing another set of $\L_p$'s amounts to replacing
\be\label{eq:wignerambiguity}
	\cO_m(p)\to e^{-i m\f(p)} \cO_m(p)
\ee
for some $p$-dependent angle $\f(p)$. The quantities that we will work with are going to be invariant under this transformation.

Finally, note that this only defines $\cO_m(p)$ for future-directed timelike $p$. For past-directed $p$ we use the same definition but replacing $q_0$ by $-q_0$. The new standard $\L'_p$ which take past-directed $p$ to $-q_0$ are defined by $\L'_p=\L_{-p}$. With this definition if follows, using the reality property of $|j,m\>$, that
\be\label{eq:Omconjugation}
	[\cO_m(p)]^\dagger=e^{i\pi j}\cO_{-m}(-p).
\ee
We will not require a definition of $\cO_m(p)$ for spacelike $p$.

\subsection{Two-point function}
\label{sec:two-point-function}

We would like to study the inner product of the ket states
\be\label{eq:generalmomentumstate}
	|\cO_m(p)\>\equiv \cO_m(p)|\O\>.
\ee
For the bra state we have
\be\label{eq:conjugatedefinition}
\qquad \<\cO_m(p)|=\p{|\cO_m(p)\>}^\dagger=\<\O|\p{\cO_m(p)}^\dagger=e^{i\pi j}\<\O|\cO_{-m}(-p),
\ee
where we used equation~\eqref{eq:Omconjugation}. The inner product of states~\eqref{eq:generalmomentumstate} is then related to the Wightman two-point function in momentum space,
\be\label{eq:innerprodtocompute}
	\<\cO_m(p)|\cO_{m'}(q)\>=e^{i\pi j}\<\O|\cO_{-m}(-p)\cO_{m'}(q)|\O\>.
\ee
Using the definitions~\eqref{eq:smomentumoperator},~\eqref{eq:standardmomentumoperator} and~\eqref{eq:momentumoperator}, we can in principle compute~\eqref{eq:innerprodtocompute} by taking the Fourier transform of the position space Wightman two-point function
\be
	\<\O|\cO(x_1,s_1)\cO(x_2,s_2)|\O\>,
\ee
the expression for which is fixed by conformal symmetry and is given explicitly by~\eqref{eq:standard2pt}.\footnote{Equation~\eqref{eq:standard2pt} is for spacelike separation, as a Wightman function it should be understood with the appropriate $i\e$ prescription.} However, a significant part of the answer is fixed by simple symmetries. Indeed, we can immediately write~\cite{Mack:1975je,Karateev:2018oml}
\be
\label{eq:innerprod-in-terms-of-Bm}
	\<\cO_m(p)|\cO_{m'}(q)\>=(2\pi)^3\de^3(p-q)\de_{m,m'}(-p^2)^{\De-\frac{3}{2}}\theta(p)\cB_m(\De,j).
\ee
Here, the $\de^3(p-q)$ comes from translation in invariance, $\de_{m,m'}$ comes from invariance under little group of $p$,\footnote{Recall that the delta function $\de^3(p-q)$ sets $q=p$.} the power of $p^2$ is fixed by scaling invariance, and $\theta(p)$ restricts $p$ to be in the forward null cone as required by positivity of energy. 

The only undetermined part here are the coefficients $\cB_m(\De,j)$. Correspondingly, the only symmetry that we have not yet utilized are the special conformal generators. They act rather non-trivially in momentum space, but it is possible to show~\cite{Mack:1975je} that special conformal invariance implies the following recursion relation for the coefficients $\cB_m(\De,j)$,
\be
	\cB_m(\De,j)=\frac{\De+m-2}{\De-m-1}\cB_{m-1}(\De,j).
\ee
We do this in appendix~\ref{sec:details-fourier-space-2pt}. This recursion relation can be easily solved as
\be
	\cB_m(\De,j)=\frac{\G(\De+m-1)\G(\De-m-1)}{\G(\De+j-1)\G(\De-j-1)}\cB_j(\De,j).
\ee
The coefficient $\cB_j(\De,j)$ can be chosen arbitrarily, which reflects the freedom of normalizing the two-point function. In appendix~\ref{sec:details-fourier-space-2pt} we show that 
\be
	\cB_j(\De,j)=c_\cO\frac{2\pi^2\G(\De+j-1)}{\G(\De+j)\G(2\De-2)},
\ee
and thus
\be\label{eq:Bvalue}
	\cB_m(\De,j)=c_\cO\frac{2\pi^2\G(\De+m-1)\G(\De-m-1)}{\G(\De+j)\G(2\De-2)\G(\De-j-1)}{},
\ee
provided the position space two-point function is normalized as in~\eqref{eq:standard2pt}. Note that the values~\eqref{eq:Bvalue} can be used to derive the unitarity bounds~\cite{Minwalla:1997ka} in $d=3$ following the $d=4$ derivation in~\cite{Mack:1975je}.

\subsubsection{Completeness relation and conformal blocks}
\label{sec:completeness-relation}

With the knowledge of the inner products~\eqref{eq:innerprodtocompute}, we can now write down the spinning analog of the completeness relation \eqref{eq:completeness-scalars} for the conformal multiplet $\cH_\cO$ of $\cO$, 
\be\label{eq:generalcompleteness}
\sum_{\Psi\in\cO}|\Psi\>\<\Psi|=\sum_{m=-j}^j \cB_m(\De,j)^{-1}\int_{p>0}\frac{d^3p}{(2\pi)^3}(-p^2)^{\frac{3}{2}-\De}|\cO_m(p)\>\<\cO_m(p)|.
\ee
Using this inside a four-point function, we find
\be\label{eq:generalcb}
	&\sum_{I}\<\cO_1(x_1)\cO_2(x_2)\cO_3(x_3)\cO_4(x_4)\>^{(I)} g_{\De,j,I}^{a,b}(z,\bar z)=\nn\\
	&=\sum_{m=-j}^j \cB_m(\De,j)^{-1}\int_{p>0}\frac{d^3p}{(2\pi)^3}(-p^2)^{\frac{3}{2}-\De}\<0|\cO_1(x_1)\cO_2(x_2)|\cO_m(p)\>^{(a)}\<\cO_m(p)|\cO_3(x_3)\cO_4(x_4)|0\>^{(b)}.
\ee
As in the scalar example in the beginning of this section, we can now use this relation to analyze the poles of the conformal block, which come from the factor $\cB_m(\De,j)^{-1}$. We start with the classification of such poles.

\subsubsection{Null states}
\label{sec:null-states}

\begin{figure}[t]
	\centering
	\begin{subfigure}{0.45\textwidth}
		\begin{tikzpicture}
	\begin{pgfonlayer}{nodelayer}
		\node [style=none] (1) at (6, 0) {};
		\node [style=none] (2) at (-6, 0) {};
		\node [style=none] (3) at (0, 0) {};
		\node [style=none] (4) at (0, -0.125) {};
		\node [style=none] (5) at (4, 0) {};
		\node [style=none] (6) at (4, -0.125) {};
		\node [style=none] (7) at (-4, 0) {};
		\node [style=none] (8) at (-4, -0.125) {};
		\node [style=none] (9) at (5, 0) {};
		\node [style=none] (10) at (5, -0.125) {};
		\node [style=none] (11) at (3, 0) {};
		\node [style=none] (12) at (3, -0.125) {};
		\node [style=none] (13) at (2, 0) {};
		\node [style=none] (14) at (2, -0.125) {};
		\node [style=none] (15) at (1, 0) {};
		\node [style=none] (16) at (1, -0.125) {};
		\node [style=none] (17) at (-3, 0) {};
		\node [style=none] (18) at (-3, -0.125) {};
		\node [style=none] (19) at (-2, 0) {};
		\node [style=none] (20) at (-2, -0.125) {};
		\node [style=none] (21) at (-1, 0) {};
		\node [style=none] (22) at (-1, -0.125) {};
		\node [style=none] (23) at (-5, 0) {};
		\node [style=none] (24) at (-5, -0.125) {};
		\node [style=none] (25) at (0, -0.5) {};
		\node [style=none] (26) at (0, -0.5) {$0$};
		\node [style=none] (27) at (5, -0.5) {$j+1$};
		\node [style=none] (28) at (-4, -0.5) {$-j$};
		\node [style=zero] (29) at (5, 0.25) {};
		\node [style=zero] (30) at (4, 0.25) {};
		\node [style=zero] (31) at (3, 0.25) {};
		\node [style=zero] (32) at (2, 0.25) {};
		\node [style=zeropole] (33) at (1, 0.25) {};
		\node [style=zeropole] (34) at (0, 0.25) {};
		\node [style=zeropole] (35) at (-1, 0.25) {};
		\node [style=zeropole] (36) at (-2, 0.25) {};
		\node [style=zeropole] (37) at (-3, 0.25) {};
		\node [style=zeropole] (38) at (-4, 0.25) {};
		\node [style=zeropole] (39) at (-5, 0.25) {};
		\node [style=zeropole] (40) at (-6, 0.25) {};
		\node [style=zero] (41) at (0, 0.675) {};
		\node [style=zero] (42) at (-1, 0.675) {};
		\node [style=zero] (43) at (-2, 0.675) {};
		\node [style=zeropole] (44) at (-3, 0.675) {};
		\node [style=zeropole] (45) at (-4, 0.675) {};
		\node [style=zeropole] (46) at (-5, 0.675) {};
		\node [style=zeropole] (47) at (-6, 0.675) {};
		\node [style=zero] (48) at (-4, 1.1) {};
		\node [style=zero] (49) at (-5, 1.1) {};
		\node [style=zero] (50) at (-6, 1.1) {};
		\node [style=zero] (51) at (1, 0.675) {};
		\node [style=none] (52) at (6.5, 0) {$\Delta$};
	\end{pgfonlayer}
	\begin{pgfonlayer}{edgelayer}
		\draw [style=new edge style 0] (2.center) to (1.center);
		\draw (3.center) to (4.center);
		\draw (5.center) to (6.center);
		\draw (7.center) to (8.center);
		\draw (9.center) to (10.center);
		\draw (11.center) to (12.center);
		\draw (13.center) to (14.center);
		\draw (15.center) to (16.center);
		\draw (17.center) to (18.center);
		\draw (19.center) to (20.center);
		\draw (21.center) to (22.center);
		\draw (23.center) to (24.center);
	\end{pgfonlayer}
\end{tikzpicture}
		\caption{}
		\label{fig:bosonic_families}
	\end{subfigure}
	~
	\begin{subfigure}{0.45\textwidth}
		\begin{tikzpicture}
	\begin{pgfonlayer}{nodelayer}
		\node [style=none] (0) at (6, 0) {};
		\node [style=none] (1) at (-6, 0) {};
		\node [style=none] (2) at (0, 0) {};
		\node [style=none] (3) at (0, -0.125) {};
		\node [style=none] (4) at (4, 0) {};
		\node [style=none] (5) at (4, -0.125) {};
		\node [style=none] (6) at (-4, 0) {};
		\node [style=none] (7) at (-4, -0.125) {};
		\node [style=none] (8) at (5, 0) {};
		\node [style=none] (9) at (5, -0.125) {};
		\node [style=none] (10) at (3, 0) {};
		\node [style=none] (11) at (3, -0.125) {};
		\node [style=none] (12) at (2, 0) {};
		\node [style=none] (13) at (2, -0.125) {};
		\node [style=none] (14) at (1, 0) {};
		\node [style=none] (15) at (1, -0.125) {};
		\node [style=none] (16) at (-3, 0) {};
		\node [style=none] (17) at (-3, -0.125) {};
		\node [style=none] (18) at (-2, 0) {};
		\node [style=none] (19) at (-2, -0.125) {};
		\node [style=none] (20) at (-1, 0) {};
		\node [style=none] (21) at (-1, -0.125) {};
		\node [style=none] (22) at (-5, 0) {};
		\node [style=none] (23) at (-5, -0.125) {};
		\node [style=none] (25) at (0, -0.75) {$\frac{1}{2}$};
		\node [style=none] (26) at (5, -0.5) {$j+1$};
		\node [style=none] (27) at (-4, -0.5) {$-j$};
		\node [style=zero] (28) at (5, 0.25) {};
		\node [style=zero] (29) at (4, 0.25) {};
		\node [style=zero] (30) at (3, 0.25) {};
		\node [style=zero] (31) at (2, 0.25) {};
		\node [style=zeropole] (32) at (1, 0.25) {};
		\node [style=zeropole] (33) at (0, 0.25) {};
		\node [style=zeropole] (34) at (-1, 0.25) {};
		\node [style=zeropole] (35) at (-2, 0.25) {};
		\node [style=zeropole] (36) at (-3, 0.25) {};
		\node [style=zeropole] (37) at (-4, 0.25) {};
		\node [style=zeropole] (38) at (-5, 0.25) {};
		\node [style=zeropole] (39) at (-6, 0.25) {};
		\node [style=zero] (40) at (0, 0.675) {};
		\node [style=zero] (41) at (-1, 0.675) {};
		\node [style=zero] (42) at (-2, 0.675) {};
		\node [style=zeropole] (43) at (-3, 0.675) {};
		\node [style=zeropole] (44) at (-4, 0.675) {};
		\node [style=zeropole] (45) at (-5, 0.675) {};
		\node [style=zeropole] (46) at (-6, 0.675) {};
		\node [style=zero] (47) at (-4, 1.1) {};
		\node [style=zero] (48) at (-5, 1.1) {};
		\node [style=zero] (49) at (-6, 1.1) {};
		\node [style=none] (51) at (6.5, 0) {$\Delta$};
	\end{pgfonlayer}
	\begin{pgfonlayer}{edgelayer}
		\draw [style=new edge style 0] (1.center) to (0.center);
		\draw (2.center) to (3.center);
		\draw (4.center) to (5.center);
		\draw (6.center) to (7.center);
		\draw (8.center) to (9.center);
		\draw (10.center) to (11.center);
		\draw (12.center) to (13.center);
		\draw (14.center) to (15.center);
		\draw (16.center) to (17.center);
		\draw (18.center) to (19.center);
		\draw (20.center) to (21.center);
		\draw (22.center) to (23.center);
	\end{pgfonlayer}
\end{tikzpicture}
		\caption{}
		\label{fig:fermionic_families}
	\end{subfigure}
	\caption{Zeros and poles coming from $\G$-functions in~\eqref{eq:Blocal}, in bosonic (a) and fermionic (b) cases. The total number of circles indicates the order of the zero coming from denominator $\G$-functions, while the number of filled circles indicates the minimal (over $m$) order of the pole coming from the numerator $\G$-functions.}
\end{figure}
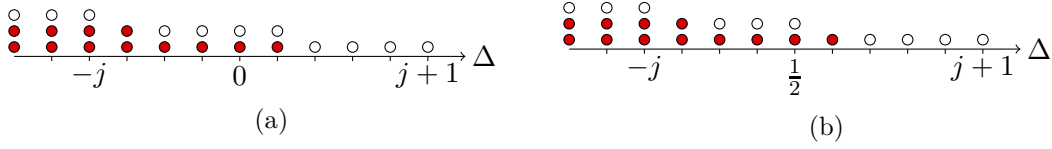

From~\eqref{eq:generalcb} we see that $g_{\De,j,I}^{a,b}(z,\bar z)$ has a pole at $\De=\De_*$ if $\cB_m(\De_*,j)=0$ for any $m\in \{-j,-j+1,\cdots,j\}$. We reproduce the expression~\eqref{eq:Bvalue} for $\cB_m(\De,j)$ here for convenience
\be\label{eq:Blocal}
		\cB_m(\De,j)=c_\cO\frac{2\pi^2}{\G(\De+j)\G(2\De-2)\G(\De-j-1)}{\G(\De+m-1)\G(\De-m-1)}.
\ee
The zeros can come from any of the three $\G$-functions in denominator, and some of them may be canceled by poles of $\G$-functions in the numerator. The $\G$-functions in the denominator lead to four families of potential zeros,
\be
	\mathrm{I}: \quad& \De=1-j-k,\\
	\mathrm{II}: \quad& \De=2+j-k,\\
	\mathrm{III}: \quad& \De=\tfrac{3}{2}-k,\\
	\mathrm{IV}: \quad& \De=2-k,
\ee
where $k\geq 1$ is an integer, and families $\mathrm{III}$ and $\mathrm{IV}$ both come from $\G(2\De-2)$. These families overlap, so in general we get higher-order zeros coming from the denominator $\G$-functions. Similarly, there could be higher-order poles coming from the numerator $\G$-functions. Note that the order of the pole of the conformal block is determined by the maximal order of the zero of $\cB_m(\De,j)$ among different $m$. 

\paragraph{Bosons} First we discuss the case of integer $j$. We are interested in the maximal order of the zero of $\cB_m(\De,j)$ and thus the minimal order of the pole that $\G(\De+m-1)\G(\De-m-1)$ has among all possible values of $m$. Because of the symmetry $m\to -m$, we can restrict to $m\geq 0$. Unless there is a pole at $m=0$, the minimal pole order is $0$. If there is a pole at $m=0$, the minimal order is at least $1$, because $\G(\De-m-1)$ will then also have a pole for all $m>0$. If $\G(\De+m-1)$ has a pole at $m=j$, then the minimal order is $2$. We therefore conclude that $\G(\De+m-1)\G(\De-m-1)$ has at least a simple pole for all $m$ iff $\De=2-k$, and at least a double pole iff $\De=2-j-k$, where in both cases $k$ is a positive integer.

The family $\mathrm{III}$ of zeros only contains half-integer $j$, and thus does not overlap with the poles in the numerator or with the extra zeros from the other $\G$-functions. This family therefore leads to simple zeros in $\cB_m(\De,j)$ for all $k>0$. To see what happens to the other families, it is convenient to represent them graphically as in figure~\ref{fig:bosonic_families}. In this figure, the total number of circles above a given value of $\De$ gives the order of the zero coming from the denominator, while the number of filled circles gives the minimal order of the pole coming the numerator. The maximal order of the zero of $\cB_m(\De,j)$ over different $m$ is therefore given by the number of unfilled circles. We see that at each value of $\De$ this order is at most $1$.\footnote{Recall that the family $\mathrm{III}$ is excluded from this figure. It would appear at half-integer values of $\De$.} We can describe the resulting set of zeros by the following families
\be
\mathrm{I}: \quad& \De=1-j-k,\\
\mathrm{II}: \quad& \De=2+j-k,\quad k\leq j\\
\mathrm{III}: \quad& \De=\tfrac{3}{2}-k,\\
\mathrm{IV}: \quad& \De=2-k,\quad k\leq j.
\ee
We will soon see that this organization of zeros is indeed natural.

\paragraph{Fermions} Now consider the case of half-integer $j$. Again the minimal pole order is $0$ unless $\G(\De+m-1)\G(\De-m-1)$ has a pole at $m=\half$, in which case the order is at least $1$. If in addition $\G(\De+m-1)$ has a pole at $m=j$, then the minimal pole order is $2$. We therefore conclude that  $\G(\De+m-1)\G(\De-m-1)$ has at least a simple pole for all $m$ iff $\De=\tfrac{5}{2}-k$, and at least a double pole iff $\De=2-j-k$, where in both cases $k$ is a positive integer.

This time, the family $\mathrm{IV}$ has integer $\De$, whereas all other families and the pole positions described above have half-integer $\De$. Therefore the family $\mathrm{IV}$ leads to simple zeros in $\cB_m(\De,j)$ for all $k>0$. We again represent the remaining families graphically in figure~\ref{fig:fermionic_families}. We see that we get at most simple zeros, which can be described by the families
\be
\mathrm{I}: \quad& \De=1-j-k,\\
\mathrm{II}: \quad& \De=2+j-k,\quad k\leq j\\
\mathrm{III}: \quad& \De=\tfrac{3}{2}-k,\quad k\leq j,\\
\mathrm{IV}: \quad& \De=2-k.
\ee
Note that here $k\leq j$ effectively means $k\leq j-\thalf$, since $j$ is half-integer.

\paragraph{Notation}

Now that we have a classification of poles, let us formally introduce some notation. We label poles by $\bi=(A,k)$ where $A$ is the family number, $A\in\{\mathrm{I},\mathrm{II},\mathrm{III},\mathrm{IV}\}$, and $k$ satisfies the constraints listed above for the given values of $A$ and $j$. We denote the position of the pole by $\De_\bi$ and spin by $j$.

\subsection{Differential operators}
\label{sec:differential-ops}

As we explained in the beginning of section~\ref{sec:correlators-mom-space}, zeros $\De_\bi$ in $\cB_m(\De,j)$ signal existence of null states, which can alternatively be described as differential equations satisfied by primary operators. These differential equations have to be conformally-invariant. In other words, if $\cO$ is a primary, and $\cD$ is a differential operator of this type, it should be that $\cD\cO$ is a primary as well. Otherwise, we would be able act with an appropriate combination of special conformal generators on the equation
\be
	(\cD\cO)(x=0,s) = 0
\ee
and conclude that $\cO(x=0,s)=0$, which is too strong.\footnote{Another option is that we will find a conformally-invariant equation of which the one at hand is a consequence. In either case, we conclude that only conformally-invariant equations are interesting.} On the other hand, if $\cD\cO$ is itself a primary, all special conformal generators annihilate the left-hand side of the above equation, and we cannot derive stronger equations from it.

It is easy to see that conformally-invariant operators are relatively rare. Let us collect some facts. By translation invariance, $\cD$ is a polynomial in $\frac{\ptl}{\ptl x^\mu}$ with polynomial coefficients. By scaling invariance, this polynomial has to be homogeneous of some degree, $n>0$. Therefore, if $\cO$ has scaling dimension $\De$ and spin $j$, then $\cO'\equiv \cD\cO$ has scaling dimension $\De'=\De+n$ and some spin $j'$.\footnote{It is easy to see that $|j-j'|\leq n$, since each derivative transforms in a vector irrep.} Since $\cD$ is conformally-invariant, it means that the conformal Casimirs of $\cO$ and $\cO'$ should be the same. For example, we must have the equality of quadratic Casimirs,

\be\label{eq:c2equality}
	\De(\De-3)+j(j+1)=\De'(\De'-3)+j'(j'+1)=(\De+n)(\De+n-3)+j'(j'+1).
\ee
Since $\De^2$ cancels on both sides, we can see that for a given choice of $n>0$, $j$ and $j'$, $\cD$ can only exist for a single value of $\De$. Taking into account also the existence of quartic Casimir of the $\mathfrak{so}(5)$ conformal algebra given by
\be
	(\De-1)(\De-2)j(j+1)
\ee
it is easy to show that $(\De',j')$ must in fact take one of the following values\footnote{Formally it could also be $(\De,-1-j)$ or $(3-\De,-1-j)$, but $j'$ should be a non-negative (half-)integer. There is also the option of having $(\De',j')=(1-j,\De-2)$, but because of $\De'=\De+n$ this would imply $\De=1-j-n\leq 0$, and thus $j'=\De-2<0$.}
\be\label{eq:not-obviously-bad-reflections}
	(3-\De,j),\quad (1-j,1-\De), \quad (j+2, 1-\De), \quad (j+2,\De-2).
\ee
However, it turns out that the mere equality of conformal Casimir eigenvalues is not enough for existence of $\cD$, and a more careful representation-theoretic analysis is needed for a complete classification of such operators. It has been known to mathematicians for a some time~\cite{MR1255551} and in the physics literature the relevant analysis was recently performed in~\cite{Penedones:2015aga}. In this subsection we will reproduce this classification by more elementary methods, and obtain useful explicit expressions for $\cD$ in momentum space.

We start by describing the most general possible form for $\cD$ in momentum space using the spin basis defined in section~\ref{eq:standardmomentumoperator}. Simply from symmetries we must have, assuming future-directed timelike $p$,
\be\label{eq:generalD}
	\cO'_m(p)=(\cD\cO)_m(p)=(-p^2)^{\frac{n}{2}}\cD_m \cO_m(p),\quad\text{(no summation)}
\ee
where $\cD_m$ are some constants that characterize $\cD$, similarly to how $\cB_m(\De,j)$ characterize the two-point function in section~\ref{sec:two-point-function}.\footnote{This equation may appear somewhat strange for odd values of $n$ (which are, as we will see, possible), because above we stated that $\cD$ is a polynomial in $\frac{\ptl}{\ptl x^\mu}$, and thus must be polynomial in $p^\mu$ in momentum space. The resolution is that our definition~\eqref{eq:momentumoperator} of $\cO_m(p)$ involves acting with rotation generators on spin indices, and these rotation generators are not, generally speaking, themselves polynomial. In other words, it is only if we express the action in terms of $\cO(p,s)$ that $\cD$ should be explicitly polynomial in $p^\mu$.} In this equation, the momentum is the same on both sides due to translation invariance, and the power of $p^2$ is fixed by scaling invariance. Finally, the fact that $m$ is the same on both sides follows from the invariance under little group of $p$: more generally, $\cD$ could act by a matrix $\cD_{m,m'}$, but applying little group rotation on both sides, under which $\cO_m$ and $\cO'_m$ get phase $e^{-i\phi m}$, we must conclude that $\cD_{m,m'}$ is diagonal. The fact that the action of $\cD$ on $\cO_m(p)$ is diagonal will turn out to be extremely useful for computation of the residue matrices in section~\ref{sec:residue-matrices}.

For future reference, we also quote the following formula for action of $\cD$ for past-directed timelike momentum $p$,
\be\label{eq:generalDpastdirected}
	(\cD\cO)_m(p)=(-1)^{j'-j}(-p^2)^{\frac{n}{2}}\cD_{-m} \cO_m(p),
\ee
where $j'$ is spin of $\cO'=\cD\cO$. We derive this from~\eqref{eq:generalD} in appendix~\ref{app:Dparity}.

Consider now the standard conformally-invariant two-point function of the operator $\cO$, 
\be
	\<0|\cO_m(p)\cO_{m'}(q)|0\>.
\ee
We can compute the two-point function of $\cO'=\cD\cO$ with $\cO$ as
\be
	\<0|\cO'_m(p)\cO_{m'}(q)|0\>=(-p^2)^{\frac{n}{2}}\cD_m\<0|\cO_m(p)\cO_{m'}(q)|0\>.
\ee
We know that $\cO'$ is a primary of dimension $\De'=\De+n\neq \De$, and thus we can conclude that this two-point function must vanish identically, since two-point functions can only be non-zero only between primaries of equal scaling dimensions. But this is only possible if 
\be
	\cD_m\cB_m(\De,j)=0,\quad \forall m\in\{-j,j+1,\cdots,j\}.
\ee
This immediately shows that differential operator $\cD$ can only exist if $\cB_m(\De,j)=0$ for some $m$. Moreover, $\cD_m$ can only be only non-zero for those $m$ for which $\cB_m(\De,j)$ vanishes. We have classified pairs $(\De_\bi,j)$ for which some $\cB_m(\De_\bi,j)$ vanish in section~\ref{sec:null-states}, and so for $\cD$ to exit we must have $\De=\De_\bi$ for some $\bi$. We will henceforth label $\De,\De',j'$ and $n$ by $\bi$, i.e.\ write $\De_\bi,\De'_\bi,j'_\bi$ and $n_\bi$. Similarly, we will write $\cD_\bi$ to distinguish operators corresponding to different $\bi$.\footnote{This implicitly assumes there is a single differential operator for each $\De_\bi$. This is indeed true in odd dimensions~\cite{Penedones:2015aga}.} Let us now see how that classification compares to the possibilities~\eqref{eq:not-obviously-bad-reflections}. This will yield the values of $(\De'_\bi,j'_\bi)$ corresponding to the families identified in section~\ref{sec:null-states}.

Let us start with the first option $(\De'_\bi,j')=(3-\De_\bi,j)$. We conclude that $j'_\bi=j$ and $\De_\bi=\tfrac{3}{2}-\tfrac{n_\bi}{2}$. This corresponds to families $\mathrm{III}$ and $\mathrm{IV}$ in the classification of section~\ref{sec:two-point-function}, with $n_\bi=2k$ and $n_\bi=2k-1$ respectively. We know from that classification that in one of these families $k$ is restricted by $k\leq j$, a restriction which it is not possible to derive from the equality of Casimirs. Considering the case $(\De'_\bi,j'_\bi)= (1-j,1-\De_\bi)$ we find family $\mathrm{I}$ with $\De_\bi=1-j-n_\bi$, $j'=j+n_\bi$, and $n_\bi=k$. In the case $(\De'_\bi,j'_\bi)=(j+2, \De_\bi-2)$ we obtain family $\mathrm{II}$ with $\De_\bi=2+j-n_\bi$, $j'_\bi=j-n_\bi$, and $n_\bi=k$. The case $(\De'_\bi,j'_\bi)=(j+2, 1-\De_\bi)$ yields $\De_\bi=2+j-n_\bi$ and $j'_\bi=n_\bi-j-1$, implying that $n_\bi\geq j+1$. This case turns out not to lead to conformally-invariant differential operators. Of course, in all cases $\De'_\bi=\De_\bi+n_\bi$ by definition of $n_\bi$. We will summarize these results in subsection~\ref{sec:summary-diff-ops} below.

The above identification is actually ambiguous, because we could, for example, say that the case $(\De'_\bi,j'_\bi)= (1-j,1-\De_\bi)$ instead corresponds to family $\mathrm{II}$ with $k=2j+n_\bi+1$. To remove this ambiguity we can proceed to determine $\cD_{\bi,m}$ as below, and check that our identifications indeed give rise to $\cD_\bi$ polynomial in $p$ when \eqref{eq:generalD} is expressed in terms of $\cO(p,s)$, while the other possible identifications don't. Similarly, in this way we can prove that the case $(\De'_\bi,j'_\bi)=(j+2, 1-\De_\bi)$ should be dismissed. While this is possible to do, we will simply refer to classification of the differential operators $\cD_\bi$ in~\cite{Penedones:2015aga}, which confirms the above identifications.

Now that we have identified all cases in which $\cD_\bi$ exist, we need to find the corresponding constants $\cD_{\bi,m}$. Similarly to the two-point function, the constants $\cD_{\bi,m}$ are constrained by special conformal invariance. We perform the detailed analysis in appendix~\ref{sec:differential-ops-details}, and here give only the summary. Special conformal symmetry, applied to~\eqref{eq:generalD}, implies first, that the equation~\eqref{eq:c2equality} must be satisfied and, second, the two recursion relations,
\be
	{(1-\De_\bi-m)\sqrt{(j-m)(j+m+1)}}\cD_{\bi,m+1}&={(1-\De'_\bi-m)\sqrt{(j'_\bi-m)(j'_\bi+m+1)}}\cD_{\bi,m},\\
	{(2-\De'_\bi+m)\sqrt{(j'_\bi+m)(j'_\bi+m+1)}}\cD_{\bi,m+1}&={(2-\De_\bi+m)\sqrt{(j-m)(j+m+1)}}\cD_{\bi,m}.
\ee
These two relations are compatible iff for every $m$ such that $\cD_m\neq 0$ or $\cD_{m+1}\neq 0$ we have
\be
	(1-\De_\bi-m)(2-\De_\bi+m)(j-m)(j+m+1)=(1-\De'_\bi-m)(2-\De'_\bi+m)(j'_\bi-m)(j'_\bi+m+1).
\ee
Note that everything in the list~\eqref{eq:not-obviously-bad-reflections} is generated by affine Weyl reflections $(\De,j)\to (3-\De,j)$ and $(\De,j)\to(1-j,1-\De)$. But the product $(1-\De-m)(2-\De+m)(j-m)(j+m+1)$ is invariant under such reflections, and so the above equality is satisfied identically for all options in~\eqref{eq:not-obviously-bad-reflections} (which is also not hard to check explicitly). These recursion relations then allow us to determine $\cD_{\bi,m}$ up to normalization and a small subtlety to be discussed below.

\subsubsection{Relation between $\cD_m$ and $\cB_m(\De,j)$}
\label{sec:relation-Dm-and-BM}

Before summarizing the expressions for $\cD_{\bi,m}$ for each family, let us consider the following question. Above we have noted that the two-point function $\<\cO'\cO\>$ vanishes. Therefore, so does also the two-point function $\<\cO'\cO'\>$, since it is given by applying $\cD_\bi$ to $\<\cO'\cO\>$. In other words,
\be\label{eq:2ptzero}
	\<(\cD_\bi\cO)(\cD_\bi\cO)\>=0.
\ee
As we discussed above, $\cO$ has to have some very specific scaling dimension $\De=\De_\bi$ for $\cD_\bi$ to exist and generate the primary $\cO' = (\cD_\bi \cO)$. Let $\cO_\De$ be a generic primary of scaling dimension $\De$ and $\<\cO_\De\cO_\De\>$ be the corresponding standard two-point function. Even though the differential operator $\cD_\bi$ is only conformally-invariant when applied to $\cO=\cO_{\De_\bi}$, we can still consider the expression
\be
\<(\cD_\bi\cO_{\De})(\cD_\bi\cO_{\De})\>,
\ee
which is simply $\cD_\bi$ applied twice to $\<\cO_\De\cO_\De\>$. For generic $\De$ this is not a conformally-invariant two-point function, and at $\De=\De_\bi$ it vanishes due to~\eqref{eq:2ptzero}. However, we can consider the limit
\be
	\lim_{\De\to \De_\bi} (\De-\De_\bi)^{-1}\<(\cD_\bi\cO_{\De})(\cD_\bi\cO_{\De})\>.
\ee 
We claim that this expression is proportional to the standard two-point function of $\cO'$ (which is different from the actual two-point function of $\cO'$ computed from that of $\cO$; the latter is equal to zero). To see this we only need to check special conformal invariance, i.e. prove that
\be
\label{eq:order-after-special-conformal-transf-action}
\<(\cK_\mu\cD_\bi\cO_{\De})(\cD_\bi\cO_{\De})\>+\<(\cD_\bi\cO_{\De})(\cK_\mu\cD_\bi\cO_{\De})\>=O((\De-\De_\bi)^2),
\ee
where $\cK_\mu$ is the differential operator that implements the infinitesimal action of special conformal transformation.\footnote{The precise form of $\cK_\mu$ of course depends on what it acts on. When we write $\cK_\mu\cD_\bi\cO_{\De}$ we use the same expression for $\cK_\mu$ as if $\cD_\bi\cO_{\De}$ were a primary of dimension $\De_\bi+n_\bi$.} Using conformal invariance of $\<\cO_\De\cO_\De\>$ we can write
\be\label{eq:Kintermediate}
	\<(\cK_\mu\cD_\bi\cO_{\De})(\cD_\bi\cO_{\De})\>+\<(\cD_\bi\cO_{\De})(\cK_\mu\cD_\bi\cO_{\De})\>=
	\<([\cK_\mu,\cD_\bi]\cO_{\De})(\cD_\bi\cO_{\De})\>+\<(\cD_\bi\cO_{\De})([\cK_\mu,\cD_\bi]\cO_{\De})\>.
\ee
Statement of conformal invariance of $\cD_\bi$ at $\De=\De_\bi$ is that $[\cK_\mu,\cD_\bi]\cO_{\De}=O(\De-\De_\bi)$. But, on the other hand, we know that
\be
	\<\cO(\cD_\bi\cO)\>=0,
\ee
and thus
\be
	\<\cO_\De(\cD_\bi\cO_{\De})\>=O(\De-\De_\bi).
\ee
Therefore,
\be
\<([\cK_\mu^\De,\cD_\bi]\cO_{\De})(\cD_\bi\cO_{\De})\>=O((\De-\De_\bi)^2),
\ee
and similarly for the second term in~\eqref{eq:Kintermediate}, and we recover \eqref{eq:order-after-special-conformal-transf-action}.

Thus, the expression $(\De-\De_\bi)^{-1}\<(\cD\cO_{\De})(\cD\cO_{\De})\>$ approaches, up to a normalization, the standard two-point function $\<\cO' \cO'\>$  as $\De$ approaches $\De_\bi$. In terms of $\cD_m$ and $\cB_m(\De,j)$ this implies tha
\be
\cD_{\bi,m}^2\cB_m(\De,j)=C(\De-\De_\bi)\cB_m(\De'_\bi,j'_\bi)+O((\De-\De_\bi)^2)\,,
\ee
for some constant $C$. We will normalize $\cD_m$ so that the following relation holds
\be\label{eq:Dmnormalization}
	\cD_{\bi,m}^2\cB_m(\De,j)=(\De-\De_\bi)e^{-i\pi (j'_\bi-j)}\cB_m(\De'_\bi,j'_\bi)+O((\De-\De_\bi)^2)\,.
\ee
The factor $e^{-i\pi (j'_\bi-j)}$ is for future convenience. A form of this relation that is going to be useful for our purposes is 
\be\label{eq:residueform}
	\mathrm{res}_{\De=\De_\bi}\cB_m(\De,j)^{-1}=\cD_{\bi,m}^2 e^{i\pi (j'_\bi-j)}\cB_m(\De'_\bi,j'_\bi)^{-1}\,.
\ee

\subsubsection{Summary of differential operators}
\label{sec:summary-diff-ops}

With the normalization condition we can now finally give the solutions for the coefficients $\cD_{\bi,m}$.

\paragraph{Family $\mathrm{I}$}

Family $\mathrm{I}$ has $\De_\bi=1-j-k$ with $k\geq 1$, $j'_\bi=j+k$, $\De'_\bi=\De+k$, $n_\bi=k$. All $m$ in $\cB_m$ develop a zero, and so all $\cD_{\bi,m}$ are non-zero. We find
\be\label{eq:Dfamily1}
\cD_{\bi,m}=\sqrt{\frac{c_{\cO'}}{c_\cO}\frac{-1}{(k-1)!k!}\frac{(1+j-m)_k(1+j+m)_k}{(1+2j)_{2k}}}
\ee
These satisfy $\cD_{\bi,m}=\cD_{\bi,-m}$ so we get a parity-even operator.

\paragraph{Family $\mathrm{II}$}

Family $\mathrm{II}$ has $\De_\bi=2+j-k$ with $1\leq k\leq j$, $j'_\bi=j-k$, $\De'_\bi=\De+k$, $n_\bi=k$. Only $|m|\leq j-k=j'_\bi$ in $\cB_m$ develop a zero and so only $\cD_{\bi,m}$ with $|m|\leq j-k$ are non-zero. We find
\be
\cD_{\bi,m}=\sqrt{\frac{c_{\cO'}}{c_\cO}\frac{-1}{(k-1)!k!}\frac{(1+j-k-m)_k(1+j-k+m)_k}{(2+2j-2k)_{2k}}}
\ee
These satisfy $\cD_{\bi,m}=\cD_{\bi,-m}$ so we get a parity-even operator.

\paragraph{Family $\mathrm{III}$}

Family $\mathrm{III}$ has $\De_\bi=\tfrac{3}{2}-k$ for $k\geq 1$, $j'_\bi=j$, $\De'_\bi=\De+2k$, $n_\bi=2k$.

For bosons all $\cB_m$ develop a $0$, and so all $\cD_{\bi,m}$ are non-zero. For fermions, only $\cB_m$ with $|m|\geq k+\half$ develop a zero and so only the corresponding $\cD_{\bi,m}$ are non-zero; we furthermore have the restriction that $k\leq j-\half$.

In either case we find
\be
\cD_{\bi,m}=\sqrt{-\frac{c_{\cO'}}{c_\cO}k\frac{\half+j-k}{\half+j+k}}\frac{(\half+m-k)_{2k}}{(2k)!(\half+j-k)_{2k}}
\ee
These satisfy $\cD_{\bi,m}=\cD_{\bi,-m}$ so we get a parity-even operator. The above equation does not strictly follow recursion relations in the fermionic case, since then the $\cD_{\bi,m}$ with $m\leq -k -\half$ and with $m\geq k+\half$ are not connected by the recursion relation. (Recall that the intermediate $\cD_{\bi,m}$ are zero.) Instead we show in appendix~\ref{app:Dparity} that a spin-preserving $\cD_\bi$ of even order must be parity-even thus linking the two sets by requiring $\cD_{\bi,m}=\cD_{\bi,-m}$.

\paragraph{Family $\mathrm{IV}$ }

Family $\mathrm{IV}$ has $\De_\bi=2-k$ for $k\geq 1$, $j'_\bi=j$, $\De'_\bi=\De+2k-1$, $n_\bi=2k-1$. 

For bosons, only $\cB_m$ with $|m|\geq k$ develop a zero, so only the corresponding $\cD_{\bi,m}$ are non-zero; we also get a restriction $k\leq j$. For fermions all $\cB_m$ develop a zero, and all $\cD_{\bi,m}$ are non-zero and there is no upper bound on $k$.

In either case we find
\be
\cD_{\bi,m}=\sqrt{\frac{c_{\cO'}}{c_\cO}\frac{(1+j-k)(2k-1)}{2(j+k)}}\frac{(1-k+m)_{2k-1}}{(2k-1)!(1+j-k)_{2k-1}}
\ee
These satisfy $\cD_{\bi,m}=-\cD_{\bi,-m}$ so we get a parity-odd operator. There is again a subtlety for bosons that $\cD_{\bi,m}$ with $m\geq k$ and $m\leq-k$ are not linked together by the recursion relation. We again fix the relation $\cD_{\bi,m}=-\cD_{\bi,-m}$ by requiring that the operator be parity-odd (c.f.\ appendix~\ref{app:Dparity}).

\subsection{Derivation of the recursion relation} 
\label{sec:derivation-of-Zamolodchikov-recursion}

We are now finally in the position to derive the residue recursion relation. We start with the equation~\eqref{eq:generalcb}, which we reproduce for convenience here
\be
&\sum_{I}\<\cO_1(x_1)\cO_2(x_2)\cO_3(x_3)\cO_4(x_4)\>^{(I)} g_{\De,j,I}^{a,b}(z,\bar z)=\nn\\
&=\sum_{m=-j}^j \cB_m(\De,j)^{-1}\int_{p>0}\frac{d^3p}{(2\pi)^3}(-p^2)^{\frac{3}{2}-\De}\<0|\cO_1(x_1)\cO_2(x_2)|\cO_m(p)\>^{(a)}\<\cO_m(p)|\cO_3(x_3)\cO_4(x_4)|0\>^{(b)}.
\ee
Let us pick some pole of the conformal block, at $\De=\De_\bi$. Using~\eqref{eq:residueform} we find that
\be
	&\sum_{I}\<\cO_1(x_1)\cO_2(x_2)\cO_3(x_3)\cO_4(x_4)\>^{(I)}\mathrm{res}_{\De=\De_\bi} g_{\De,j,I}^{a,b}(z,\bar z)\nn\\
	&=\sum_{m=-j}^j \cD_{\bi,m}^2 e^{i\pi (j'_\bi-j)}\cB_m(\De'_\bi,j'_\bi)^{-1}\int_{p>0}\frac{d^3p}{(2\pi)^3}(-p^2)^{\frac{3}{2}-\De_\bi}\<0|\cO_1(x_1)\cO_2(x_2)|\cO_m(p)\>^{(a)}\nn \\ &\hspace{7cm} \times \<\cO_m(p)|\cO_3(x_3)\cO_4(x_4)|0\>^{(b)}\nn\\
	&=\sum_{m=-j}^j \cB_m(\De'_\bi,j'_\bi)^{-1}\int_{p>0}\frac{d^3p}{(2\pi)^3}(-p^2)^{\frac{3}{2}-\De'_\bi}\<0|\cO_1(x_1)\cO_2(x_2)|(\cD_\bi\cO)_m(p)\>^{(a)}\nn\\
	&\hspace{7cm}\times(-1)^{j'_\bi-j}\<(\cD_\bi\cO)_m(p)|\cO_3(x_3)\cO_4(x_4)|0\>^{(b)},\nn\\
	&=\sum_{I}\<\cO_1(x_1)\cO_2(x_2)\cO_3(x_3)\cO_4(x_4)\>^{(I)}(\cL_\bi)^a_{a'}(\cR_\bi)^b_{b'} g_{\De'_\bi,j'_\bi,I}^{a',b'}(z,\bar z).
\ee 
where we used~\eqref{eq:generalD},~\eqref{eq:generalDpastdirected} and the fact that $\De'_\bi-\De_\bi=n_\bi$. Here $\<(\cD_\bi\cO)_m(p)|=e^{i\pi j'}\<0|(\cD_\bi\cO)_{-m}(-p)$ similarly to~\eqref{eq:conjugatedefinition}. By construction, the matrices $\cL$ and $\cR$ are identified as
\be\label{eq:LRdefinition}
	(\cL_\bi)^a_{a'}\<\cO_1(x_1)\cO_2(x_2)\cO'(x)\>^{(a')}&=\<\cO_1(x_1)\cO_2(x_2)(\cD_\bi\cO)(x)\>^{(a)},\nn\\
	(\cR_\bi)^b_{b'}\<\cO_4(x_4)\cO_3(x_3)\cO'(x)\>^{(b')}&=(-1)^{j'_\bi-j}\<\cO_4(x_4)\cO_3(x_3)(\cD_\bi\cO)(x)\>^{(b)},
\ee
where on the left hand side we use the standard three-point structures for the quantum numbers $\De'_\bi,j'_\bi$ of $\cO'=\cD_\bi\cO$.

This concludes our momentum-space derivation of the residue recursion relations. This, of course, is not our main result -- while the details are new, the general idea and the equivalents of formulas~\eqref{eq:LRdefinition} have been known before~\cite{Zamolodchikov:1987,Penedones:2015aga}. In principle, we could have started this paper with formulas~\eqref{eq:LRdefinition} and then proceeded to derive the closed-form expressions for $(\cL_\bi)^a_{a'}$ and $(\cR_\bi)^b_{b'}$ as we do below. However, we would still need all the same preparatory work in sections~\ref{sec:tensor-struct}-\ref{sec:deriving-residue-matrices}, and seeing that with it the above derivation only takes a couple of lines, we have chosen to include it in order to provide a unified perspective.

\section{The residue matrices}
\label{sec:residue-matrices}

Using the momentum-space expressions for the differential operators $\cD_\bi$ in terms of the coefficients $\cD_{\bi,m}$, we can find a closed-form expression for the residue matrices $(\cL_\bi)^{a}_{a'}$ and $(\cR_\bi)^{a}_{a'}$. The idea is essentially the same one as behind the computation of shadow-transform in~\cite{Karateev:2018oml}, which we now review in the context of this paper.

Our goal is to compute the residue matrices using equation~\eqref{eq:LRdefinition}. For concreteness, let us focus on $(\cL_\bi)^a_{a'}$. First, starting with a three-point tensor structure
\be
\<\cO_1\cO_2\cO\>^{(a)}
\ee
we determine the corresponding value of
\be
\<0|\cO_1(x_1,s_1)\cO_2(x_2,s_2)\cO(x,s)|0\>^{(a)}.
\ee
Then, we can apply Fourier transform to $x$ and, using the definitions~\eqref{eq:standardmomentumoperator} and~\eqref{eq:momentumoperator} we compute 
\be
\<0|\cO_1(x_1,s_1)\cO_2(x_2,s_2)\cO_m(p)|0\>^{(a)}.
\ee
We then act on this three-point function with $\cD_\bi$ using~\eqref{eq:generalD}, and reverse the above steps to find right-hand side of~\eqref{eq:LRdefinition}. All of the above steps seem rather non-trivial, and we have to use some tricks to perform them. 

The first simplification is that in the entire course of the calculation we can set $x_1$ and $x_2$ to any values we like. Indeed, a three-point structure is completely fixed by its value for any non-coincident configuration of the three operator insertions, since any two such configurations are related to each other by a conformal transformation. In particular, we will set $x_1=0$ and $x_2=\oo$, where $\oo$ is the spatial infinity of the Minkowski space. 
The second simplification comes from our careful choice of tensor structures. Specifically, we will define a basis of three-point tensor structures, called the $\SO(3)$ basis below, in which the tensor structures take the form 
\be
\label{eq:intro-to-q}
&\<0|\cO_1(0,s_1)\cO_2(\oo,s_2)\cO(x,s)|0\>^{(a)}
=\Q^{(a)}_{\mu_1\cdots \mu_{n_a}}(s_1,s_2,s)(x^{\mu_1}\cdots x^{\mu_{n_a}}-\text{traces}) (x^2)^{-\frac{\De+\De_{12}+{n_a}}{2}},
\ee
where $\De_{12}=\De_1-\De_2$. This is valid for $x$ spacelike, for timelike $x$ an appropriate $i\e$-prescription should be used. Here $\Q^{(a)}_{\mu_1\cdots \mu_{n_a}}(s_1,s_2,s)$ are some Lorentz-invariant expressions built out of $s_1,s_2,s$, and are traceless-symmetric in $\mu_i$. This form is intuitively clear: due to Lorentz invariance, the only dependence on $x$ can be through $x^2$ or through $x_\mu$ contracted with some invariants built out of $s_i$ and $s$. Scaling invariance then allows us to fix the power of $x^2$ for each invariant. What is special about our basis is that~\eqref{eq:intro-to-q} only involves tensors $\Q$ of a single spin $n_a$ for each structure. Had we chosen some generic basis,~\eqref{eq:intro-to-q} for a fixed $a$ would in general depend on several tensors $\Q$ with different numbers of indices.

The usefulness of this choice comes from the following formula for Fourier transform
\be\label{eq:fourierintro}
&\int d^3 x e^{ipx} (x^{\mu_1}\cdots x^{\mu_{n_a}}-\text{traces})  (x^2)^{-\frac{\De+\De_{12}+{n_a}}{2}}=\nn\\
&=e^{-i \frac{\pi}2 (\De+\De_{12})}\hat\cF_{\De+\De_{12},n_a}\,(p^{\mu_{1}} \dots p^{\mu_{n_{a}}} - \text{traces}) (-p^2)^{\frac{\De+\De_{12}-{n_a}-3}{2}}\theta(p),
\ee
where coefficient $\hat \cF_{\De,j}$ will be given below in~\eqref{eq:cF-coeff-definition}. Up to the coefficient $\hat \cF_{\De,j}$, the right-hand side is fixed by symmetries: the left hand side is a traceless-symmetric tensor of fixed scaling dimension, and there is only one way to construct such a tensor from $p$. The theta-function $\theta(p)$ comes from the positivity of energy, which in this formula is reflected in the $i\e$-prescription for the integrand in left-hand side. Thanks to~\eqref{eq:fourierintro}, it is straightforward to compute the value of
\be\label{eq:momentumstruct}
\<0|\cO_1(0,s_1)\cO_2(\oo,s_2)\cO(p,s)|0\>^{(a)}
\ee
in the $\SO(3)$ basis.

Finally, note that thanks to Lorentz symmetry, the value of~\eqref{eq:momentumstruct} is completely determined by its value for $p=q_0=\hat e_0$. It is then a matter of simple manipulations, aided by our careful construction of the tensors $\Q$, to extract the value of
\be
\<0|\cO_1(0,s_1)\cO_2(\oo,s_2)\cO_m(q_0)|0\>^{(a)},
\ee
where $q_0=\hat e_0$ is the standard momentum, to which~\eqref{eq:generalD} can be applied, resulting in
\be\label{eq:preresult}
\<0|\cO_1(0,s_1)\cO_2(\oo,s_2)\cD_{\bi,m}\cO_m(q_0)|0\>^{(a)}.
\ee
Using the identity~\eqref{eq:LRdefinition}, the residue matrix $(\cL_\bi)^a_{a'}$ can be extracted by comparing~\eqref{eq:preresult} with
\be
\<0|\cO_1(0,s_1)\cO_2(\oo,s_2)\cO'_m(q_0)|0\>^{(a')},
\ee
which itself can be obtained using similar methods.

In the next subsections we implement this strategy. We start by defining the appropriate tensor structures in subsection~\ref{sec:tensor-struct}. We then derive closed-from expressions for residue matrices in subsection~\ref{sec:deriving-residue-matrices}. In subsection~\ref{sec:properties-residue-matrices} we give some comments about the result.

\subsection{Three-point tensor structures}
\label{sec:tensor-struct}

The first set of three-point tensor structures that we use is the $q$-basis defined in~\cite{Kravchuk:2016qvl},  with structures labeled by $[q_1 q_2 q]$. The structures in this basis are defined by specifying their value in the following configuration of three points (different from  above):
\be
\label{eq:conformal-frame-3pt-def}
\lim_{L\to +\oo} \<\cO_1(0,s_1)\cO_2( \hat e_2 ,s_2)\cO(L \hat e_2,s)\>^{[q_1q_2q]}L^{2\De}=[q_1q_2q]\,,
\ee
where from here on we will simplify notation to $\<\dots \,\cO(\oo \hat e_i, s) \> \equiv \lim_{L \to \infty } L^{2\De} \< \dots\, \cO(L \hat e_i, s) \> $.\footnote{Regardless of choice of $i$, the operators at $\oo\hat e_i$ are at spatial infinity of Minkowski space. Only the action of Lorentz transformations on these operators depends on the choice of $i$.} The objects $[q_1 q_2 q]$ are functions of polarizations $s_i$ and $s$ defined by 
\be
\label{eq:q-basis-3-pt-function-definition}
[q_1 q_2 q] \equiv \prod_{i=1, 2, \cO}\xi_i^{j_i+q_i}(\bar\xi_i)^{\,j_i-q_i},\qquad 	s_{i, \a} \equiv \begin{pmatrix}
	\xi_i \\ \bar \xi_i
\end{pmatrix}\,,
\ee
where it is understood that $s_\cO=s$, $j_\cO = j$ and $q_\cO=q$. In~\eqref{eq:conformal-frame-3pt-def} we have $\sum_i q_i=0$ due to little group invariance~\cite{Kravchuk:2016qvl} and $q_i$ range over $\{-j_i, \cdots j_i\}$. These structures form a complete linearly-independent basis of three-point tensor structures.  The physical three-point functions can then be written as
\be
\<\cO_1(x_1,s_1)\cO_2(x_2 ,s_2)\cO(x,s)\>_\O=\sum_{q_i}\l_{[q_1q_2q]}\<\cO_1(x_1,s_1)\cO_2(x_2 ,s_2)\cO(x,s)\>^{[q_1q_2q]}\,,
\ee
and, in particular, 
\be
\label{eq:OPE-coeff-intro}
\<\cO_1(0,s_1)\cO_2( \hat e_2 ,s_2)\cO(\oo \hat e_2,s)\>_\O=\sum_{q_i}\l_{[q_1q_2q]}[q_1q_2q]\,.
\ee
The reality properties of $\l_{[q_1q_2q]}$, as well as of the OPE coefficients in other bases to be defined below, are discussed in appendix \ref{sec:reality-of-conf-blocks-and-OPE-coeff}. Finally, here and below, the definition of tensor structure applies to general choice of operators, and not necessarily the one in~\eqref{eq:conformal-frame-3pt-def}. In particular, for the three-point structures $\<\cO_4\cO_3\cO\>^{[q_4q_3q]}$ one simply replaces $1\to 4$ and $2\to 3$ everywhere above.

Notice that the quantum numbers $q_i$ are essentially spin projections. For this reason, it will be useful to interpret the functions $[q_1q_2q]$ in terms of some functions $|j,m\>$ which transform in the standard way under $\mathfrak{su}(2)$.
We first introduce a new set of spinor polarizations, 
\be
\chi^1 = -\frac{\bar\xi+i \xi}{2}\,, \qquad \chi^2 = \frac{\xi+i \bar \xi}{2}\,.
\ee
Using $\chi^\a$ we define the functions $|j, m\>$,
\be
\label{eq:j,m-to-spinor-polarizations}
|j,m\>\equiv (-1)^{j-m}\binom{2j}{j+m}^{1/2}(\chi^1)^{j+m}(\chi^2)^{j-m}\,.
\ee
By acting with rotation generators on this definition,\footnote{The objects $|j,m\>$ should be understood as functions of polarizations $s_i,s$. The action on functions is defined in the standard way in terms of the action on $s$, i.e.\ for a function $f$ and an element $g\in\SL(2,\R)$ we have $(gf)(s)\equiv f(g^{-1}s)$.} one can check that the generators $J_\mu$, defined by
\be
J_\mu=\frac{-i}{2}\e_{\mu\nu\l}M^{\nu\l}_E\,,
\ee
where $\epsilon_{\mu \nu \rho}$ is the Levi-Civita symbol in Euclidean signature and $M^{\mu\nu}_E$ are the Euclidean rotation generators defined in appendix~\ref{sec:conventions}, act on $|j,m\>$ in the standard way. Explicitly, if we set $J_\pm=J_1\pm iJ_2$, then
\be \label{eq:standardJaction}
J_0 |j, m\> = m |j, m\>\,, \qquad J_\pm |j, m\> = \sqrt{(j\mp m)(j\pm m+1)} |j, m\pm 1\>\,.
\ee  

The functions $|j, m\>$ can be used to define a basis of three-point tensor structures that is closely related to $q$-basis. We will call this new basis the $\SO(2)$ basis. The structures in this basis are labeled by $(m_1m_2m)$ and are again defined by specifying their values in a (different) standard configuration of the three points
\be
\label{eq:SO(2)-basis-def}
\<0|\cO_1(0,s_1)\cO_2(\oo \hat e_2,s_2)\cO(\hat e_0,s)|0\>^{(m_1 m_2 m)}= |j_1,m_1;\, j_2,m_2; \,j,m\>_r\,,
\ee
where we have defined the functions\footnote{As we explain in appendix \ref{sec:reality-of-conf-blocks-and-OPE-coeff} the phase appearing in \eqref{eq:SO(2)-basis-to-states} is chosen such that the OPE coefficients associated to the three-point tensor structures defined in this section have simple reality properties.}
\be 
\label{eq:SO(2)-basis-to-states}
|j_1,m_1;\, j_2,m_2; \,j,m\>_r \equiv e^{\frac{i\pi}{2}(\De+\De_{12})}|j_1,m_1\>| j_2,m_2\>|j,m\>\,.
\ee
Similarly to the $q$-basis, the projections $m_i$ are restricted by $m_1+m_2+m=0$ due to little group invariance. The relation between the $\SO(2)$ basis and the standard conformal frame basis can be derived using \eqref{eq:j,m-to-spinor-polarizations} and analytic continuation from the space-like configuration of~\eqref{eq:q-basis-3-pt-function-definition} to the time-like configuration of~\eqref{eq:SO(2)-basis-def}. We find
\be
\label{eq:q-basis-to-SO(2)-basis}
\<\cO_1(x_1) \cO_2(x_2) \cO(x)\>^{[q_1q_2q]}= 	\<\cO_1(x_1) \cO_2(x_2) \cO(x)\>^{(q_1q_2q)}(-1)^{j_1-j+q_2}\prod_{i={1,2, \cO}} \binom{2j_i}{j_i+q_i}^{-1/2}.
\ee

We can now define the $\SO(3)$ basis with the property~\eqref{eq:intro-to-q}, which is useful for calculation of the Fourier transform. It is in this basis in which we will derive the expressions for the residue matrices. To construct a tensor structure with the property~\eqref{eq:intro-to-q} we first have to define the functions $\Q^{(a)}_{\mu_1\cdots \mu_{n_a}}(s_1,s_2,s)$. We will do so in terms of functions~\eqref{eq:SO(2)-basis-to-states}. Note that the functions $\Q$ by definition transform irreducibly under $\SL(2,\R)$ while the basis elements~\eqref{eq:SO(2)-basis-to-states} do not. We therefore start by extracting the irreducible components of~\eqref{eq:SO(2)-basis-to-states}. 
For $j_{12}\in j_1\otimes j_2$, we define 
\be\label{eq:so(2)-to-so(2.5)}
	|j_{12},m_{12};j,m\>_r &\equiv \sum_{\substack{m_1,m_2\\m_1+m_2=m_{12}}} \<j_1,m_1;j_2,m_2|j_{12},m_{12}\>|j_1,m_1;j_2,m_2;j,m\>_r\\ 
	&\equiv \sum_{\substack{m_1,m_2\\m_1+m_2=m_{12}}} \<j_1,m_1;j_2,m_2|j_{12},m_{12}\> e^{\frac{i\pi}{2}(\De+\De_{12})} |j_1,m_1\>|j_2,m_2\>|j,m\>\,,
\ee
where $\<j_1,m_1;j_2,m_2|j_{12},m_{12}\>$ are the Clebsch-Gordan coefficients.\footnote{\label{footnote:CG-convention}In what follows we will use the same convention for the Clebsch-Gordan coefficients as those in \texttt{Mathematica 11}, which are all real, with \texttt{ClebschGordan[$\{j_1, m_1\}, \{j_2, m_2\}, \{j_{12}, m_{12}\}$]} = $\<j_1,m_1;j_2,m_2|j_{12},m_{12}\>$.}${}^,$\footnote{In summations involving Clebsch-Gordan coefficient, the sums run over the values of parameters for which the Clebsch-Gordan coefficients are non-zero.}
Then, for $j_{12\cO}\in j_{12}\otimes j$ we define the functions
\be\label{eq:j12j12Odefinition}
|j_{12},j_{12\cO}\>_r&\equiv \sum_{m} \<j_{12},-m;j,m|j_{12\cO},0\>|j_{12},m_{12}; j,m\>_r \nn \\ &= \sum_{m} \<j_{12},-m;j,m|j_{12\cO},0\>e^{\frac{i\pi}{2}(\De+\De_{12})} |j_{12}, m_{12}\>|j, m\>\,.
\ee
Note that we omit the spin projection label one the left hand side because due to the constraint $m_1+m_2+m=0$ it is always equal to $0$. We now identify the label $(a)$ of $\Q$ to be $(a)=(j_{12},j_{12\cO})$, and we define $\Q$ by requiring that
\be\label{eq:Qdefinition}
\Q^{(j_{12},j_{12\cO})}_{0\cdots 0}(s_1,s_2,s)=|j_{12},j_{12\cO}\>_r,
\ee
and that $\Q^{(j_{12},j_{12\cO})}$ is a Lorentz-invariant function of $s_1,s_2$ and $s$ with $n_a=n_{j_{12},j_{12\cO}}=j_{12\cO}$ traceless-symmetric indices.\footnote{Specifying $\Q$ for all $\mu_i=0$ is sufficient to determine it for general $\mu_i$. Indeed, this value is sufficient to compute $\Q^{(j_{12},j_{12\cO})}_{\mu_1\cdots\mu_{j_{12\cO}}}(s_1,s_2,s) x^{\mu_1}\cdots x^{\mu_{j_{12\cO}}}$ for $x\propto\hat e_0$. Then the values for general timelike $x$ are determined by Lorentz invariance, and $\Q$ can be extracted by differentiating with respect to $x$.} The $\SO(3)$ basis structures can then be defined by requiring~\eqref{eq:intro-to-q} to hold, or more directly by demanding
\be
\label{eq:SO(3)-basis-definition}
\<0|\cO_1(0,s_1)\cO_2(\oo \hat e_2,s_2)\cO(\hat e_0,s)|0\>^{(j_{12},j_{12\cO})}= |j_{12},j_{12\cO}\>_r\,.
\ee
Since the functions $|j_{12},j_{12\cO}\>$ with 
\be
	j_{12}&\in j_1\otimes j_2=|j_1-j_2|\oplus \cdots\oplus (j_1+j_2),\nn\\
	j_{12\cO}&\in j_{12}\otimes j=|j_{12}-j|\oplus \cdots\oplus (j_{12}+j),
\ee
span the same space of functions as~\eqref{eq:SO(2)-basis-to-states}, and the latter span the same space as $[q_1q_2q]$, it follows that the $\SO(3)$ basis is a complete and linearly-independent basis of three-point structures. As can be checked by using the relation of $\SO(3)$ basis to $q$-basis and the results of~\cite{Kravchuk:2016qvl}, the $\SO(3)$ structures have definite space parity (see appendix~\ref{app:spaceparity}) given by
\be
(-1)^{j_1-j_2+j-j_{12\cO}}.
\ee

\subsection{Deriving the residue matrices}
\label{sec:deriving-residue-matrices}

We now proceed to derive the residue matrices following the algorithm described in the beginning of this section. The first step is to compute the Fourier-transformed structure
\be
	\<0|\cO_1(0,s_1)\cO_2(\oo \hat e_2,s_2)\cO(p,s)|0\>^{(j_{12},j_{12\cO})}.
\ee
By construction, we have
\be
\<0|\cO_1(0,s_1)&\cO_2(\oo \hat e_2,s_2)\cO(x,s)|0\>^{(j_{12},j_{12\cO})} = \mathbb Q_{\mu_1, \dots, \mu_{j_{12\cO}}}^{(j_{12},j_{12\cO})}(s_1, s_2, s) \nn \\ &\times (x^{\mu_{1}} \dots x^{\mu_{j_{12\cO}}} - \text{traces}) x^{-\De - \De_{12} - j_{12\cO}}\,.
\ee 
To define the right hand side for timelike $x$ the $i\e$ prescription $x^0\to x^0+i\e$ has to be used. As discussed around~\eqref{eq:fourierintro}, the Fourier transform of the right-hand side is easy to compute. In particular, we find
\be 
\label{eq:F-transform-3pt-func}
\<0|\cO_1(0,s_1)&\cO_2(\oo \hat e_2,s_2)\cO(p,s)|0\>^{(j_{12},j_{12\cO})} = e^{-i \frac{\pi}2 (\De+\De_{12})}\hat\cF_{\De+\De_{12},j_{12\cO}} \mathbb Q^{(j_{12},j_{12\cO})}_{\mu_1, \dots, \mu_{j_{12\cO}}}(s_1, s_2, s) \nn \\ &\times (p^{\mu_{1}} \dots p^{\mu_{j_{12\cO}}} - \text{traces}) (-p^2)^{\frac{ \De+\De_{12}   - j_{12\cO} - 3}2}\theta(p^0)\,,
\ee 
where the coefficient $\hat\cF_{\De,J}$ is given by
\be
\label{eq:cF-coeff-definition}
\hat \cF_{\De,J}\equiv \frac{\pi^{{5}/{2}} 2^{4-\De}}{\Gamma(\frac{\De+J}{2})\Gamma(\frac{\De-1-J}{2})}\,.
\ee 
Choosing $p = q_0= \hat e_0$ and using the definition~\eqref{eq:Qdefinition} of $\Q^{(j_{12},j_{12\cO})}_{\mu_1, \dots, \mu_{j_{12\cO}}}(s_1, s_2, s)$ we find 
\be
\<0|\cO_1(0,s_1)\cO_2(\oo \hat e_2,s_2)\cO(q_0,s)|0\>^{(j_{12},j_{12\cO})}  = e^{-i \frac{\pi}2 (\De+\De_{12})}\hat\cF_{\De+\De_{12},j_{12\cO}}|j_{12},j_{12\cO}\>\,.
\ee
In order to determine the action of the differential operators on $\cO$ we would like to compute the value of
\be
\<0|\cO_1(0,s_1)\cO_2(\oo \hat e_2,s_2)\cO_m(p)|0\>^{(j_{12},j_{12\cO})}.
\ee
For this we first use~\eqref{eq:j12j12Odefinition} to write
\be
\label{eq:interm-basis-calc}
\<0|\cO_1(0,s_1)\cO_2(\oo,s_2)\cO(q_0,s)|0\>^{(j_{12},j_{12\cO})}  &= \sum_{m}\<j_{12},-m;j, m|j_{12\cO},0\>\nn \\ & \times \hat\cF_{\De+\De_{12},j_{12\cO}}\,  |j_{12},-m\>|j,m\>\,.
\ee
Comparing this with the definition~\eqref{eq:standardmomentumoperator} of $\cO_m(q_0)$, we find
\be
\<0|\cO_1(0,s_1)\cO_2(\oo,s_2)\cO_m(q_0)|0\>^{(j_{12},j_{12\cO})} =  & \hat\cF_{\De+\De_{12},j_{12\cO}}\<j_{12},-m;j, m|j_{12\cO},0\>|j_{12},-m\>\,.
\ee
As discussed in section~\ref{sec:differential-ops}, when acting on the Fourier transformed operators $\cO_m(q_0)$, the differential operators $\cD_\bi$ are diagonalized. In particular, using~\eqref{eq:generalD} we find
\be 
\label{eq:res-matr-deriv-after-applying-D}
&\<0|\cO_1(0,s_1)\cO_2(\oo,s_2) (\cD_\bi \cO)_m(q_0)|0\>^{(j_{12},j_{12\cO})}=\nn\\&\qquad \cD_{\bi, m} \hat\cF_{\De_\bi+\De_{12},j_{12\cO}}\,\<j_{12},-m;j,m|j_{12\cO},0\>  |j_{12},-m\>\,,
\ee
where we have specialized to $\De=\De_\bi$ so that the action of $\cD_\bi$ is conformally-invariant. Finally, repeating the above states in reverse, now with $\cO'=\cD_\bi\cO$ instead of $\cO$, we find 
\be
\label{eq:res-matr-deriv-final-step}
\<0|&\cO_1(0,s_1)\cO_2(\oo,s_2)(\cD_\bi \cO)(x=\hat e_0,s)|0\>^{(j_{12},j_{12\cO})}  = \nn \\ &=\sum_{m,j'_{12\cO}} \hat\cF_{\De'_\bi+\De_{12},j'_{12\cO}}^{-1} \<j_{12\cO}',0|j_{12},-m;j'_\bi,m\> \cD_{\bi, m} \<j_{12},-m;j,m|j_{12\cO},0\>  \hat\cF_{\De_\bi+\De_{12},j_{12\cO}}|j_{12},j'_{12\cO}\>\,,
\ee
where the sums over $m$ and $j_{12\cO}'$ are constrained by the selection rules of Clebsch-Gordan coefficients and $\Delta'_\bi$ and $j'_\bi$ are the scaling dimension and spin of $\cO'=(\cD_\bi \cO)$. Comparing this with~\eqref{eq:SO(3)-basis-definition}, we find
\be
\<\cO_1(x_1)\cO_2(x_2 )(\cD_\bi\cO)(x)\>^{(j_{12},j_{12\cO})} =\sum_{j_{12\cO}'} ( \cM_\bi^{\De_{12}})^{(j_{12},j_{12\cO})}_{(j_{12},j'_{12\cO})}\<\cO_1(x_1)\cO_2(x_2 )\cO(x)\>^{(j_{12},j'_{12\cO})}\,,
\ee
where,
\be
\label{eq:value-residue-matrix-M}
&( \cM_\bi^{\De_{12}})^{(j_{12},j_{12\cO})}_{(j_{12},j'_{12\cO})}=\nn\\
&=\hat\cF_{\De'_\bi+\De_{12},j'_{12\cO}}^{-1}\hat\cF_{\De_\bi+\De_{12},j_{12\cO}}\sum_{m}  \<j_{12\cO}',0|j_{12},-m;j'_\bi,m\> \cD_{\bi, m}  \<j_{12},-m;j,m|j_{12\cO},0\>  \,.
\ee
The corresponding result for $\<  \cO_4  \cO_3 \cO\>$ can be obtained by simply replacing $1\to4$ and $2\to 3$ everywhere above. Therefore, using \eqref{eq:LRdefinition}, we find that the left- and right- residue matrices are given by, 
\be
\label{eq:residue-matrix-L-and-R}
(\cL_\bi)^{(j_{12},j_{12\cO})}_{(j_{12},j'_{12\cO})}=(\cM_\bi^{\De_{12}})^{(j_{12},j_{12\cO})}_{(j_{12},j'_{12\cO})}, \qquad 
(\cR_\bi)^{(j_{43},j_{43\cO})}_{(j_{43},j'_{43\cO})}=(-1)^{j'_\bi-j}(\cM_\bi^{\De_{43}})^{(j_{43},j_{43\cO})}_{(j_{43},j'_{43\cO})}\,.
\ee
The equations~\eqref{eq:value-residue-matrix-M} and~\eqref{eq:residue-matrix-L-and-R}, together with the computation of $h_{\oo,j,I}^{a,b}$ in section~\ref{sec:leading-term-h}, represent the main result of this paper. Before moving on to compute $h_{\oo,j,I}^{a,b}$, let us briefly discuss some of the key features of~\eqref{eq:value-residue-matrix-M}.

\subsection{Properties of residue matrices}
\label{sec:properties-residue-matrices}

The first comment relates to the reality of the matrices $\cL_\bi$ and $\cR_\bi$. In~\eqref{eq:value-residue-matrix-M} all ingredients are real except possibly for the coefficients $\cD_{\bi,m}$. Using the explicit expressions in section~\ref{sec:summary-diff-ops} one can check that these coefficients are either real or pure imaginary. Therefore, the matrices $\cL_\bi$ and $\cR_\bi$ are either both real or both purely imaginary, so that the product $\cL_\bi \cR_\bi$ is always real.

These matrices are also block-diagonal: they do not change $j_{12}$ or $j_{43}$. This represents an advantage for a numerical implementation of the residue recursion relations. For example, in stress-tensor four-point function, $j_{12}$ takes 5 possible values ranging from $0$ to $4$. For generic intermediate spin $j$, these give blocks in $\cL$ of sizes $(2j_{12}+1)\times (2j_{12}+1)$. These blocks have total 165 elements, as opposed to 625 if we had a dense matrix. This should furthermore reduce to roughly half of that, $\sim 80$, due to space parity selection rules.

The expression~\eqref{eq:value-residue-matrix-M} involves a sum over $m$. This sum ranges over values of $m$ allowed by selection rules of the Clebsch-Gordan coefficients. This is 
\be
	m\in \{-\min(j_{12},j,j'_\bi),\cdots,\min(j_{12},j,j'_\bi)\}\subseteq \{-j_{12},\cdots, j_{12}\}.
\ee
We see that always $|m|\leq j_{12}\leq j_1+j_2$. This implies two things. 

First, for any fixed four-point function, this sum involves only a few terms. For example, for a stress-tensor four-point function, this sum involves at most 9 terms. Furthermore, for fixed $j_{12}$, $j-j_{12\cO}$, $j'_\bi-j'_{12\cO}$ and $m$, the Clebsch-Gordan coefficients in~\eqref{eq:value-residue-matrix-M} can be written as closed-form functions of $j$ and $j'_\bi$.

Second, if $\cD_{\bi,m}$ happens to vanish for all $m$ in the range of summation, the pole $\bi$ does not appear in the conformal block. Since the above range always includes $m=0$ or $m=\half$, it follows that this never happens for families $\mathrm{I}$ or $\mathrm{II}$. Similarly, this doesn't happen for bosonic family $\mathrm{III}$ poles or fermionic family $\mathrm{IV}$ poles. However, for bosonic family $\mathrm{IV}$ or fermionic family $\mathrm{III}$, the coefficients $\cD_{\bi,m}$ vanish for $|m|<k$, and so these families do not contribute for $k > j_{12}$, regardless of the value of $j$. Taking into account also the $\cR_\bi$ matrix, we find the selection rule for these families that $k\leq \min(j_{12},j_{34})\leq \min(j_1+j_2,j_3+j_4)$. This explains why there are no family $\mathrm{IV}$ poles in scalar blocks or family $\mathrm{III}$ poles in fermion-scalar blocks.

Our final comment concerns a simplification of~\eqref{eq:value-residue-matrix-M} for family $\mathrm{I}$ and $\mathrm{II}$ poles. For this note that in momentum space the differential operator $\cD_\bi$ acting on $\cO(p,s)$ has to be a Lorentz-invariant expression that changes spin of $\cO$ from $j$ to $j'_\bi$ using $n_\bi$ powers of momentum $p$. If $j'_\bi=j\pm n_\bi$, there is a unique way to do so. In other words, there is a unique copy of $j'_\bi$ in the tensor product of $j$ with $n_\bi$ vector irreps. The condition $j'_\bi=j\pm n_\bi$ is satisfied for families $\mathrm{I}$ and $\mathrm{II}$. For concreteness, let us focus on family $\mathrm{I}$. Using this logic, one can argue that the coefficients $\cD_{\bi,m}$ have to be proportional to
\be
\cD_{\bi,m}\propto \<j+k,m|j,m;k,0\>=\frac{2^k}{\pi^{\frac{1}{4}}}\sqrt{\frac{
		(2j)!(k-\thalf)!(j-m+k)!(j+k-m)!
	}{
		(2(j+k)!)k!(j-m)!(j+m)!
	}}.
\ee
We indeed find
\be
	\cD_{\bi,m}=\sqrt{\frac{c_{\cO'}}{c_\cO}\frac{-1}{2(2k-1)!}}\<j+k,m|j,m;k,0\>.
\ee
The right-hand side of~\eqref{eq:value-residue-matrix-M} then involves the following sum
\be
	&\sum_{m}  \<j_{12\cO}',0|j_{12},-m;j+k,m\> \<j+k,m|j,m;k,0\>  \<j_{12},-m;j,m|j_{12\cO},0\>=\nn\\
	&=(-1)^{j-j_{12}+k}\sqrt{(2j_{12\cO}+1)(2j_{12\cO}'+1)(2(j+k)+1)}
	\left\{
	\begin{matrix}
		j_{12\cO} & k & j_{12\cO}'\\
		j+k & j_{12} & j
	\end{matrix}
	\right\} 	\left(
	\begin{matrix}
		k & j_{12\cO}' & j_{12\cO}\\
		0 & 0 &0
	\end{matrix}
	\right),
\ee
where $\{\cdots\}$ is the $SU(2)$ 6-j symbol and $(\cdots)$ is the  $SU(2)$ 3-j symbol.\footnote{Our convention for the 3-j and 6-j symbols are the same as in \texttt{Mathematica 11} using the commands $\texttt{ThreeJSymbol}$ and  $\texttt{SixJSymbol}$ respectively.} This specific 6-j symbol has the property that one of its arguments ($j+k$) is the sum of two others ($j$ and $k$). In this case the 6-j symbol can be expressed in terms of square root of products of factorials of the arguments~\cite{VarshalovichAngularMomentum}. Therefore, we can express $\cM_\bi$ for family $\mathrm{I}$ as a square root of a product of simple factors. The same applies to family $\mathrm{II}$, where we encounter the 6-j symbol
\be
	\left\{
\begin{matrix}
	j_{12\cO} & k & j_{12\cO}'\\
	j-k & j_{12} & j
\end{matrix}
\right\},
\ee
in which case now $j$ is the sum of $j-k$ and $k$. We couldn't find a similar simplification for families $\mathrm{III}$ and $\mathrm{IV}$. In particular, evaluating the elements of $\cM_\bi$ with $\bi$ in these families in explicit examples, we see that, unlike in the case of families $\mathrm{I}$ and $\mathrm{II}$, it is not equal to a square root of a product of simple factors. Furthermore, the coefficients $\cD_{\bi,m}$ do not appear to be related to Clebsch-Gordan coefficients in a simple way.

\section{The leading term of the conformal block at $\Delta\to \infty$}
\label{sec:leading-term-h}

In this section we derive the function $h_{\oo, j, I}^{a, b}(z, \bar z)$, the last ingredient that we need in order to write down a closed-form expression for the residue recursion relation for general spinning blocks. To obtain $h_{\oo, j, I}^{a, b}(z, \bar z)$ we study the Casimir differential equations for conformal blocks~\cite{DO1,Kos:2013tga, Kos:2014bka, Kravchuk:2017dzd}. In order to obtain this equation, we apply the quadratic conformal Casimir operator $\hat C_2$ to the first two operators, $\cO_1$ and $\cO_2$, in the four-point function $\<\cO_1(x_1)\cO_2(x_2)\cO_3(x_3)\cO_4(x_4)\>_\O$. Considering the contribution of a single conformal block to this four-point function, we can use \eqref{eq:generalcb}, together with the conformal invariance of the three-point function $\<\cO_1 \cO_2 \cO\>$ to move the action of $\hat C_2$ from $\cO_1$ and $\cO_2$ to the exchanged operator, $\cO$. Since $\hat C_2$ acts on a single operator by multiplication, we find that the conformal blocks have to satisfy the eigenvalue equation  
\be
\label{eq:Casimir-equation}
(\hat C_2 - C_{\De, j})\left[\<\cO_1(x_1)\cO_2(x_2) \cO_3(x_3) \cO_4(x_4)\>^{(I)} g_{\De, j, I}^{a, b}(z, \bar z)\right] = 0\,,
\ee
where
\be
C_{\De, j}= \De(\De-3)+j(j+1)\,.
\ee
The action of the Casimir on the contribution of a conformal block to the four-point function can be rewritten as,   
\be 
\hat C_2 &\left[\<\cO_1(x_1)\cO_2(x_2) \cO_3(x_3) \cO_4(x_4)\>^{(I)} g_{\De, j, I}^{a, b}(z, \bar z)\right] = \nn \\ &\qquad \qquad =\<\cO_1(x_1)\cO_2(x_2) \cO_3(x_3) \cO_4(x_4)\>^{(I)} (C_2)_I^J  g_{\De, j, J}^{a, b}(z, \bar z)
\ee 
where $ (C_2)_I^J$ is a matrix of second order differential operators in $z$ and $\bar z$ which we will explicitly write below. Thus, the Casimir equation \eqref{eq:Casimir-equation} can be rewritten as a set of second order differential equations that the conformal blocks $g_{\De, j, I}^{a, b}(z, \bar z)$ have to satisfy,
\be
\label{eq:Casimir-eq-specific}
  (C_2)^J_I  g_{\De, j, J}^{a, b}(z, \bar z)= C_{\De, j} g_{\De, j, I}^{a, b}(z, \bar z)
\ee
The number of differential equations is equal to the number of four-point tensor structures $I$ and each equation generically involves the contribution of conformal blocks $g_{\De, j, I}^{a, b}(z, \bar z)$ from multiple four-point tensor structures.  By studying the large $\De$ limit of this (generally, large) system of differential equations one can in principle determine $h_{\oo, j, I}^{a, b}(z, \bar z)$~\cite{Kos:2013tga, Kos:2014bka, Kravchuk:2017dzd}.

Once again, in order to solve this rather non-trivial problem we have employed several tricks. While so far, everything in our derivation has been independent of the exact basis of the four-point tensor structures, $h_{\oo, j, I}^{a, b}(z, \bar z)$ depends explicitly on this choice of basis. By carefully choosing the basis of four-point tensor structures we will be able to rewrite the Casimir equation as a system of differential equations for which all derivative terms act diagonally on the tensor structures $I$. By using this basis we will thus be able to solve for $\log\,g_{\De, j, I}^{a, b}(z, \bar z)$ to the first two orders in $\De$ and determine $h_{\oo, j, I}^{a, b}(z, \bar z)$.  Therefore, before discussing the derivation of these differential equations we first define the appropriate basis of tensor structures in the subsection below.

\subsection{Four-point tensor structures}
\label{sec:four-point-tensor-struct}

To define four-point tensor structures we use the  conformal frame basis~\cite{Kravchuk:2016qvl}, which we now review. By using a conformal mapping, we can bring any four points $x_i$  into the configuration $y_i = x_i$:
\be
\label{eq:4-pt-coordinate-config-conformal-frame}
y_1&=(0,0,0)\,, \qquad y_2&=(\tfrac{\bar z-z}{2},\tfrac{z+\bar z}{2},0)\,, \qquad 
y_3&=(0,1,0)\,, \qquad 
y_4&=(0,\oo,0)\,.
\ee
The corresponding value of the four-point function can be expanded as
\be
\label{eq:4-pt-function-g(z,zb)}
 \<\cO_1(x_1,s_1)\cO_2(x_2,s_2)\cO_3(x_3,s_3)\cO_4(x_4,s_4)\>=\sum_{q_i = -j_i}^{j_i} g_{[q_1q_2q_3q_4]}(z,\bar z)[q_1q_2q_3q_4]\,.
\ee
where we use the convention described below \eqref{eq:q-basis-3-pt-function-definition} for the point at infinity.\footnote{The reality properties of the functions $g_{[q_1q_2q_3q_4]}(z,\bar z)$ are discussed in appendix \ref{sec:reality-of-conf-blocks-and-OPE-coeff}. }  The functions $[q_1q_2q_3q_4]$ of polarizations $s_i$ are defined similarly to the functions $[q_1q_2 q_3]$ appearing in the three-point tensor structure basis \eqref{eq:q-basis-3-pt-function-definition},   
\be
\label{eq:q-struct-4pt-definition}
[q_1 q_2 q_3 q_4]\equiv\prod_{i=1}^4\xi_i^{\ell_i+q_i}\bar\xi_i^{\ell_i-q_i}.
\ee
Correspondingly, we define the basis of four-point tensor structures labeled by $[q_1q_2q_3 q_4]$ by requiring that they take the following values in the standard configuration~\eqref{eq:4-pt-coordinate-config-conformal-frame},
\be
\label{eq:q-struct-4-pt-struct-placement}
\<\cO_1(x_1,s_1)\cO_2(x_2,s_2)\cO_3(x_3,s_3)\cO_4(x_4,s_4)\>^{[q_1q_2q_3q_4]} =  [q_1q_2q_3q_4]\,.
\ee
The structures~\eqref{eq:q-struct-4-pt-struct-placement} with $q_i\in\{-j_i,\cdots,j_i\}$ form a complete basis of four-point tensor structures~\cite{Kravchuk:2016qvl}.

It is important to note that the four-point tensor structures~\eqref{eq:q-struct-4-pt-struct-placement} form a canonical basis, in the following sense. Given any generic four points $x_i$ in $\R^3$, one can draw a unique two-dimensional sphere $S$ through them.  Since conformal transformations map spheres to spheres, this sphere is naturally associated to the points $x_i$. For each $i$ there is a unique rotation generator $R_i$ that preserves the tangent space $T_{x_i}S$ to the sphere at $x_i$. It is then natural to decompose the operators $\cO_i$ into spin components that are diagonal under rotations $R_i$. The tensor structures~\eqref{eq:q-struct-4-pt-struct-placement} perform precisely such a decomposition (wick-rotated to Lorentzian signature). Indeed, since the decomposition is natural, we can compute it in particular in the configuration~\eqref{eq:4-pt-coordinate-config-conformal-frame}. In this configuration the sphere degenerates into the plane $01$, all $R_i$ are equal to $M_{01}$, and $M_{01}$ acts diagonally on~\eqref{eq:q-struct-4pt-definition}. This fact implies that the functions $g_{[q_1q_2q_3 q_4]}(z, \bar z)$ behave nicely under an appropriate action of the two-dimensional conformal group, and this ultimately leads to the simplifications in the Casimir equation that we will discuss momentarily.

With the above definitions for the four-point tensor structures, we are now prepared to rewrite the Casimir equation \eqref{eq:Casimir-equation} as a system of second-order differential equations for $g_{[q_1q_2q_3 q_4]}(z, \bar z)$.
\subsection{Casimir equation}

To start, it proves convenient to define 
\be
\label{eq:tilde-g-definition}
 	\tilde g_{\De, j, [q_1q_2q_3q_4]}^{a, b}(z,\bar z)\equiv z^{h_1+h_2}\bar z^{\bar h_1+\bar h_2}g_{\De, j, [q_1q_2q_3q_4]}^{a, b}(z,\bar z)\,,
\ee
and 
\be
	h_i\equiv \frac{\De_i-q_i}{2},\quad \bar h_i\equiv \frac{\De_i+q_i}{2}\,,	\quad \a \equiv h_2-h_1,\quad \bar \a \equiv \bar h_2-\bar h_1, \quad \b \equiv h_3-h_4,\quad \bar \b\equiv \bar h_3-\bar h_4\,,
\ee
where we note that $\a$, $\bar \a$, $\b$ and $\bar \b$ depend on the external scaling dimensions of the four-point function via the scaling dimension differences $\De_{12}$ and $\De_{43}$. In this basis, the Casimir equation  \eqref{eq:Casimir-eq-specific} can be rewritten as a set of second-order differential equations on $\tilde g_{\De, j, [q_1q_2q_3q_4]}^{a, b}(z,\bar z)$, 
\be
\label{eq:Casimir-eq-explicit}
(\widetilde C_2)^{[p_1p_2p_3p_4]}_{[q_1 q_2 q_3 q_4]} 	\tilde g_{\De, j, [p_1p_2p_3p_4]}^{a, b}(z,\bar z) = C_{\De, j} \tilde g_{\De, j, [q_1q_2q_3q_4]}^{a, b}(z,\bar z)  \,,
\ee
where $(\widetilde C_2)^{[p_1p_2p_3p_4]}_{[q_1 q_2 q_3 q_4]}$ is a matrix of second-order differential operators, related to $(C_2)^{[p_1p_2p_3p_4]}_{[q_1 q_2 q_3 q_4]}$ by accounting for the different powers of $z$ and $\bar z$ in \eqref{eq:tilde-g-definition}, and summation over the allowed values of $p_i$ is understood. The differential operator $\widetilde C_2$ can be explicitly written as 
\be
\label{eq:C2-differential-equation-from-expanded}
	(\widetilde C_2)^{[p_1p_2p_3p_4]}_{[q_1q_2q_3q_4]}&=\left(D_2+D_1\right)\delta^{[p_1p_2p_3p_4]}_{[q_1q_2q_3q_4]}+(\text{non-derivative terms})\,,
\ee
where we have only written explicitly the terms that are second order ($D_2$) and first order ($D_1$) in derivatives. We emphasize that these terms are diagonal in the basis of four-point tensor structures $[q_1q_2q_3q_4]$. 

Explicitly, the differential operator $D_2$ is give by
\be
	D_2=2z^2(1-z)\ptl_z^2+2\bar z^2(1-\bar z)\ptl_{\bar z}^2\,,
\ee
while $D_1$ is given by
\be
	D_1=-2(1+\a+\b)z^2\ptl_z-2(1+\bar \a+\bar \b)\bar z^2\ptl_{\bar z} +\frac{2z\bar z}{z-\bar z}\p{(1-z)\ptl_z-(1-\bar z)\ptl_{\bar z}}\,.
\ee   
Using \eqref{eq:Casimir-eq-explicit} together with \eqref{eq:C2-differential-equation-from-expanded} we find an eigenvalue equation for $\tilde g_{ \De, j, [q_1q_2q_3q_4]}^{a, b}(z, \bar z)$ for which all derivative terms are diagonal in the four-point tensor structures, while non-derivative terms are not necessarily diagonal. 

The derivative terms are, however, sufficient to fix the first two orders in the large-$\De$ expansion of the block, which are needed in order to derive $h_{\oo, j, I}^{a, b}(z, \bar z)$.  To see this  we consider the ansatz
\be
\label{eq:g(z,bz)-ansatz}
\tilde g_{\De, j, [q_1 q_2 q_3 q_4]}^{a, b}(z,\bar z)=\exp\left[{\De f_{(0), j, [q_1 q_2 q_3 q_4]}^{a, b}(z,\bar z)+f_{(1),  j, [q_1 q_2 q_3 q_4]}^{a, b}(z,\bar z)+O(\De^{-1})}\right]\,.
\ee
and we note that when applying the differential operators from \eqref{eq:C2-differential-equation-from-expanded} only $D_2$ and $D_1$ contribute to order $\De^2$ and $\De$. Using   \eqref{eq:Casimir-eq-explicit} we thus find a set of diagonal differential equations which we can solve for $ f_{(0), j, [q_1 q_2 q_3 q_4]}^{a, b}(z,\bar z)$ and $f_{(1),  j, [q_1 q_2 q_3 q_4]}^{a, b}(z,\bar z)$. Since we are working with a set of differential equations diagonal in all tensor structures, in what follows we will drop the $q_i$,~$\De$,~$j$ subscripts and the $a,\, b$ superscripts, keeping all structures fixed. 

At order $\De^2$ we get the equation
\be
	2z^2(1-z)(\ptl_z f_0(z,\bar z))^2+2\bar z^2(1-\bar z)(\ptl_{\bar z} f_0(z,\bar z))^2=1\,.
\ee
Using a change of variables from $z$ and $\bar z$ to $\rho$ and $\bar \rho$, defined in \eqref{eq:rho-brho-defintion},  we get
\be\label{eq:f0equation}
	\p{\frac{\ptl f_0}{\ptl\log \r}}^2+\p{\frac{\ptl f_0}{\ptl\log \bar\r}}^2=\frac{1}{2}\,.
\ee
The relevant solution can be fixed by considering the OPE limit $\r,\bar \r\to 0$,\footnote{One could also fix the coefficients of \eqref{eq:f0-solution} up to relative $\pm$-signs by studying the leading contribution for large $\De$ solution of quartic conformal Casimir.   }
\be
\label{eq:f0-solution}
	f_0=\thalf \log\r+\thalf\log \bar\r+\log a_0,
\ee
where $a_0$ enters the normalization of the two-point function~\eqref{eq:b-j-normalization}.

Similarly, by looking at the first order in $\De$, it is straightforward determine the equation for $f_1$, which we will not explicitly show here due to the large number of terms. The solution of this equation is given by
\be
\label{eq:f1-solution}
	f_1=-\thalf \log(1-\r\bar\r)&-(\thalf- \a- \b)\log(1+\r)-(\thalf-\bar \a-\bar \b)\log(1+\bar\r)\nn\\
	&-(\thalf+\a+\b)\log(1-\r)-(\thalf+\bar \a+\bar \b)\log(1-\bar \r)+\log f(\theta)\,,
\ee
where $f(\theta)$ is a yet undetermined function of $\theta$, with $\cos\theta\equiv \frac{\r+\bar\r}{2\sqrt{\r\bar\r}}$. Thus, using the ansatz \eqref{eq:g(z,bz)-ansatz} together with the solutions \eqref{eq:f0-solution} and \eqref{eq:f1-solution}, we find that the large-$\De$ asymptotic of $\tilde g_{\De, j, [q_1q_2q_3q_4]}^{a, b}(\rho, \bar \r)$ is given by
\be
\label{eq:hat-g-solution-r-rb}
	\tilde g_{\De, j, [q_1q_2q_3q_4]}^{a, b}(\r, \bar \r)&\sim \frac{ a_0^\De(\r\bar\r)^{\frac{\De}{2}}(1-\r\bar\r)^{-\half}}{
		(1+\r)^{\half-\a-\b}
		(1+\bar\r)^{\half-\bar \a-\bar \b}
		(1-\r)^{\half+\a+\b}
		(1-\bar\r)^{\half+\bar \a+\bar \b}
	}f^{a,b}_{j, [q_1q_2q_3q_4]}(\theta)\,,
\ee
where we have reintroduced the four-point and three-point tensor structure subscripts and superscripts.\footnote{This can be compared to the standard result for the scalar case obtained in \cite{DO1}
	\be
	\label{eq:scalar-conformal-block-inf}
	\tilde g^\text{Scalar}_{\De, j}(z, \bar z)
	&=\frac{a_0^\De(\r\bar\r)^{\frac{\De}{2}}(1-\r\bar\r)^{-\half}}{
		(1+\r)^{\half-\a-\b}
		(1+\bar\r)^{\half- \a- \b}
		(1-\r)^{\half+\a+\b}
		(1-\bar\r)^{\half+ \a+\b}
	}f_j(\theta)\,,
	\ee
	with $\a = \De_{21}/2$ and $\b = \De_{34}/2$.
	We see that the only generalization is to replace $\a\to\bar \a$ and $\b\to \bar \b$ for the terms involving $\bar \rho$, where for external spinning operators $\a$ and $\b$ take into account the 2D spins $q_i$.}

What remains is to determine the function $f^{a,b}_{j, [q_1q_2q_3q_4]}(\theta)$. From the residue recursion relation \eqref{eq:recursionintro} we note that the leading-$r$ (with $r^2 = \rho \bar \rho$) behavior of the conformal block is captured by $h_{\oo, j, [q_1q_2q_3q_4]}^{a, b}(z, \bar z)$. Thus, we can determine $f^{a,b}_{j, [q_1q_2q_3q_4]}(\theta)$ by looking at the leading $r$ term of the conformal block and reading-off the angular dependence. 
The leading-$r$ terms of general conformal blocks were recently studied in~\cite{Kravchuk:2017dzd}. We derive $f(\theta)$ in appendix~\ref{sec:deriving-f} using results of~\cite{Kravchuk:2017dzd}. In this section we only state the result for two different bases of three-point tensor structures. We find that when using the conformal frame basis introduced in section \ref{sec:tensor-struct} for the three-point tensor structures, $a = [p_1 p_2 p]$ and $b = [ p_4 p_3p']$, $f_{ j,[q_1 q_2 q_3 q_4]}^{[p_1 p_2p],\,[p_4 p_3p']}(\theta)$ is given by, 
\be 
\label{eq:f-th-q-basis}
f_{ j,[q_1 q_2 q_3 q_4]}^{[p_1 p_2p],\,[p_4 p_3p']}(\th) & = b_j^{-1} (w_{j_1})_{q_1}^{p_1}(w_{j_2})_{q_2}^{p_2}(w_{j_3})_{q_3}^{-p_3}(w_{j_4})_{q_4}^{-p_4}\,\, i^{F_{34\cO}-2(j_3+j_4)+p'-p}\nn \\ &\times\binom{2j}{j-p}^{-\half}\binom{2j}{j-p'}^{-\half}d^j_{-p',-p}(-\theta)\,,
\ee
where, 
\be
(w_j)^p_q\equiv \binom{2j}{j+p}^{-\half}\binom{2j}{j+q}^\half d_{p,q}^j(\tfrac{\pi}{2})\,,
\ee
and $d^{j}_{p, q}(\phi)$ is a Wigner $d$-matrix element and $F_{43\cO}$ is the number of fermionic operators among $\cO$, $\cO_3$ and $\cO_4$.\footnote{Here we used the Wigner $d$-matrix element $d^{j}_{m,m'}(\theta)$ using the same conventions as in \texttt{Mathematica 11}, with \texttt{WignerD[\{$j$,$m_1$,$m_2$\},$\theta$]}$\,=d^{j}_{m,m'}(\theta)$.} Here, $b_j$ is the contribution of the normalization coefficient of the two-point function defined in \eqref{eq:b-j-normalization}. 

Since the residue matrices are expressed in the $\SO(3)$ basis of three-point tensor structures, it is convenient to convert the three-point $q$-basis tensor   structures in \eqref{eq:f-th-q-basis} to $\SO(3)$ structures. We find that we can relate $f_{[q_1q_2q_3q_4], j}^{( j_{12}, j_{12\cO}), \,( j_{43}, j_{43\cO})}(\theta)$ to $f_{[q_1q_2q_3q_4], j}^{[p_1p_2p],\,[p_3p_4p']}(\theta)$ (see appendix \ref{sec:deriving-f} for details) as,
\be
\label{eq:f-th-SO(3)-basis}
 f_{ j, [q_1q_2q_3q_4]}^{( j_{12}, j_{12\cO}), \,( j_{43}, j_{43\cO})}(\theta) = \sum_{\substack{p_1, p_2, p\\p_3, p_4, p'}}\cW^{( j_{12}, j_{12\cO}), \,( j_{43}, j_{43\cO})}_{[p_1p_2p],\,[p_4p_3p']} f_{j, [q_1, q_2, q_3, q_4]}^{[p_1p_2p],\,[p_4p_3p']}(\theta)\,,
 \ee
 where, 
 \be \label{eq:w-def}
 \cW^{( j_{12}, j_{12\cO}), \,( j_{43}, j_{43\cO})}_{[p_1p_2p],\,[p_4p_3p']} &= (-1)^{j_1+j_4-2j+p_2+p_3} \<j_1, p_1; j_2, p_2|j_{12}, -p\>\<j_{12}, -p; j, p|j_{12\cO}, 0\>\nn \\ &\times \<j_4, p_4; j_3, p_3|j_{43}, -p'\>\<j_{43}, -p'; j, p'|j_{43\cO}, 0\> \nn \\ &\times\left[\binom{2j}{j+p} \binom{2j}{j+p'}\prod_{i=1}^4 \binom{2j_i}{j_i+p_i}\right]^{1/2}\,.
 \ee

 To summarize, using \eqref{eq:hat-g-solution-r-rb} together with the solutions for $f(\th)$, \eqref{eq:f-th-q-basis} or \eqref{eq:f-th-SO(3)-basis}, we find that $h^{a,b}_{\oo,j, I}$ is given by
 \be
 \label{eq:hinf-final}
	h^{a,b}_{\oo,j , [q_1q_2q_3q_4]}(z,\bar z)=\frac{z^{-h_1-h_2}\bar z^{-\bar h_1-\bar h_2}(1-\r\bar\r)^{-\half}}{
		(1+\r)^{\half-\a-\b}
		(1+\bar\r)^{\half-\bar\a-\bar\b}      
		(1-\r)^{\half+\a+\b}
		(1-\bar\r)^{\half+\bar\a+\bar\b}
	}f^{a, b}_{j, [q_1q_2q_3q_4]}(\theta)\,.
 \ee

\section{Summary of results}
\label{sec:summary-results}

\begin{table}\begin{center}
\begin{tabular}{|c|c|c|c|c|c|}
\hline
Family$\,\, \bi$  & $\De_\bi$ & P & $j'_\bi$ & $n_\bi$ & k\\
\hline \hline
I & $1-j-k$ &+ & $j+k$ & $k$ & $1\leq k$\\\hline
II& $2+j-k$ &+ & $j-k$ & $k$ & $1\leq k \leq j$\\\hline
III& $\tfrac{3}{2}-k$ & +& $j$ & $2k$ & $\begin{cases}
1\leq k,&j\in \Z \\ 1\leq k \leq \min(j,j_{12},j_{43}), & j\in \Z+\half
\end{cases}$\\\hline
IV& $2-k$ &  -& $j$ & $2k-1$ & $\begin{cases}
1\leq k \leq \min(j,j_{12},j_{43}),&j\in \Z \\ 1\leq k, & j\in \Z+\half
\end{cases}$\\
\hline
\end{tabular}\end{center}
\vspace{-0.5cm}
\caption{Summary for the location of the poles $\De_\bi$ in family $\bi$ and of the properties of the differential operator associated to each pole. Specifically, we show the parity $P$ of the differential operator, the spin $j'_\bi$ of the resulting operator $\cO' = (\cD \cO)$, the difference in scaling dimension $n_\bi = \De'_\bi-\De_\bi$, together with the restriction on $k$ showing when such poles and differential operators exist.  Here, $j_{12}$ and $j_{43}$ are parts of the labels of $\SO(3)$ basis three-point tensor structures defined in section~\ref{sec:tensor-struct}. \label{table:summary}}
\end{table}

In this section we summarize our results for the ease of future reference. We have started by defining the conformal blocks  
\be 
g^{a,b}_{\De,j,I}(z,\bar z)\,,
\ee
in section \ref{sec:definition-conf-blocks}, where the blocks are functions of the cross-ratios $z$ and $\bar z$ defined in \eqref{eq:z-barz-def}. As mentioned in section \ref{sec:definition-conf-blocks}, the conformal blocks depend on a choice of three- and four-point tensor structures: $a$ and $b$ label the three-point tensor structures  $\<\cO_1\cO_2 \cO\>^a$ and $\<\cO_4 \cO_3 \cO\>^b$, which we either take to be in the conformal frame basis defined in \eqref{eq:conformal-frame-3pt-def} or in the $\SO(3)$ basis defined by (\ref{eq:SO(2)-basis-def}$\,$--$\,$\ref{eq:SO(3)-basis-definition}), while $I$ labels the four-point tensor structures $\<\cO_1 \cO_2 \cO_3 \cO_4\>^{[q_1q_2q_3q_4]}$, defined by \eqref{eq:q-struct-4pt-definition} and \eqref{eq:q-struct-4-pt-struct-placement}. The blocks also depend on the normalization of the two-point function~\eqref{eq:standard2pt}.

To write down the residue recursion relation we have introduced the functions
\be
h^{a,b}_{\De,j,I}(z,\bar z)\equiv(a_0 r)^{-\De}g^{a,b}_{\De,j,I}(z,\bar z),
\ee
where $a_0$ is a normalization-dependent constant. In terms of $h^{a,b}_{\De,j,I}(z,\bar z)$ we have derived the residue recursion relation
\be
\label{eq:recursion-summary}
	h^{a,b}_{\De,j,I}(z,\bar z)=h^{a,b}_{\oo,j,I}(z,\bar z)+\sum_{\bi}\frac{(a_0r)^{n_\bi}}{\De-\De_\bi}(\cL_\bi)^a_{a'}(\cR_\bi)^{b}_{b'}h^{a',b'}_{\De'_\bi,j'_\bi,I}(z,\bar z)\,.
\ee 
In table \ref{table:summary}, we have listed the classification of all the poles of the function $h^{a,b}_{\De,j,I}(z,\bar z)$ appearing in  \eqref{eq:recursion-summary} at $\De = \De_\bi$. Here $\bi = (A,k)$, where $A=\mathrm{I}$, II, III or IV is the family label associated to each pole, and $k$ labels poles inside a family. The residue at $\De = \De_\bi$ contains a conformal block whose associated spin $j'_\bi$ and scaling dimension $\De '_\bi = \De_\bi +n_{\bi} $ is also listed in table \ref{table:summary}.

The function $h^{a,b}_{\oo,j,I}(z,\bar z)$ was explicitly derived in section \ref{sec:leading-term-h} and is given by \eqref{eq:hinf-final}.  The residue matrices appearing in \eqref{eq:recursion-summary} can be expressed in the $\SO(3)$ basis as 
\be 
	(\cL_\bi)^{(j_{12},j_{12\cO})}_{(j_{12},j'_{12\cO})}=(\cM_\bi^{\De_{12}})^{(j_{12},j_{12\cO})}_{(j_{12},j'_{12\cO})}, \qquad 
	(\cR_\bi)^{(j_{43},j_{43\cO})}_{(j_{43},j'_{43\cO})}=(-1)^{j'_\bi-j}(\cM_\bi^{\De_{43}})^{(j_{43},j_{43\cO})}_{(j_{43},j'_{43\cO})}\,,
\ee
with
\be 
	 ( \cM_\bi^{\De_{12}})^{(j_{12},j_{12\cO})}_{(j_{12},j'_{12\cO})}=\sum_{m} \hat\cF_{\De'_\bi+\De_{12},j'_{12\cO}}^{-1} &\<j_{12\cO}',0|j_{12},-m;j'_\bi,m\>\nn \\ \times \cD_{\bi, m}& \<j_{12},-m;j_3,m|j_{12\cO},0\>  \hat\cF_{\De+\De_{12},j_{12\cO}}\,,
\ee
where $\hat \cF_{\De, j}$ is defined in \eqref{eq:cF-coeff-definition}, $\<j_{12},-m;j_3,m|j_{12\cO},0\> $ are Clebsch-Gordan coefficients$^\text{\ref{footnote:CG-convention}}$ and $\cD_{\bi, m}$ are the eigenvalues of the differential operators listed in section \ref{sec:differential-ops} and derived in appendix \ref{sec:differential-ops-details}.  Note that these matrices are diagonal in the labels $j_{12}$ and $j_{43}$. With these results in mind, we now present several useful examples and checks.

\section{Examples and checks}
\label{sec:examples-and-checks}
\subsection{A warm-up: scalar blocks}
\label{sec:scalar-check}
 
 This warm-up example serves as an initial guidebook to using the results of this paper and as a verification of our results by recovering the well-known recursion relation for scalar blocks \cite{Kos:2014bka,Kos:2013tga}.  In the case of external scalars there is a unique three-point tensor structure, defined in embedding space with the conventions of \cite{Costa:2011mg} as
 \be 
\<\cO_1\cO_2 \cO\>^{(1)} = \frac{\left(( X_1\cdot Z)(X_2\cdot X)-(X_1 \cdot X)(X_2 \cdot Z)\right)^j}{(X_{12})^{\frac{\De_1+\De_2-\De+j}2} (X_{20})^{\frac{\De_2+\De-\De_1+j}2} (X_{01})^{\frac{\De+\De_1-\De_2+j}2}}\,,
 \ee
 where we have used the convention $X_{i0}=-2X_i\.X$. This tensor structure can be related to an $\SO(3)$ basis structure by a proportionality factor, 
 \be 
 \label{eq:scalars-tensor-struct-convert}
\<\cO_1\cO_2 \cO\>^{(1)}  = 2^{j/2}\binom{2j}{j}^{-\half} \<\cO_1\cO_2 \cO\>^{(0, j)}\,.
 \ee
  Therefore the residue matrix has a single element and it is computed using \eqref{eq:value-residue-matrix-M}.
 
  Following the rules presented in table \ref{table:summary} only poles belonging to families I-III can appear in the recursion relation, since $j_{12}=j_{43}=0$. For simplicity we compute the residue coefficient for poles belonging to family I. The selection rules for the Clebsch-Gordan coefficient only allow the term $m=0$ to appear in the sum in  \eqref{eq:value-residue-matrix-M} and, therefore, using the expression~\eqref{eq:cF-coeff-definition} for $\cF$ coefficients and equation~\eqref{eq:Dfamily1} for $\cD_{\bi,m}$ we find
  \be 
  (\cM_\bi^{\De_{12}})_{0,j}^{0,j'} &= \cD_{\bi, 0}\frac{\hat \cF_{1-j-k+\De_{12}, j}}{\hat \cF_{1-j+\De_{12}, j+k}} \nn \\ &= 2^k \small{\left(\frac{1-k + \De_{12}}2\right)_{\frac{k}2}\left(\frac{-k-2j + \De_{12}}2\right)_{\frac{k}2}}\sqrt{\frac{-c_{\cO'}(1+j)^2_k}{c_\cO k!\Gamma(k)(1+2j)_{2k}}}\,,
  \ee
with $\bi=(k, \,\text{I})$. Using the normalization for the two-point function with $c_{\cO'} = c_\cO=1$, and accounting for the relation between the three-point tensor structures \eqref{eq:scalars-tensor-struct-convert}, we recover the result for the residue coefficient obtained in \cite{Kos:2013tga, Kos:2014bka}. Finally, the result for the $h_{\infty, j}(z, \bar z)$ can  be obtained by using \eqref{eq:hinf-final} together with the formula for $f_j(\th)$ from \eqref{eq:f-th-q-basis}. The result for the related function $\tilde g^\text{Scalar}_{\De, j}(z, \bar z)$ is listed in \eqref{eq:scalar-conformal-block-inf}. The $\th$-dependent Wigner-d function appearing in  $f_j(\th)$ is simply related to Legendre polynomials and, our results for $h_{\infty, j}(z, \bar z)$ agree with those of~\cite{Kos:2013tga, Kos:2014bka}.

\subsection{Example: scalar-fermion blocks}
\label{sec:scalar-fermion-check}
We will now work through how to use our formulas to write down the residue recursion relations in the case of the fermion-scalar-scalar-fermion correlator, comparing with the results of \cite{Iliesiu:2015akf}. We will assume that $\cO_2$ and $\cO_3$ are scalars, while $\cO_1$ and $\cO_4$ are spin-$\half$ fermions.

In this situation, there are two possible parity-definite three-point structures, which were defined in~\cite{Iliesiu:2015akf} using embedding formalism as
\be
	\<\cO_1\cO_2\cO\>^{(+)} &= \frac{
		\<S_1 S\> \<S X_1 X_2 S\>^{j-\frac{1}{2}}
		}{
		X_{12}^{\half(\De_1+\De_2-\De-3j+\frac{3}{2})}X_{20}^{\half(\De_2+\De-\De_1+j-\half)}X_{01}^{\half(\De+\De_1-\De_2+j+\half)} 
		}\,,\\
	\<\cO_1\cO_2\cO\>^{(-)} &= \frac{\<S_1 X_2 S\> \<S X_1 X_2 S\>^{j-\half}}{X_{12}^{\half(\De_1+\De_2-\De-3j+\half)}X_{20}^{\half(\De_2+\De-\De_1-j+\frac32)}X_{01}^{\half(\De+\De_1-\De_2+j-\half)}} \,.
\ee
The structures for the three-point function $\<\cO_4\cO_3\cO\>$ are obtained by replacing $1\to 4$ and $2\to 3$. The label $\pm$ of the structure determines its space parity.

These structures can be written succinctly in the \(q\)-basis by evaluating these structures in the conformal frame \(x_1 = 0, x_2 = \hat{e}_2, x_3 = \oo\hat{e}_2\) and comparing with the definition~\eqref{eq:conformal-frame-3pt-def}
\begin{equation}
	\<\cO_1\cO_2\cO\>^{(\pm)} = -2^{j-\half}{\textstyle\left(\<\cO_1\cO_2\cO\>^{\left[\half,0,-\half \right]}\pm \<\cO_1\cO_2\cO\>^{\left[-\half,0,\half \right]}\right)}\,,
\end{equation}
and similarly in the SO(3) basis, by using the relations~\eqref{eq:q-basis-to-SO(2)-basis},~\eqref{eq:so(2)-to-so(2.5)}, and~\eqref{eq:j12j12Odefinition}
\begin{equation}\label{eq:EF-SO3-relation}
	\<\cO_1\cO_2\cO\>^{(\pm)} = -\pi^{1/4}(-1)^{j-\half}\sqrt{\frac{\Gamma(j + \frac{3}{2})}{\Gamma(j+1)}}{ \<\cO_1\cO_2\cO\>^{(\half, j \pm \half) }}\,.
\end{equation}

With these relations in hand, we can calculate our residue matrices following the methods of this paper and compare with the results of \cite{Iliesiu:2015akf} by changing bases. For the purpose of demonstration, we will calculate the matrices for poles of Family I.

To calculate our residue matrix, we use equation \eqref{eq:value-residue-matrix-M}. The selection rules for the Clebsch-Gordon coefficients restrict the summation to \(m=\pm\half\) (recall that \(j_{12}=\half\)). Note that $j'=j+k$, and so $j'_{12\cO}=j+k\pm\thalf$, while $j_{12\cO}=j\pm\thalf$. Let us focus on the case $j_{12\cO}'=j_{12\cO}+k$ first. Using the expression~\eqref{eq:cF-coeff-definition} for $\cF$ coefficients, equation~\eqref{eq:Dfamily1} for $\cD_{\bi,m}$, and the standard expressions for the Clebsch-Gordan coefficients, we find
\be
	&(\cM_\bi^{\De_{12}})_{\half,j_{12\cO}+k}^{\half,j_{12\cO}} =\frac{\widehat{\cF}_{\Delta+\Delta_{12},j_{12\cO}}}{\widehat{\cF}_{1-j-k+\Delta_{12},j_{12\cO}+k}} \bigg({\textstyle\<j_{12\cO}+k,0|\half,\half;j+k,-\half\> \cD_{\bi,\half}\<\half,\half;j,-\half|j_{12\cO},0\> } \nn\\
	&\hspace{6.5cm}+{ \textstyle\<j_{12\cO}+k,0|\half,-\half;j+k,\half\> \cD_{\bi,-\half}\<\half,-\half;j,\half|j_{12\cO},0\>}\bigg) \nn\\ 
	&= 2^{2-j-k-\Delta}\frac{\Gamma(\frac{1-j+\Delta_{12} +j_{12\cO}}{2})\Gamma(\frac{-j+\Delta_{12} -j_{12\cO}}{2})}{\Gamma(\frac{\Delta+\Delta_{12}+j_{12\cO}}{2})\Gamma(\frac{\Delta+\Delta_{12}-j_{12\cO}-1}{2})} \sqrt{\frac{c_{\cO'}}{c_\cO}\frac{-1}{(k-1)!k!}\frac{(\frac{3}{2}+j)_k(\half+j)_k}{(1+2j)_{2k}}}\,,
\ee
with $\bi=(k,\, \text{I})$. For the cases when $j_{12\cO}'\neq j_{12\cO}+k$ it turns out that the terms $m=+\thalf$ and $m=-\thalf$ cancel each other, making the corresponding $(\cM_\bi^{\De_{12}})_{\half,j_{12\cO}'}^{\half,j_{12\cO}}$ vanish.

Given the above, we can convert to the basis from \cite{Iliesiu:2015akf} using~\eqref{eq:EF-SO3-relation} to confirm our results and we find exact  agreement.

We can then compute the leading term of the block, \(h_\infty\). There are four four-point structures in the language of \cite{Iliesiu:2015akf}:
\begin{align}
\<\cO_1 \cO_2 \cO_3 \cO_4\>^{(1)} &=  i\frac{\<S_1S_4\>}{\sqrt{X_{14}}}K_4\,, \nn \\
\<\cO_1 \cO_2 \cO_3 \cO_4\>^{(2)} &=  -i\frac{\<S_1X_2X_3S_4\>}{\sqrt{X_{12}X_{34}X_{23}}} K_4\,,\nn \\
\<\cO_1 \cO_2 \cO_3 \cO_4\>^{(3)} &=  -i\frac{\<S_1X_2S_4\>}{\sqrt{X_{12}X_{24}}} K_4\,,\nn \\
\<\cO_1 \cO_2 \cO_3 \cO_4\>^{(4)} &=  -i\frac{\<S_1X_3S_4\>}{\sqrt{X_{13}X_{34}}} K_4\,,
\end{align}
with the prefactor
\begin{equation}
K_4 \equiv \p{\frac{X_{24}}{X_{14}}}^{\frac{\Delta_{12}}{2}} \p{\frac{X_{13}}{X_{14}}}^{\frac{\Delta_{34}}{2}} \frac{1}{X_{12}^{\frac{\Delta_1+\Delta_2}{2}} X_{34}^{\frac{\Delta_3+\Delta_4}{2}}}\,.
\end{equation}
We can express each of these in the \(q\)-basis by evaluating these structures on the embedding nullcone in the spacelike-separated frame of \eqref{eq:4-pt-coordinate-config-conformal-frame} to get the following basis relation:
\begin{align}
\label{eq:fermion-scalar-4-pt-function-basis}
\<\cO_1 \cO_2 \cO_3 \cO_4\>^{(1)} &=  i(z \bar{z})^{\frac{\Delta_1+\Delta_2}{2}}{\textstyle\p{[\half,0,0,\half] - [-\half,0,0,-\half]} },\nn \\
\<\cO_1 \cO_2 \cO_3 \cO_4\>^{(2)} &= i(z \bar{z})^{\frac{\Delta_1+\Delta_2}{2}}{\textstyle\p{\sqrt{\frac{z(\bar{z}-1)}{\bar{z}(z-1)}}[\half,0,0,\half] - \sqrt{\frac{(z-1)\bar{z}}{(\bar{z}-1)z}}[-\half,0,0,-\half]} },\nn \\
\<\cO_1 \cO_2 \cO_3 \cO_4\>^{(3)} &=  i(z \bar{z})^{\frac{\Delta_1+\Delta_2}{2}}{\textstyle\p{\sqrt{\frac{z}{\bar{z}}}[\half,0,0,-\half] - \sqrt{\frac{\bar{z}}{z}}[-\half,0,0,\half]} },\nn \\
\<\cO_1 \cO_2 \cO_3 \cO_4\>^{(4)} &=  i(z \bar{z})^{\frac{\Delta_1+\Delta_2}{2}}{\textstyle\p{[\half,0,0,-\half] - [-\half,0,0,\half]} }.
\end{align}
Here we use the shorthand \eqref{eq:q-struct-4-pt-struct-placement} of \([q_1q_2q_3q_4]=\<\cO_1 \cO_2 \cO_3 \cO_4\>^{[q_1q_2q_3q_4]}. \) 

Then our task is to calculate the leading terms corresponding to these \(q\)-structures and then convert them into the desired basis by the above relations. For the purpose of demonstration, we will explain how to calculate \(h_\infty\) for the structure \(\<\cO_1 \cO_2 \cO_3 \cO_4\>^{(1)}\). Using~\eqref{eq:fermion-scalar-4-pt-function-basis}, the two \(q\)-structures that contribute to \(\<\cO_1 \cO_2 \cO_3 \cO_4\>^{(1)}\) are \([\half,0,0,\half]\) and \([-\half,0,0,-\half].\) 

Below, we will solely explicitly list the leading term corresponding to \([\half,0,0,\half]\). By \eqref{eq:hinf-final}, we must determine \(f_{j,[q_1,q_2,q_3,q_4]}^{a,b}(\theta)\), whose three-point-structure indices can be \((\half,j\pm\half)\). Then by \eqref{eq:f-th-SO(3)-basis} we must determine the terms in the \(q\)-basis that contribute in the sum. We can observe the selection rules of the Clebsch-Gordon coefficients and the definition of \(\cW_{[p_1p_2p],[p_4p_3p']}^{(j_{12},j_{12\cO}),(j_{43},j_{43\cO})}\) in \eqref{eq:w-def} to then see that the only terms that survive are those that have \(p_1=-p=\pm\half\), \(p_4=-p'=\pm\half\), and \(p_2=p_3=0\). Given that, we can calculate the relevant results of \(h_{\infty,j,I}^{a,b}\). Here, we present an example of one of such expression for one choice of three- and four-point tensor structures:
\begin{align}
h_{\infty,j,[\half,0,0,\half]}^{[\half,0,-\half],[\half,0,-\half]}(z, \bar z)&=  \frac{(z\bar{z})^{\frac{\Delta_1+\Delta_2}{2}}}{\p{1-\r\bar{\r}}^{3/2}}\p{\frac{z}{\bar{z}}}^{\frac{1}{4}} \p{\frac{(1-\r)(1-\bar{\r})}{(1+\r)(1+\bar{\r})}}^{\frac{\De_{12}+\De_{43}}{2}} \!f_{j,[\half,0,0,\half]}^{[\half,0,-\half],[\half,0,-\half]}(\theta)\,,\nn\\
f_{j,[\half,0,0,\half]}^{[\half,0,-\half],[\half,0,-\half]}(\theta) &= -\frac{i}{2}\frac{\cos \frac{\theta }{2} \,\,\Gamma \left(\ell +\frac{1}{2}\right) \Gamma \left(\ell +\frac{3}{2}\right)}{(2 \ell )!\, c_\cO(\Delta ,\ell )} P_{\ell -\frac{1}{2}}^{(0,1)}(\cos \theta)\,.
\end{align}
From there, it's just a matter of putting all of our expressions together to get an expression for \(h_{\oo,j,[q_1,q_2,q_3,q_4]}^{(j_{12},j_{12\cO}),(j_{43},j_{43\cO})}\), all of which match that of \cite{Iliesiu:2015akf} given the correct conventions.

\subsection{Checks from differential operators}
We would additionally like to check that the residue matrices are correct across all families of poles. To do so, we calculated various residue matrices using both the approach of this paper and using the methods presented in~\cite{Dymarsky:2017xzb, Karateev:2017jgd, Penedones:2015aga}. 

To start, we will review how the weight-shifting operators can be used to recover the residue matrices on a case-by-case basis   \cite{ Dymarsky:2017xzb, Karateev:2017jgd}.  To start, we note that in the space of three-point structures, there are two convenient bases for this calculation: the monomial basis and the differential basis, first introduced in \cite{Costa:2011dw}. The idea is to construct a set of conformally-invariant differential operators $\cD^{(\a)}$ which act (simultaneously) on the coordinates of two fictitious scalar\footnote{We for simplicity focus on the case when we start with two scalars, which in 3d is appropriate for the case when the product $\cO_1\cO_2$ is bosonic.} operators $\f_1$ and $\f_2$ and produce objects which transform as $\cO_1$ and $\cO_2$. Acting with these differential operators on the standard three-point structure of $\f_1,\f_2$ and $\cO$ we find
\be
    \cD^{(\alpha)}\<\f_1 \f_2 \cO\> = M_a^\alpha(\Delta) \<\cO_1 \cO_2 \cO\>^{(a)}\,,
\ee
for some transformation matrix $M(\De)$, where  $ \<\cO_1 \cO_2 \cO\>^{(a)}$ are the standard three-point tensor structures.\footnote{To be precise, the scaling dimensions of $\f_1$ and $\f_2$ in general depend on the index $\a$ of the differential operator. For the simplicity of discussion, we will ignore this subtlety. } To distinguish the basis $\<\cO_1 \cO_2 \cO\>^{(a)}$ and the basis in the left hand side of the above equation, we will refer to the former as the monomial and to latter as the differential bases.

Given these, we can express our spinning conformal block in terms of the appropriate transformation matrices for the left- and right-bases and the appropriate set of differential operators acting on a set of scalar conformal blocks.  Explicitly,
\be\label{eqn:spintoscalar}
    g^{a,b}_{\De,j,I}(z,\bar z) = (M_L^{-1}(\De))^a_\alpha(M_R^{-1}(\De))^b_\beta (\cD_L)^\alpha (\cD_R)^\beta g^\text{scalar}_{\De,j,I}(z,\bar z)\,.
\ee
We wish to  find the poles and residues of the spinning blocks. In the above expression, there are only two sources of poles \cite{Karateev:2017jgd}: either the transformation matrices (of which there are a finite number of poles) or the scalar blocks, whose poles and residues are known. The source of the poles directs, then, how to calculate the residue matrices.

\subsubsection{Poles arising from the scalar blocks} 
\label{sec:poles-from-scalar-blocks}
First we consider the poles arising from the scalar blocks. The classification of those poles and expressions for their residues are known \cite{Kos:2013tga, Kos:2014bka}. We have the following pole behavior:
\be
	g^\text{scalar}_{\De,j,I}(z,\bar z) \underset{\Delta\to\Delta_\bi}{\sim} \frac{R_\bi}{\Delta-\Delta_\bi}	g^\text{scalar}_{\De_\bi',j'_\bi,I}(z,\bar z)
\ee
where \(R_\bi\) is the set of scalar residue coefficients. Applying this equation to \eqref{eqn:spintoscalar}, we get the following as the block approaches a pole:
\be
    g^{a,b}_{\De,j,I}(z,\bar z) \underset{\Delta\to\Delta_\bi}{\sim}& \frac{R_\bi}{\Delta-\Delta_\bi} (M_L^{-1}(\De_\bi))^a_\alpha(M_R^{-1}(\De_\bi))^b_\beta (\cD_L)^\alpha (\cD_R)^\beta g^\text{scalar}_{\De_\bi',j',I}(z,\bar z)\\
    &=\frac{R_\bi}{\Delta-\Delta_\bi} (M_L^{-1}(\De_\bi))^a_\alpha(M_R^{-1}(\De_\bi))^b_\beta (M_L(\De_\bi'))^\alpha_{a'} (M_R(\De_\bi'))^\beta_{b'} g^{a',b'}_{\De_\bi',j',I}(z,\bar z)\\
    &=\frac{1}{\Delta-\Delta_\bi} (\cL)^a_{a'} (\cR)^b_{b'} g^{a',b'}_{\De_\bi',j',I}(z,\bar z)\,,
\ee
which gives the following expression for the residue matrices:
\be
    (\cL)_{a'}^a \propto&~ (M_L^{-1}(\De_\bi))^a_\alpha(M_L(\De_\bi'))^\alpha_{a'}\,,\\
    (\cR)_{b'}^b \propto&~ (M_R^{-1}(\De_\bi))^b_\beta(M_R(\De_\bi'))^\beta_{b'}\,.
\ee
Note that these are up to a proportionality constant: there is some freedom as to where the overall factor of \(R_\bi\) comes from, so as a result we compare the tensor product of the residue matrices, rather than individually. This result implies that the residue matrices corresponding to the poles of the scalar block only need the basis transformations to and from the differential basis. We can in principle repeat the same process for fermionic exchanged operators by starting with fermion-scalar-scalar-fermion blocks instead of blocks with external scalar operators. 

To test the results of this paper, we calculated the residue matrices for vector-vector-scalar-scalar as well as vector-scalar-vector-scalar conformal blocks. We found exact agreement between the results of this paper and the differential operator method for all three families, for three point structures of both even and odd parities.

\subsubsection{Poles arising from the transformation matrices}

We can attempt to proceed in the same manner as above for poles arising from the transformation matrices themselves. However, after some analysis one can check that the residue matrices associated to such poles cannot be determined from \eqref{eqn:spintoscalar}; rather, \eqref{eqn:spintoscalar} leads to a  consistency condition that the residue matrices $\cL$ and $\cR$ must satisfy.

Instead, in order to determine the residues associated to these poles, we will directly follow the prescription described in \cite{Penedones:2015aga}. Specifically, we will determine the position space differential operators $\cD_\bi$ (discussed in section \ref{sec:differential-ops} in momentum space) corresponding to such poles. The residue matrices are then determined by solving the following equations 
\be 
\label{eq:diff-ops-and-residues}
\<(\cD_\bi\cO_\De)(\cD_\bi\cO_\De)\> &\hspace{-0.25cm}\underset{\Delta\to\Delta_\bi}{\sim} (\De-\De_\bi)Q_\bi\<\cO'\cO'\>\,,\nn\\
\<\cO_1\cO_2(\cD_\bi\cO_\De)\>^{(a)}&=\sqrt{Q_\bi}(\cL_\bi)_{a'}^a \<\cO_1\cO_2\cO'\>^{(a')}\,,\nn\\
\<\cO_4\cO_3(\cD_\bi\cO_\De)\>^{(b)}&=\sqrt{Q_\bi}(\cR_\bi)_{b'}^b \<\cO_4\cO_3\cO'\>^{(b')}\,,
\ee
where \(Q_\bi\) is some constant. 

So far, through the checks performed in sections \ref{sec:scalar-check}, \ref{sec:scalar-fermion-check} and \ref{sec:poles-from-scalar-blocks}, we have verified in specific cases the expressions for the residue matrices associated to families I, II and III for bosonic exchanges and I, II and IV for fermionic exchanges. The poles corresponding to family IV in for bosonic exchanges and family III in fermionic exchanges come precisely from the transformation matrices in \eqref{eqn:spintoscalar}. Therefore, we will explicitly check the residues of these families by using \eqref{eq:diff-ops-and-residues}.  

For family IV in bosonic exchanges we will consider the case of the four-point function vector-scalar-scalar-vector. In this case $j_{12}$ and $j_{43}$ can both be equal to $1$ and therefore, according to table \ref{table:summary}, the $k=1$ family IV pole exists.  To determine $\cD_\bi$, we should search for a first-order parity-odd differential operator that does not change spin and is conformally-invariant when acting on operators of dimension $\De_\bi=1$. It is easy to check using the standard rules~\cite{Costa:2011dw,Karateev:2017jgd} that the following embedding-space expression (in conventions of \cite{Costa:2011mg}) satisfies these conditions,
\be
    \cD_\bi = \left(Z\cdot \pdr{}{Z}\right)\left(W\cdot \pdr{}{X}\right) - \left(Z\cdot \pdr{}{X}\right)\left(W\cdot \pdr{}{Z}\right).
\ee
Here \(W\) is a polarization vector for the second row of the $\SO(3)$ Young diagram, satisfying all the conditions of orthogonality and gauge equivalence as \(Z\) to both \(X\) and \(Z\). This operator changes spin by adding a second row box to the representation of $\cO$, which, in $d=3$ can be then removed by using the Hodge star. Since the Hodge star is parity-odd, this makes $\cD_\bi$ parity-odd.

The Hodge star operation in embedding space is somewhat non-trivial to describe. We will not go into the details and instead note that it amounts to performing the following replacement:
\be 
    \star : W_i^{[a}Z_i^bX_i^{c]} \mapsto \epsilon\indices{^{abc}_{de}}Z_i^dX_i^e.
\ee
This rule suffices since due to the gauge equivalences of $W$, the expressions can depend on it only through the left hand side of the above equation.

Using this differential operator in~\eqref{eq:diff-ops-and-residues}, we obtained $k=1$ residue matrices that are in perfect agreement with the results of this paper.

Additionally, we considered the case of vector-fermion-fermion-vector, which has an exchanged fermion. The pole coming from the transformation matrices here is the family III \(k=1\) pole, whose residues can be derived in a manner analogous to the above, using the following differential operator:
\begin{equation}
\cD_\bi = 	\pdr{}{X}\cdot \pdr{}{X} - \left\<S,\pdr{}{X},\pdr{}{S}\right\>^2\,,
\end{equation}
where we have used the embedding formalism conventions of \cite{Iliesiu:2015akf}. 
We confirmed that the residues derived with this differential operator match our results.

\section{Discussion}
\label{sec:discussion}

In this paper we have derived closed-form expressions for all the ingredients which are necessary for an implementation of a residue recursion relation for general 3-dimensional conformal blocks. This result opens several possible future directions. 

First of all, it is an important problem to develop a practical computational implementation of these recursion relations. Such an implementation would greatly facilitate future numerical bootstrap studies of spinning correlators and of systems of such correlators.\footnote{Besides the applications for the spinning conformal bootstrap that we have already mentioned in the introduction, there are numerous other theories in $d=3$ and $d=4$ where spinning operators play an important role and, consequently, one that the bootstrap could greatly constrain their operator spectra. For instance, when using the bootstrap to constrain gauge theories in $d=3$ \cite{Chester:2016wrc, Chester:2017vdh, Li:2018lyb}, in the presence of a Chern-Simons term, one might make use of the fact that monopole operators exhibit a charge-spin relation, while when constraining the conformal window of QCD in $d=4$, it oftentimes happens that the lowest dimensional baryon operator is not a scalar. } It appears that our results are well-suited for this purpose: in particular, our expressions for residue matrices have several attractive features that are discussed in section~\ref{sec:properties-residue-matrices}. We are working in this direction and hope to report on it in near future~\cite{Erramilli:toAppear}.

Secondly, by using the closed-form expression for the residue recursion relation we expect to be able to improve the understanding of superconformal blocks, for superconformal field theories (SCFTs) with various amounts of supersymmetry.\footnote{We thank S.~Pufu for this suggestion.} While previous studies have determined the superconformal blocks for external scalar superconformal primaries that usually lie in short supermultiplets (for example, see \cite{Khandker:2014mpa, Bobev:2015jxa, Buric:2019rms}),  one can hope to use our result to determine superconformal blocks for external superconfromal primaries with arbitrary spin. Towards that end, one can use superconformal Ward identities and the action of the superconformal Casimir to fix the relative ratios between OPE coefficients of operators within the same superconformal multiplet and, consequently,  express a superconformal block as a linear combination of the regular conformal blocks whose recursion relation we have determined in this paper. It would also be interesting to adapt our methods to directly derive a residue recursion relation for general superconformal blocks. For example, a classification for all the poles of the superconformal blocks was derived in \cite{Sen:2018del}.

Finally, we can generalize our results to a higher number of dimensions $d$ by following a similar derivation to that presented in this paper. Here, we should consider two separate cases: even $d$, and odd (or generic non-integer) $d$. 

In the case of odd \(d\), conformal blocks have only simple poles and the theory behind the recursion relations is exactly the same as described in this paper. One can still write the momentum-space versions of the two-point function and of the differential operators. If the basis of functions $|j,m\>$ is replaced by a Gelfand-Tseitlin (G-T) basis (see~\cite{Kravchuk:2017dzd} for a recent review), the calculation in appendices~\ref{sec:details-fourier-space-2pt} and~\ref{sec:differential-ops-details} of the recursion relations~\eqref{eq:app-recursion-relation} and~\eqref{eq:recursion-Dm-+} for the coefficients $\cB_m$ and $\cD_m$ can be generalized.\footnote{Indeed we have obtained preliminary results for the generalization of $\cB_m$.} Similarly, the tensor structures of section~\ref{sec:tensor-struct} can be generalized appropriately using G-T basis, if we assume that the Clebsch-Gordan coefficients of $\mathrm{Spin}(d)$ are known. The computation of residue matrices can therefore proceed in a fashion similar  to the derivation in section~\ref{sec:deriving-residue-matrices}. The derivation of $h_{\infty,j,I}^{a,b}(z,\bar z)$ is based on a special property of $q$-basis four-point tensor structures described in section~\ref{sec:four-point-tensor-struct}. Structures with this property can be constructed in any number of dimensions.\footnote{For example, the $d=4$ \(q\)-structures defined in~\cite{Cuomo:2017wme} also have this property.} In such a basis, the resulting form for $h_{\infty,j,I}^{a,b}(z,\bar z)$ from equation~\eqref{eq:hat-g-solution-r-rb} can therefore be easily generalized. 
The function $f$ in the right-hand side of~\eqref{eq:hat-g-solution-r-rb} can again be fixed by studying the OPE limit, which was analyzed in general dimension in~\cite{Kravchuk:2017dzd}. To summarize, all the steps in our derivation can be successfully applied in any odd $d$. 

In the case of even $d$ the conformal blocks can have higher-order poles, and a general theory of recursion relations is not known. However, when deriving scalar conformal blocks in even $d$, it is possible to obtain the recursion relations by carefully taking the limit of generic $d$ recursion relations~\cite{Kravchuk:unpublished}. The objects that appear in the right-hand side of the recursion relations involve derivatives of conformal blocks with respect to $\De$, as well as the regular parts of conformal blocks at a pole in $\De$. The representation-theoretic interpretation of this is at the moment unclear, and so it is an interesting conceptual question whether the methods of the present paper can help shed some light on the even-dimensional case. It is worth noting that this includes \(d=4\), which is perhaps the next most interesting case beyond \(d=3\).

\section*{Acknowledgements}
We thank Silviu Pufu, Slava Rychkov, Emilio Trevisani, and especially Walter Landry, David Poland, and David Simmons-Duffin for valuable discussions.  LVI is supported in part by the US NSF under Grant No. PHY-1820651 and by the Simons Foundation Grant No. 488653.  PK is supported by DOE grant No.\ DE-SC0009988. This research was supported in part by Perimeter Institute for Theoretical Physics. Research at Perimeter Institute is supported by the Government of Canada through the Department of Innovation, Science and Economic Development and by the Province of Ontario through through the ministry of Research and Innovation.

\appendix

\section{Conventions}
\label{sec:conventions}

\subsection{3d spinor and operators conventions}
\label{sec:3d-spinor-conventions}
The Lorentz group in $d= 3$ is $\mathrm{Spin}(2, 1)\simeq \SL(2, \mathbb R)$. The anti-Hermitian generators of the Lorentz group satisfy the commutation relations
\be
\label{eq:Lorentz-group-algebra-comm}
	[M^{\mu\nu},M^{\r\s}]=\eta^{\nu\r}M^{\mu\s} + \eta^{\mu \s} M^{\nu \r}-\eta^{\mu\r}M^{\nu\s} -\eta^{\nu\s}M^{\mu\r} \,,
\ee
where we choose the Lorentzian metric signature to be	$\eta^{\mu\nu}=\eta_{\mu\nu}=\text{diag}(-1, 1, 1)$. By performing a Wick rotation
\be
x_{\text{Euclidean}}^0=ix_{\text{Lorentzian}}^0,\qquad x_{\text{Euclidean}}^k=x_{\text{Lorentzian}}^k \quad(k=1,2),
\ee
which extends straightforwardly to general-rank tensors with upper or lower indices, one obtains the Euclidean generators $M^{\mu\nu}_E = M_{\mu\nu}^E$, which we use extensively throughout the paper. 

To construct the spinor representations it is convenient to define the gamma-matrices satisfying the usual relations
\be
	\g^\mu \g^\nu + \g^\nu \g^\mu = 2\eta^{\mu\nu}\,.
\ee
Explicitly, we chose
\be
\label{eq:gamma-matrices-convention}
	(\gamma^0)^\a{}_\b=\begin{pmatrix}
		0 & 1 \\ -1 & 0
	\end{pmatrix},
	(\gamma^1)^\a{}_\b=\begin{pmatrix}
	0 & 1 \\ 1 & 0
	\end{pmatrix},
	(\gamma^2)^\a{}_\b=\begin{pmatrix}
	1 & 0 \\ 0 & -1
\end{pmatrix}\,.
\ee
In spinor irrep the Lorentz generators are represented by the matrices
\be
\label{eq:lorentz-generators}
	\p{\cM^{\mu\nu}}^\a{}_\b=\frac{1}{4}\p{[\g^\mu,\g^\nu]}^\a{}_\b\,.
\ee
These generators satisfy the same commutation relations as~\eqref{eq:Lorentz-group-algebra-comm} and preserve the symplectic form
\be
\label{eq:symplectic-form-convention}
	\O_{\a\b}=\O^{\a\b}=\begin{pmatrix}
			0 & 1\\
			-1 & 0
	\end{pmatrix}\,.
\ee
The latter property identifies these matrices with elements of Lie algebra of $\SL(2,\R)=\mathrm{Sp}(2,\R)$. The elements of the spinor irrep are real two-dimensional vectors with upper indices $s^\a$. The dual irrep has representation matrices
\be
	\p{\cM^{\mu\nu}_\text{dual}}_\a{}^\b\equiv-([\cM^{\mu\nu}]^T)_\a{}^\b=-(\cM^{\mu\nu})^\b{}_\a,
\ee
and its elements are real two-dimensional vectors with lower indices $s_\a$. The spinor irrep and its dual are isomorphic, with the isomorphism established by the symplectic form $\O$. We identify the elements of these two irreps by index raising/lowering rules
\be
	s_\a=\O_{\a\b}s^\b,\qquad s^\a=s_\b \O^{\b\a}.
\ee
The list of finite-dimensional irreps of $\SL(2,\R)$ is exhausted by spin-$j$ irreps, where $j$ is a non-negative (half-)integer, the elements of which are symmetric tensors of the form
\be
	T^{\a_1\cdots\a_{2j}}.
\ee
All these irreps are real since the spinor irrep is.

We now discuss how the above formulas are interpreted in the context of primary operators. Since for now we are concerned only with the spin of the operator, we will mostly consider primary operators $\cO(x=0)$ inserted at the origin $x=0$ of Minkowski space, which is preserved by $M^{\mu\nu}$. In such situations we will often omit the positional argument of primary operators, i.e.\ write $\cO\equiv\cO(x=0)$.

A primary operator of spin $j$ is a tensor
\be
	\cO^{\a_1\cdots\a_{2j}},
\ee
symmetric in indices $\a_i$, which satisfies (as far as only the Lorentz group is concerned) the commutation relation
\be\label{eq:MOcommutator0}
	[M^{\mu\nu},\cO^{\a_1\cdots\a_{2j}}]=-\sum_{k=1}^{2j} (\cM^{\mu\nu})^{\a_k}{}_\b \cO^{\a_1\cdots \a_{k-1}\b\a_{k+1}\cdots \a_{2j}}.
\ee
The minus sign is required for the commutation relations of $M$ and $\cM$ to be consistent. It is convenient to use the index-free notation~\eqref{eq:define-polarizations}
\be
	\cO(s)=s_{\a_1}\cdots s_{\a_{2j}}\cO^{\a_1\ldots \a_{2j}}\,,
\ee
where $s$ is a real spinor variable, whose components we will often denote by
\be
s_{\a} \equiv \begin{pmatrix}
	\xi \\ \bar \xi
\end{pmatrix}.
\ee
Whenever we talk about $s$ as an argument of $\cO$, we mean $s_\a$, i.e.\ $s$ with a lower index.  
With this definition,~\eqref{eq:MOcommutator0} implies
\be
	e^{\half w_{\mu\nu}M^{\mu\nu}}\cO(s)e^{-\half w_{\mu\nu}M^{\mu\nu}}=\cO(e^{\half w_{\mu\nu}\cM^{\mu\nu}_{\text{dual}}}s)
\ee
for any real matrix $w_{\mu\nu}$. The generalization of this to $x\neq 0$ is
\be\label{eq:finitespinor}
	e^{\half w_{\mu\nu}M^{\mu\nu}} \cO(x,s) e^{-\half w_{\mu\nu}M^{\mu\nu}}=\cO(e^{\half w_{\mu\nu}\cM^{\mu\nu}_\text{vector}}x, e^{\half w_{\mu\nu}\cM^{\mu\nu}_\text{dual}}s)\,,
\ee
where
\be
\p{\cM^{\mu\nu}_\text{vector}}^\r{}_\s=\eta^{\r\mu}\de^{\nu}_\s-(\mu\leftrightarrow\nu)\,.
\ee

For bosonic operators (integer $j$), it is often convenient to use vector polarizations $z_\mu$, and correspondingly represent the operators by tensors
\be
	\cO^{\mu_1\cdots \mu_j},
\ee
or functions
\be
	\cO(z)\equiv z_{\mu_1}\cdots z_{\mu_j}\cO^{\mu_1\cdots \mu_j}.
\ee
The analogue of~\eqref{eq:finitespinor} is
\be
	e^{\half w_{\mu\nu}M^{\mu\nu}} \cO(x,z) e^{-\half w_{\mu\nu}M^{\mu\nu}}=\cO(e^{\half w_{\mu\nu}\cM^{\mu\nu}_\text{vector}}x, e^{\half w_{\mu\nu}\cM^{\mu\nu}_\text{vector}}z)\,.
\ee

In the case when $j=1$, we can identify
\be
	\cO^\mu = \frac{1}{\sqrt{2}}\gamma^\mu_{\a\b}\cO^{\a\b}\,, \qquad z^\mu=-\frac{1}{\sqrt{2}}\gamma^\mu_{\a\b} s^\a s^\b\,,
\ee
which ensures that 
\be
	\cO(z)=\cO(s).
\ee
This identification straightforwardly extends to $j>1$. 

Since both spinor and vector polarizations are real, as are all representation matrices, we can have Hermitian primaries which satisfy
\be
\p{\cO^{\a_1\ldots \a_{2j}}}^\dagger(x) = \cO^{\a_1\ldots \a_{2j}}\,,\qquad \cO(x,s)^\dagger = \cO(x,s)\,,\qquad \cO(x,z)^\dagger = \cO(x,z)\,,
\ee
for real Lorentzian $x$.

Finally, we will need the action of the Lorentz group on functions of polarizations $s_\a$. Note that for any group $G$ acting on the left on a space $X$, the natural action on functions $f$ on $X$ is defined by
\be
	(gf)(x)=f(g^{-1}x),\quad g\in G,x\in X.
\ee
The inverse on the right hand side is needed to be consistent with group multiplication, i.e.\ that action of $g_1g_2$ on $f$ is the same as first acting with $g_2$ and then with $g_1$. We thus define for functions of $s_\a$ the action
\be
	e^{\half w_{\mu\nu}M^{\mu\nu}}\.f(s)=f(e^{-\half w_{\mu\nu}\cM^{\mu\nu}_\text{dual}}s),
\ee
where we use $\cdot$ to distinguish this action, and it is understood that $\cdot$-action has higher precedence then evaluating $f$ at $s$. By expanding in $w_{\mu\nu}$, we get the infinitesimal action
\be\label{eq:functionaction}
	M^{\mu\nu}\cdot f(s)=-(\cM^{\mu\nu}_\text{dual})_{\b}{}^{\a}s_\a\frac{\ptl}{\ptl s_\b} f(s).
\ee

\subsection{Space parity}
\label{app:spaceparity}

In this section we discuss the action of space parity. In Lorentzian signature in 3d we are forced to consider reflections $R_i$ ($i=1,2$) along spatial directions, i.e.
\be
(R_i x)^i=-x^i,\quad (R_i x)^\mu=x^\mu\quad (\mu\neq i).
\ee
It will be convenient to write this more compactly as
\be
	(R_i x)^\mu=(r_i)^\mu_\nu x^\nu,
\ee
where $r_i$ is a reflection matrix.
On the Hilbert space $R_i$ is represented by a unitary operator, which we will also denote by $R_i$. This operator should only have the following properties,
\be	
R_i X^\mu R_i^{-1}&=(r_i)^\mu_\nu X^\nu,\quad (X=P,K),\\
R_i M^{\mu\nu} R_i^{-1}&=(r_i)^\mu_{\mu'}(r_i)^\nu_{\nu'} M^{\mu'\nu'}.\label{eq:Mparity}
\ee
It follows that the square $R_i^2$ commutes with the conformal generators and so is an internal symmetry, but is otherwise unconstrained.

When acting on a local operator $\cO$, $R_i$ produces a new local operator $\cO'$. Due to~\eqref{eq:Mparity}, if we simply define
\be
	\cO'(R_i x,s)=R_i \cO(x,s) R_i^{-1},\qquad {\color{gray}(\text{wrong})}
\ee
then $\cO'(s)$ will transform under conformal transformations in a non-standard way. For this reason, we define instead
\be
	\cO'(R_i x,\gamma_i s)=R_i \cO(x,s) R_i^{-1},
\ee
which transforms in the standard way. It is easy to check that the definition of $\cO'$ does not depend on choice of $i$ and that $\cO'$ is Hermitian if $\cO$ is. In principle, $\cO'$ can be independent of $\cO$. We will consider the simplest situation when $\cO$ is Hermitian and
\be
	\cO'(x,s)=\kappa \cO(x,s)
\ee
for some parity $\kappa\in \R$. In other words,
\be
	R_i \cO(x,s) R_i^{-1}=\kappa \cO(R_i x,\gamma_i s).
\ee
By requiring $R_i$-invariance of the two-point function of $\cO$, we can get a constraint on $\kappa^2$, which turns out to be $\kappa^2=1$, and so $\kappa=\pm 1$. We can refer to the operators with $\kappa=1$ as parity-even and with $\kappa=-1$ as parity-odd. Invariant expressions of $s$ and $x$ can be similarly classified into parity-even or parity odd depending on how they transform under $x\to R_i x, s\to \gamma_i s$.

For our purposes it is also convenient to use the Euclidean space parity $\cP_E$, which simply sends Euclidean $x$ to $-x$. Such a parity transformation commutes with $\mathrm{Spin}(3)$ generators, and so it must act in $\mathrm{Spin}(3)$ irreps by multiplication by a constant. For example, in vector representation this constant is $-1$. It is convenient to choose the constant in spinor irreps to be $i$. In this way, $\gamma$-matrices become parity-even invariants, since they have two spinor and one vector index. This definition leads to the same notion of parity of invariant expressions as the Lorentzian definition above.

\subsection{Conformal algebra}
\label{sec:conformal-algebra}
The conformal algebra in 3d, in Lorentzian signature is $\mathfrak{so}(3, 2)$ and the generators satisfy the commutation relations 
\be
\label{eq:conformal-algebra-comm}
	&[D,P_\mu]=P_\mu, \quad [D,K_\mu]=-K_\mu,\nn\\
	&[M^{\mu\nu},X^\r]=\eta^{\nu\r}X^\mu-\eta^{\mu\r}X^\nu,\quad (X=K,P)\nn\\
	&[K_\mu, P_\nu]=2\eta_{\mu\nu}D-2M_{\mu\nu},
\ee
together with \eqref{eq:Lorentz-group-algebra-comm},  and all other commutators being trivial. All generators $D,P,K,M$ are anti-Hermitian.

\subsection{Action of conformal generators}

With the algebra conventions chosen in section \ref{sec:conformal-algebra}, we have the following action of conformal generators on local operators
\be
[D,\cO(x,s)]&=(\De+x\.\ptl_x)\cO(x,s),\nn\\
[P_\mu,\cO(x,s)]&=\ptl_\mu \cO(x,s),\nn\\
[M_{\mu\nu},\cO(x,s)]&=(x_\nu\ptl_\mu-x_\mu\ptl_\nu)\cO(x,s)-M_{\mu\nu}\.\cO(x,s),\nn\\
[K_\mu,\cO(x,s)]&=(2x_\mu x^\s-x^2\de^\s_\mu)\ptl_\s \cO(x,s)+2x_\mu \De \cO(x,s)+2x^\s M_{\mu\s}\.\cO(x,s),
\ee
where the action $M_{\mu\nu}\.$ is interpreted as action on functions of $s$, as defined above in~\eqref{eq:functionaction}.

\subsection{Reality of OPE coefficients and conformal blocks}
\label{sec:reality-of-conf-blocks-and-OPE-coeff}

For practical purposes (such as formulating a numerical bootstrap problem) it is necessary to discuss the reality of the OPE coefficients and of the conformal blocks in our set of conventions. We start by discussing the reality of the OPE coefficients. Note that
\be\label{eq:threeptreality}
	[\<\O|\cO_1(x_1,s_1)\cO_2(x_2,s_2)\cO_3(x_3,s_3)|\O\>]^*&=\<\O|\cO_3(x_3,s_3)^\dagger\cO_2(x_2,s_2)^\dagger\cO_1(x_1,s_1)^\dagger|\O\>\nn\\
	&=\<\O|\cO_3(x_3,s_3)\cO_2(x_2,s_2)\cO_1(x_1,s_1)|\O\>\nn\\
	&=i^{F_{123}}\<\O|\cO_1(x_1,s_1)\cO_2(x_2,s_2)\cO_3(x_3,s_3)|\O\>,
\ee
where we have assumed that $x_i$ are spacelike separated and that $\cO_i$ are Hermitian, and $F_{123}$ is the number of fermionic operators among the $\cO_i$. This equation determines the reality properties of physical three-point functions. Expanding the latter in a basis of tensor structures as in~\eqref{eq:physical-structure-3pt-relation}, we can determine the reality properties of the OPE coefficients from the reality properties of the three-point tensor structures.

Since it is easy to see that the $q$-basis tensor structures defined in section~\ref{sec:tensor-struct} are real, we find that the OPE coefficients in the \(q\)-basis are real if all operators are bosonic or purely imaginary if any two operators in the associated three-point function are fermionic. Similarly, since the $\SO(2)$ and $\SO(3)$ basis tensor structures are related to $q$-basis by real coefficients, it follows that the same reality properties hold for the OPE coefficients in those bases as well.

Let us now consider the reality properties of four-point function components $g_{[q_1q_2q_3q_4]}(z, \bar z)$ and of the respective conformal blocks. Similarly to~\eqref{eq:threeptreality} we find
\be
&[\<\O|\cO_1(x_1,s_1)\cO_2(x_2,s_2)\cO_3(x_3,s_3)\cO_4(x_4,s_4)|\O\>]^*\nn\\
&=i^{F_{1234}}\<\O|\cO_1(x_1,s_1)\cO_2(x_2,s_2)\cO_3(x_3,s_3)\cO_4(x_4,s_4)|\O\>,
\ee
and so the physical four-point function is real if there are 0 or 4 fermions among $\cO_i$, and purely imaginary if there are two fermions. Similarly to $q$-basis three-point structures, it is easy to see that $q$-basis four-point structures are real, and so the same reality properties hold for $g_{[q_1q_2q_3q_4]}(z, \bar z)$ when $z$ and $\bar z$ correspond to real Lorentzian spacelike-separated $x_i$, which includes real $0<z,\bar z<1$. 

The reality properties of the conformal blocks $g^{a,b}_{\De,j,[q_1q_2q_3q_4]}(z,\bar z)$ can be determined by requiring that the conformal block times the two OPE coefficients have the same reality property as $g_{[q_1q_2q_3q_4]}(z, \bar z)$. If there are 0 or 4 fermionic operators among $\cO_i$, the exchanged operator $\cO$ is bosonic, and the OPE coefficients in any of the bases discussed above are either both real or both pure imaginary. This implies that their product is real, and thus $g^{a,b}_{\De,j,[q_1q_2q_3q_4]}(z,\bar z)$ is real since $g_{[q_1q_2q_3q_4]}(z, \bar z)$ is. If there are 2 fermionic operators among $\cO_i$, it is easy to check by a similar reasoning that the block is real if $\cO$ is bosonic and pure imaginary if the exchanged operator is fermionic. 

To summarize, in all cases, the reality of the conformal block $g^{a,b}_{\De,j,[q_1q_2q_3q_4]}(z,\bar z)$ is determined by the statistics of $\cO$: it is real for bosonic $\cO$ and pure imaginary for fermionic $\cO$. This conclusion can also be reached from the following argument: a conformal block is, as far as reality properties are concerned, a product of two three-point structures divided by a four-point structure and a two-point function. Since our three- and four-point structures are real, reality properties of the conformal block are the same as those of the two-point function, which can be inferred from~\eqref{eq:standard2pt} and agree with the above conclusion.

Let us verify that our results are consistent with these reality properties. Since we have checked that the product $\cL_\bi\otimes\cR_\bi$ of residue matrices is real, the reality of the blocks is determined by reality of $h^{[p_1 p_2p],\,[p_3p_4p']}_{\oo,j,[q_1q_2q_3q_4]}$, which in turn is the same as of $f^{[p_1 p_2p],\,[p_3p_4p']}_{j,[q_1q_2q_3q_4]}$ given by~\eqref{eq:f-th-q-basis}. One can check that the $w$-matrices are all real, so the phases in that formula can only come from
\be
	i^{F_{340}-2(j_3+j_4)+m'-m}d^j_{-m',-m}(-\theta).
\ee
The $d$-matrix is real for real argument, but in our case $\theta$ is imaginary since $\r$ and $\bar \r$ are real. However, in this case $i^{m'-m}d^j_{-m',-m}(-\theta)$ can be shown to be real. Furthermore $i^{F_{34\cO}}$ is real, since zero or two operators of the three can be fermionic. Thus the reality of the conformal  block is determined by
\be
\label{eq:reality-conf-blocks}
	i^{-2(j_3+j_4)}=\pm i^{2j}\,,
\ee
which is real for bosonic $\cO$ and imaginary for fermionic $\cO$, as expected.

\section{Details about the two-point function in Fourier space}
\label{sec:details-fourier-space-2pt}

Consider the two-point function
\be
G(x,s_1,s_2)\equiv\<0|\cO(x,s_1)\cO(0,s_2)|0\>\,,
\ee
which is represented in terms of its Fourier transform as 
\be
\label{eq:two-point-function-F-transf}
G(x,s_1,s_2)=\int_{p>0} \frac{d^dp}{(2\pi)^d}e^{ip\.x}\cG(p,s_1,s_2)\,.
\ee
The requirement of invariance under special conformal transformations is
\be
((2x_\mu x^\s-x^2\de^\s_\mu)\ptl_\s+2x_\mu\De+2x^\s M_{\mu\s}^1)G(x,s_1,s_2)=0\,,
\ee
where $M_{\mu\s}^1$ represents the action on the first operator, which is interpreted as action on functions as defined in appendix~\ref{sec:3d-spinor-conventions}, and we omit the $\.$ notation.  In the following, $M_{\mu\s}^2$ gives the action on the second and $M_{\mu\s}^{12} = M_{\mu\s}^1+M_{\mu\s}^2$. Since we have $x\.\ptl G(x,s_1,s_2)=-2\De G(x,s_1,s_2)$, we can rewrite this as
\be
(-x^2\ptl_\mu-2x_\mu\De+2x^\s M_{\mu\s}^1)G(x,s_1,s_2)=0\,.
\ee
In momentum space this can be rewritten as
\be
\label{eq:special-conf-transf-action-momentum-space}
0
=&(p_\mu \ptl^2-2(\De-1)\ptl_\mu+2M^1_{\mu\s}\ptl^\s)\cG(p,s_1,s_2)\,.
\ee
If we assume Lorentz-invariance of $\cG$, it then suffices to check this equation at a single value of $p$, say $p=\hat e_0$. 
We will compute the momentum derivatives in \eqref{eq:special-conf-transf-action-momentum-space} using Lorentz invariance. Specifically, we have
\be
(p_\nu\ptl_\mu-p_\mu\ptl_\nu-M_{\mu\nu}^{12})\cG(p,s_1,s_2)=0\,.
\ee
Plugging $p=\hat e_0$ and $\nu=0$ and $\mu\neq 0$ we find
\be
\label{eq:first-derivative-momentum-space}
\ptl_\mu \cG(\hat e_0,s_1,s_2)=-M_{\mu0}^{12}\cG(\hat e_0,s_1,s_2)\,.
\ee
If before plugging in these values we differentiate with respect to $\ptl_\l$, $\l\neq 0$, we find then 
\be
\label{eq:double-space-derivative-momentum-space}
\ptl_\mu\ptl_\l\cG(\hat e_0,s_1,s_2) = \p{-\eta_{\mu\l}(2\De-3)+M^{12}_{\mu0}M^{12}_{\l0}}\cG(\hat e_0,s_1,s_2)\,.
\ee
From this equation, using the commutation relations for $M$, we see that in order for the two derivatives to commute we need $M_{\mu\l}^{12}\.\cG(\hat e_0,s_1,s_2)=0$ for all $\mu,\l\neq 0$. This simply expresses the $\mathrm{Spin}(2)$ little-group invariance of $\cG$, discussed in section \ref{sec:two-point-function}.

Using homogeneity of $\cG(p,s_1,s_2)$ in $p$ we also find
\be\label{eq:p0derivative}
\ptl_0\cG(\hat e_0,s_1,s_2)&=(2\De-3)\cG(\hat e_0,s_1,s_2)\,,\nn\\
\ptl_0^2\cG(\hat e_0,s_1,s_2)&=(2\De-4)(2\De-3)\cG(\hat e_0,s_1,s_2)\,,
\ee
which, together with \eqref{eq:double-space-derivative-momentum-space}, can be used to write the action of the Laplacian in momentum space: 
\be
\label{eq:action-of-laplacian-in-momentum-space}
\ptl^2\cK(\hat e_0,s_1,s_2) = \p{-2(2\De-3)(\De-1)+\eta^{\l \zeta} M^{12}_{\l0}M^{12}_{\zeta 0}}\cG(\hat e_0, s_1,s_2)\,.
\ee
Combining \eqref{eq:first-derivative-momentum-space},~\eqref{eq:p0derivative}, and \eqref{eq:action-of-laplacian-in-momentum-space} we find that the constraint \eqref{eq:special-conf-transf-action-momentum-space} imposed by special conformal transformations when evaluated at $p = \hat e_0$ implies:
\be
\label{eq:explicit-special-conf-generator-constraint}
\text{For }\mu = 0:& \qquad 
\eta^{\l \zeta} \p{M_{\l0}^1M_{\zeta 0}^1-M_{\l 0}^2M_{\zeta 0}^2}\cG(\hat e_0,s_1,s_2)=0\,,\nn \\ 
\text{For }\mu \neq  0:&\qquad  [
	-(\De-2)M^1_{\mu0}+(\De-1)M_{\mu0}^2 \nn \\ & \qquad -\eta^{\l \zeta} M^1_{\mu \l}M^1_{\zeta 0}+\eta^{\l \zeta}  M^2_{\l 0}M^2_{\mu \zeta}
]\cG(\hat e_0,s_1,s_2)=0\,.
\ee
The first equation can be further simplified by writing $\eta^{\l \zeta}  M_{\l0}M_{\zeta 0}=C_2^{\mathrm{Spin}(3)}-C_2^{\mathrm{Spin}(2)}$, which is a difference of quadratic Casimirs of $\mathrm{Spin}(3)$ and $\mathrm{Spin}(2)$. Using this and the $\mathrm{Spin}(2)$ invariance of $\cG(\hat e_0,s_1,s_2)$, one can show that the first equation is a consequence of Lorentz symmetry, and thus implies no extra constraints on $\cG(\hat e_0,s_1,s_2)$.

Consider now the second equation in \eqref{eq:explicit-special-conf-generator-constraint}. We can write it in a nicer form by taking \eqref{eq:explicit-special-conf-generator-constraint} for $\mu=1$ multiplied by $i$ minus that for $\mu=2$, and rewriting everything in terms of the $\SU(2)$ generators introduced in section \ref{sec:tensor-struct}, 
\be
J_0=-i M^{12}_E=-i M_{12}, \quad J_1=-i M^{20}_E=M^{20},\quad J_2=-iM^{01}_E=M^{01}\,.
\ee
We find
\be
\label{eq:first-constraint-on-G}
\p{
	-(\De-2)J^1_++(\De-1)J^2_+-J_0^1 J^1_+ + J^2_+ J_0^2 
}\cG(\hat e_0,s_1,s_2)=0\,.
\ee
Taking instead \eqref{eq:explicit-special-conf-generator-constraint} for $\mu=1$ multiplied by $-i$ minus that for $\mu=2$ we find
\be
\label{eq:second-constraint-on-G}
\p{
	-(\De-2)J^1_-+(\De-1)J^2_-+J_0^1 J^1_- - J^2_- J_0^2 
}\cG(\hat e_0,s_1,s_2)=0\,.
\ee

To analyze the constraints~\eqref{eq:first-constraint-on-G} and~\eqref{eq:second-constraint-on-G} it is convenient to rewrite the dependence of $\cG$ on $s_1$ and $s_2$ in terms for the functions $|j,m\>$ defined in~\eqref{eq:j,m-to-spinor-polarizations}. Specifically, using the definition of the two-point function $ \cG(\hat e_0, s_1, s_2)$ from \eqref{eq:two-point-function-F-transf}, its relation to the inner-product \eqref{eq:innerprodtocompute}, and the relation~\eqref{eq:innerprod-in-terms-of-Bm}, we find
\be\label{eq:2ptansatz}
\cG(\hat e_0,s_1,s_2)=& e^{-i\pi j} \sum_{m=-j}^j \cB_m(\De, j) |j,m\>\otimes |j,-m\>\,.
\ee
Here we use the tensor product notation when in reality we simply have a product of two functions. We do so to keep track which function depends on which $s_i$: by convention, we assume that the first tensor factor is a function of $s_1$, while the second is a function of $s_2$.

Combining the ansatz~\eqref{eq:2ptansatz}, with \eqref{eq:first-constraint-on-G} and using~\eqref{eq:standardJaction} we find
\be
\sum_m&(d-\De-1-m-1)\cB_m(\De, j) \sqrt{(j-m)(j+m+1)}|j,m+1\>\otimes |j,-m\>\nn\\
&+(\De-1-m)\cB_m(\De, j) \sqrt{(j+m)(j-m+1)}|j,m\>\otimes |j,-m+1\>=0\,,
\ee
which, after a shift $m\to m-1$ in the first term under the sum, leads to the recursion relation
\be
\label{eq:app-recursion-relation}
\cB_m(\De, j)=\frac{\De+m-2}{\De-m-1}\cB_{m-1}(\De, j)\,.
\ee
The second equation \eqref{eq:second-constraint-on-G} leads to the same relation. The recursion relation has the solution
\be
\cB_m(\De, j)=\frac{\Gamma(\De+m-1)\Gamma(\De-m-1)}{\Gamma(\De+j-1)\Gamma(\De-j-1)}\cB_j(\De, j)\,.
\ee

We now need to compute the overall normalization $\cB_j(\De, j)$. In order to do this, we take the two-point function defined in \eqref{eq:standard2pt}
\be\label{eq:2ptcanonical-appendix}
\<\cO(x_1,s_1)\cO(x_2,s_2)\>=c_\cO\frac{(is_{1,\a} (\gamma^\mu)^{\a\b} s_{2,\b} x_{12,\mu})^{2j}}{x_{12}^{2\De+2j}}\,,
\ee
and rewrite \eqref{eq:2ptansatz} in terms of the components of the spinor polarization  $\chi$ by using \eqref{eq:j,m-to-spinor-polarizations}: 
\be
\label{eq:expansion-in-terms-of-chi}
	\cG(\hat e_0, s_1, s_2)= \sum_{m=-j}^j e^{ i \pi j}\cB_m(\De, j) \binom{2j}{j+m}(\chi^1_1)^{j+m}(\chi^2_1)^{j-m}(\chi^1_2)^{j-m}(\chi^2_2)^{j+m}\,.
\ee
The term including $\cB_j(\De, j)$ is
\be
	 e^{ i \pi j} \cB_j(\De, j)(\chi^1_1)^{2j}(\chi^2_2)^{2j}\,.
\ee
To pick it out we can set $\chi_1^1=-1,\chi_1^2=0,\chi_2^1=0,\chi_2^2=1$, so that we find simply $\cG(\hat e_0, s_1, s_2)=e^{-i \pi j} \cB_j(\De, j)$. For such a choice of $\chi$, we can rewrite the numerator of the two-point function \eqref{eq:2ptcanonical-appendix} by using $	s_{1,\a} (\gamma^\mu)^{\a\b} s_{2,\b}=i\de^\mu_0$. From this the two point function becomes
\be
\label{eq:2ptparticular-appendix}
\<\cO(x_1,s_1)\cO(x_2,s_2)\> = 	c_\cO \frac{(x^0_{12})^{2j}}{x_{12}^{2\De+2j}}\,.
\ee

Therefore, in order to compute $\cB_j(\De, j)$ we want to compute the Fourier transform of \eqref{eq:2ptparticular-appendix} at $p=\hat e_0$. Using the $x^0\to x^0 + i \e$ prescription we find, 
\be
&\cG(\hat e_0, s_1, s_2) = \int d^dx e^{-ipx}\frac{c_\cO(x^0)^{2j}}{(x^2+i\e x^0)^{\De+j}}=c_\cO\frac{2\pi^2\Gamma(\De+j-1)}{\Gamma(\De+j)\Gamma(2\De-2)}e^{-i\pi j}\,.
\ee
This implies
\be
\label{eq:value-for-Bj}
\cB_j(\De, j)=c_\cO\frac{2\pi^2\Gamma(\De+j-1)}{\Gamma(\De+j)\Gamma(2\De-2)}\,,
\ee
and thus
\be
\cB_m(\De, j)=c_\cO\frac{2\pi^2 \Gamma(\De+m-1)\Gamma(\De-m-1)}{\Gamma(\De+j)\Gamma(2\De-2)\Gamma(\De-j-1)}\,,
\ee
as quoted in \eqref{eq:Bvalue}.

As a sanity check, from~\eqref{eq:innerprod-in-terms-of-Bm} we need
\be
\cB_m(\De, j)\geq 0
\ee
for unitary values of $(\De,j)$, and this is indeed the case.

\section{Details about differential operators for null states}
\label{sec:differential-ops-details}

In this appendix we present the derivation of the recurrence relations for the coefficients $\cD_{\bi,m}$, as well as other details on differential operators. 

We start with the derivation of the recurrence relations for $\cD_{\bi,m}$.
For general $p$, we can rewrite~\eqref{eq:generalD} in the form
\be\label{eq:appendixdiffop}
\cO'(p,s)=(-p^2)^{\frac{\De'-\De}{2}}\sum_{m=-j}^j \cD_m \Pi_m(p) \cO(p,s),
\ee
where the projector $\Pi_m(p)$ can be defined as follows. For $p=q_0=\hat e_0$ is defined by
\be
\Pi_m(q_0)=|j',m\>\<j,m|,
\ee
where $\<j,m|$ is a differential operator in $s$ with the property
\be
	\<j,m|j,m'\>=\de_{m,m'}.
\ee
For general future timelike values of $p$, we define $\Pi_m(p)$ using the invariance condition
\be\label{eq:Piinvariance}
\L\Pi_m(p)\L^{-1}=\Pi_m(\L p),\quad \Pi_m(\l p)=\Pi_m(p)\quad(\l > 0),
\ee
where $\L$ is any proper Lorentz transformation. Notice that $\Pi_m(q_0)$ is invariant under the little group of $q_0$, so this definition is self-consistent.

We can write the infinitesimal version of~\eqref{eq:Piinvariance} as
\be\label{eq:PiLorentzInvariance}
[M_{\mu\nu},\Pi_m(p)]=(p_\nu\ptl_\mu-p_\mu\ptl_\nu)\Pi_m(p).
\ee
For convenience, we also define $\hat \Pi_m(p)=(-p^2)^{\frac{\De'-\De}{2}}\Pi_m(p)$.

The action of special conformal generators in momentum space is given by
\be
[iK_\mu, \cO'(p,s)]&=(p_\mu \ptl^2+2(\De'-d)\ptl_\mu-2(p\.\ptl)\ptl_\mu+2M_{\mu\s}\ptl^\s)\cO'(p,s)\,,\nn\\
[iK_\mu, \cO(p,s)]&=(p_\mu \ptl^2+2(\De-d)\ptl_\mu-2(p\.\ptl)\ptl_\mu+2M_{\mu\s}\ptl^\s)\cO(p,s)\,.
\ee
Combining this with~\eqref{eq:appendixdiffop} we find
\be
[iK_\mu, \cO'(p,s)]=\sum_{m=-j}^j&\cD_m\Big(
[(p_\mu \ptl^2-2(d-\De')\ptl_\mu-2(p\.\ptl)\ptl_\mu+2M_{\mu\s}\ptl^\s)\hat \Pi_m(p)]\cO(p,s)\nn\\
&+\hat \Pi_m(p)(p_\mu \ptl^2-2(d-\De)\ptl_\mu-2(p\.\ptl)\ptl_\mu+2M_{\mu\s}\ptl^\s)\cO(p,s)\nn\\
&+2[M_{\mu\s},\hat\Pi_m(p)]\ptl^\s \cO(p,s)-2(\De-\De')\hat\Pi_m(p)\ptl_\mu \cO(p,s)\nn\\
&+2p_\mu \ptl_\s\hat \Pi_m(p)\ptl^\s\cO(p,s)-2[(p\.\ptl)\hat\Pi(p)] \ptl_\mu \cO(p,s)-2[\ptl_\mu \hat\Pi(p)](p\.\ptl)\cO(p,s)\Big)
\ee
After some manipulation we find,
\be
[iK_\mu, \cO'(p,s)]=\sum_{m=-j}^j&\cD_m\Big(
[(p_\mu \ptl^2-2(d-\De')\ptl_\mu-2(p\.\ptl)\ptl_\mu+2M_{\mu\s}\ptl^\s)\hat \Pi_m(p)]\cO(p,s)\nn\\
&+\hat \Pi_m(p)[-iK_\mu,\cO(p,s)]\nn\\
&+2[M_{\mu\s},\hat\Pi_m(p)]\ptl^\s \cO(p,s)-2(\De-\De')\hat\Pi_m(p)\ptl_\mu \cO(p,s)\nn\\
&+2p_\mu \ptl_\s\hat \Pi_m(p)\ptl^\s\cO(p,s)-2[(p\.\ptl)\hat\Pi(p)] \ptl_\mu \cO(p,s)-2[\ptl_\mu \hat\Pi(p)](p\.\ptl)\cO(p,s)\Big)
\ee
so that now we can cancel the left hand side to obtain
\be
0=\sum_{m=-j}^j&\cD_m\Big(
[(p_\mu \ptl^2-2(d-\De')\ptl_\mu-2(p\.\ptl)\ptl_\mu+2M_{\mu\s}\ptl^\s)\hat \Pi_m(p)]\cO(p,s)\nn\\
&+2[M_{\mu\s},\hat\Pi_m(p)]\ptl^\s \cO(p,s)-2(\De-\De')\hat\Pi_m(p)\ptl_\mu \cO(p,s)\nn\\
&+2p_\mu \ptl_\s\hat \Pi_m(p)\ptl^\s\cO(p,s)-2[(p\.\ptl)\hat\Pi(p)] \ptl_\mu \cO(p,s)-2[\ptl_\mu \hat\Pi(p)](p\.\ptl)\cO(p,s)\Big).
\ee
Two of terms with derivatives of $\cO$ cancel due to homogeneity of $\hat\Pi_m$ in $p$,
\be
0=\sum_{m=-j}^j&\cD_m\Big(
[(p_\mu \ptl^2-2(d-\De')\ptl_\mu-2(p\.\ptl)\ptl_\mu+2M_{\mu\s}\ptl^\s)\hat \Pi_m(p)]\cO(p,s)\nn\\
&+2[M_{\mu\s},\hat\Pi_m(p)]\ptl^\s \cO(p,s)\nn\\
&+2p_\mu \ptl_\s\hat \Pi_m(p)\ptl^\s\cO(p,s)-2[\ptl_\mu \hat\Pi(p)](p\.\ptl)\cO(p,s)\Big)
\ee
The other three cancel due to Lorentz invariance of $\hat\Pi_m(p)$~\eqref{eq:PiLorentzInvariance}, resulting in the equation
\be\label{eq:Dequation}
0=\sum_{m=-j}^j&\cD_m
[(p_\mu \ptl^2-2(d-\De')\ptl_\mu-2(p\.\ptl)\ptl_\mu+2M_{\mu\s}\ptl^\s)\hat \Pi_m(p)]\cO(p,s).
\ee

Lorentz invariance~\eqref{eq:PiLorentzInvariance} can be written as
\be
(p_\nu\ptl_\mu-p_\mu\ptl_\nu-\mathrm{adj} M_{\mu\nu})\hat\Pi_m(p)=0,
\ee
where 
\be
	\mathrm{adj} M_{\mu\nu}\hat\Pi_m(p)\equiv [M_{\mu\nu},\hat\Pi_m(p)].
\ee
Plugging $p=\hat e_0$, $\nu=0$, and $\mu\neq 0$ we find
\be
\ptl_\mu \hat\Pi_m(\hat e_0)=-\mathrm{adj} M_{\mu0}\.\hat\Pi_m(\hat e_0).
\ee
If before plugging in these values we differentiate with $\ptl_\l$, $\l\neq 0$, we find then 
\be
\ptl_\mu\ptl_\l\hat\Pi_m(\hat e_0) = \p{-h_{\mu\l}(\De'-\De)+\mathrm{adj}M_{\mu0}\mathrm{adj}M_{\l0}}\hat\Pi_m(\hat e_0).
\ee
Note that in order for the two derivatives to commute we need $\mathrm{adj}M_{\mu\l}\.\hat\Pi_m(\hat e_0)=0$ for $\mu,\l\neq 0$. This simply expresses $\mathrm{Spin}(2)$ stabilizer invariance of $\hat\Pi_m(p)$.
From scaling properties of $\hat\Pi_m(p)$ in $p$ we also find 
\be
\ptl_0\hat\Pi_m(\hat e_0)&=(\De'-\De)\hat\Pi_m(\hat e_0),\nn\\
\ptl_0^2\hat\Pi_m(\hat e_0)&=(\De'-\De-1)(\De'-\De)\hat\Pi_m(\hat e_0).
\ee
We then compute in momentum space
\be
\ptl^2\hat\Pi_m(\hat e_0) = \p{-2(\De'-\De)(\tfrac{\De'-\De+d}{2}-1)+\mathrm{adj}M_{\l0}\mathrm{adj}M_{\l0}}\hat\Pi_m(\hat e_0)
\ee

Consider now the equation~\eqref{eq:Dequation} at $\mu=0$ and at $p=\hat e_0$. Using the formulas above we find
\be
[&(2(\De'-\De)(\tfrac{\De'-\De+d}{2}-1)-2(\De'-\De-1)(\De'-\De)\nn\\
&-\mathrm{adj}M_{\l0}\mathrm{adj}M_{\l0}-2(d-\De')(\De'-\De)+2M_{0\s}\mathrm{adj} M_{0\s})\hat \Pi_m(\hat e_0)]\cO(\hat e_0,s)=0,
\ee
which leads to
\be
[(-\De(\De-d)+\De'(\De'-d)+M^L_{\l0}M^L_{\l0}-M^R_{\l0}M^R_{\l0})\hat \Pi_m(\hat e_0)]\cO(\hat e_0,s),
\ee
where $M^L_{\mu\nu}\Pi_m(\hat e_0)\equiv M_{\mu\nu}\Pi_m(\hat e_0)$ and $M^R_{\mu\nu}\Pi_m(\hat e_0)\equiv\Pi_m(\hat e_0)M_{\mu\nu}$. Recall that $M_{\l0}M_{\l0}=C_2^{\SO(3)}-C_2^{\SO(2)}$. Due to the equality of $\SO(2)$ Casimirs, the expressions above then combine into conformal Casimirs, leading to the condition
\be
[(C_2(\cO')-C_2(\cO))\hat \Pi_m(\hat e_0)]\cO(\hat e_0,s)=0,
\ee
which means that we need $C_2(\cO')=C_2(\cO)$ in order for $\cD$ to exist, which we already knew. 

Now consider $\mu\neq 0$ at $p=\hat e_0$
\be
0&=\sum_{m=-j}^j\cD_m
[-2(d-\De-1)\ptl_\mu+2M_{\mu\s}\ptl^\s]\hat \Pi_m(p)\nn\\
&=\sum_{m=-j}^j\cD_m
[-2(d-\De-1)\mathrm{adj}M_{0\mu}-2M_{\mu0}(\De'-\De)+2M_{\mu i}\mathrm{adj}M_{0i}]\hat \Pi_m(\hat e_0)\nn\\
&=-2\sum_{m=-j}^j\cD_m
[(d-\De'-1)M_{0\mu}^L+(d-\De-1)M_{0\mu}^R-M_{\mu i}^LM_{0i}^L+M_{0i}^RM_{\mu i}^R]\hat \Pi_m(\hat e_0)
\ee
Recall
\be
J_0=-i M^{12}_E=-i M_{12}, \quad J_1=-i M^{20}_E=M^{20}=-M_{20},\quad J_2=-iM^{01}_E=M^{01}=M_{10}.
\ee
Combining our equations for $\mu=1,2$ into $+$ component we find
\be
0=&-2\sum_{m=-j}^j\cD_m
[(d-\De'-1)J_+^L+(d-\De-1)J_+^R-J_0^LJ_+^L+J_+^RJ_0^R]\hat \Pi_m(\hat e_0)\nn\\
=&-2\sum_{m=-j}^j\cD_m
[(d-\De'-1-m-1)J_+^L+(d-\De-1-m)J_+^R]|j',m\>\<j,m|\nn\\
=&-2\sum_{m=-j}^j\cD_m\Big(
(d-\De'-m-2)\sqrt{(j'-m)(j'+m+1)}|j',m+1\>\<j,m|\nn\\
&\qquad\qquad\qquad -(d-\De-m-1)\sqrt{(j+m)(j-m+1)}|j',m\>\<j,m-1|\Big),
\ee
while the $-$ component we obtain
\be
0=&-2\sum_{m=-j}^j\cD_m
[(d-\De'-1)J_-^L+(d-\De-1)J_-^R+J_0^LJ_-^L-J_-^RJ_0^R]\hat \Pi_m(\hat e_0)\nn\\
=&-2\sum_{m=-j}^j\cD_m
[(d-\De'-1+m-1)J_-^L+(d-\De-1+m)J_-^R]|j',m\>\<j,m|\nn\\
=&-2\sum_{m=-j}^j\cD_m\Big(
(d-\De'+m-2)\sqrt{(j'+m)(j'-m+1)}|j',m-1\>\<j,m|\nn\\
&\qquad\qquad\qquad -(d-\De+m-1)\sqrt{(j-m)(j+m+1)}|j',m\>\<j,m+1|\Big).
\ee

For $+$ component this leads to the recursion relation 
\be
\label{eq:recursion-Dm-+}
\cD_m(d-\De'-m-2)\sqrt{(j'-m)(j'+m+1)}=\cD_{m+1}(d-\De-m-2)\sqrt{(j+m+1)(j-m)},
\ee
and for $-$ component to
\be
\label{eq:recursion-Dm--}
\cD_{m+1}(d-\De'+m-1)\sqrt{(j'+m+1)(j'-m)}=\cD_{m}(d-\De+m-1)\sqrt{(j-m)(j+m+1)}.
\ee
These are precisely the relations quoted in the main text.

\subsection{Operator parity and action on past-directed momentum}
\label{app:Dparity}

Let us now discuss some questions related to the parity action on differential operators.

According to the discussion in appendix~\ref{app:spaceparity}, action of $R_i$ reflection on a parity-even operator in momentum space at $p=q_0=\hat e_0$ can be written as
\be
	R_i \cO(q_0,s) R_i^{-1}=\cO(q_0,\g_i s).
\ee
Replacing $s\to \g_1 s$ in $|j,m\>$ amounts to replacing $|j,m\>\to |j,-m\>$, and so we find
\be
	R_1 \cO_m(q_0) R_1^{-1} = \cO_{-m}(q_0).
\ee
By comparing with~\eqref{eq:Dequation}, we find that a differential operator of definite parity satisfies
\be\label{eq:Dparity}
	\cD_{m}=\pm \cD_{-m},
\ee
where $+$ sign is for parity-even operator, and $-$ sign is for parity-odd.

On the other hand, the differential operator $\cD$ can be thought of as a Lorentz-invariant polynomial 
\be
	\cD(p,s,\ptl_s),
\ee
with the property
\be\label{eq:Dhomogeneity}
	\cD(\a p,\b s,\b^{-1}\ptl_s)=\a^n \b^{2(j'-j)}\cO(p,s,\ptl_s),
\ee
where $n$ is the order of the differential operator, while $j'-j$ is (integer) the amount by which $\cD$ changes spin. According to the discussion in appendix~\ref{app:spaceparity}, parity of $\cD$ can be determined by sending $p\to -p$ and $s\to is$, which amounts to using $\a=-1$ and $\b=i$ in the above equation. So we conclude that the parity of $\cD$ is determined by
\be\label{eq:DparityFromPoly}
	P_\cD=(-1)^{n+j'-j}.
\ee
From this we immediately infer that families I, II, and III have parity-even $\cD$ while the family IV has parity-odd $\cD$.\footnote{Note that this result comes from requiring $\cD$ to be polynomial in $p$ when expressed in terms of $s$ and $\ptl_s$, which is not obvious from the $\cD_m$ coefficients.} Using~\eqref{eq:Dparity} we can the determine the relation between $\cD_m$ and $\cD_{-m}$ in these families.

Finally, let us note what this implies for the action of $\cD$ on past-directed momenta. First, note that the coefficients $\cD_m$ can be defined by the equation
\be
\cD(q_0,s,\ptl_s)|j,m\>=\cD_m|j',m\>.
\ee
It is clear that the analogous coefficients $\cD_m^\text{past}$ for past-directed momenta are determined by
\be
\cD(-q_0,s,\ptl_s)|j,m\>=\cD_m^\text{past}|j',m\>.
\ee
Using~\eqref{eq:Dhomogeneity},~\eqref{eq:Dparity}, and~\eqref{eq:DparityFromPoly} we find
\be
	\cD_m^\text{past}=(-1)^n \cD_m=(-1)^{j'-j}\cD_{-m}.
\ee

\section{Deriving $f^{a, b}_{j, I}(\theta)$}
\label{sec:deriving-f}

As mentioned in the main text, we derive $f^{a, b}_{j, I}(\theta)$ by studying leading $r$-term of the conformal block, following the analysis in~\cite{Kravchuk:2017dzd}. In fact, it is more convenient to study leading $s$-term, with $s$ defined by
\be
	z=s e^{i\f},\quad \bar z=se^{-i\f},
\ee
because to the leading order in $r$ we have $s\simeq 4r$ and $\f\simeq \theta$.

To start, we define the contribution of the operator $\cO(x,s)$ to a $q$-basis structure as $\cO_q(x)$, with 
\be
	\cO(x,s)\equiv \sum_{q=-j}^j \xi^{j+q}\bar \xi^{j-q} \cO_{[q]}(x)\,.
\ee
Thus, the contribution of the structure $[q_1 q_2 q_3 q_4]$ to the four-point function is given by, 
\be
\label{eq:four-point-function-projection-q-basis}
g_{[q_1q_2q_3q_4]}(z, \bar z) = \lim_{L \to \oo}\<\cO_{1,[q_1]}(0)\cO_{2,[q_2]}(x_2)\cO_{3,[q_3]}(\hat e_1)\cO_{4,[q_4]}(L\hat e_1)\>\,.
\ee
The action of $e^{\theta M_{01}^E}$ on $\cO_{[q]}$ is given by\footnote{Where, one uses that
\be
	(e^{\theta M_{01}^E}s)_\a=(e^{i\theta \cM^{01}_\text{dual}})_\a{}^\b s_\b=
	\begin{pmatrix}
		e^{-\frac{i}{2}\theta}\xi \\ e^{\frac{i}{2}\theta}\bar \xi
	\end{pmatrix}_\a\,,
\ee
and so if we expand both sides in \eqref{eq:action-of-M01} one finds
\be
	\sum_{q=-j}^j \xi^{j+q}\bar \xi^{j-q} e^{\theta M_{01}^E}\cO_q(x)e^{-\theta M_{01}^E}=\sum_{q=-j}^j \xi^{j+q}\bar \xi^{j-q} e^{-iq\theta}\cO_q(e^{\theta M_{01}^E}x)\,,
\ee}
\be
\label{eq:action-of-M01}
	e^{\theta M_{01}^E}\cO(x,s)e^{-\theta M_{01}^E}=\cO(e^{\theta M_{01}^E}x,e^{\theta M_{01}^E}s) \quad \Rightarrow \quad 
	e^{\theta M_{01}^E}\cO_{[q]}(x)e^{-\theta M_{01}^E}=e^{-iq\theta}\cO_{[q]}(e^{\theta M_{01}^E}x)\,.
\ee
which allows us to rewrite the four-point function as
\be
\label{eq:4-pt-function-rewriting}
	&\<\cO_{1,[q_1]}(0)\cO_{2,[q_2]}(x_2)\cO_{3,[q_3]}(\hat e_1)\cO_{4,[q_4]}(L\hat e_1)\>\nn\\
	&= e^{i(q_1+q_2)\phi}s^{-\De_1-\De_2} \<\cO_{1,[q_1]}(0)\cO_{2,[q_2]}(\hat e_1) s^\De e^{\phi M_{01}^E} \cO_{3,[q_3]}(\hat e_1)\cO_{4,[q_4]}(L\hat e_1)\>,
\ee
where, we remind the reader that $z=se^{i\f}$. In particular the factor above is 
\be
	e^{i(q_1+q_2)\phi}s^{-\De_1-\De_2}=z^{\frac{q_1+q_2-\De_1-\De2}{2}}\bar z^{\frac{-q_1-q_2-\De_1-\De2}{2}}=z^{-h_1-h_2}\bar z^{-\bar h_1-\bar h_2}.
\ee

Now, to determine the leading-$s$ term it is sufficient to study the contribution of the exchanged primary operator (in radial quantization) and ignore of all its descendants. For this, it is useful to define $\cO_{\{q\}}(x)$ by
\be
\label{eq:O-m=q-definition}
	 \cO_{\{q\}}(e^{\frac{\pi M_{12}^E}2}x) \equiv e^{\frac{\pi}{2}M^{12}_E}\cO_{[q]}(x)e^{-\frac{\pi}{2}M^{12}_E}\,.
\ee
The two operators can be related when $x=0$ by expanding the operator $\cO(0, s)$ into two sets of spinor polarization variables: into the components of $s_\a$ and into the components of $(e^{-\frac{\pi}{2} M_{12}^E}s)_\a$. Using such an expansion we find that, 
\be
\label{eq:conversion-q-basis-m-basis-ops}
\cO_{[q]}(0)=\sum_m (w_j)^m_q  \cO_{\{m\}}(0)\,,\qquad \cO_{\{m\}}(0)=\sum_{q}(w_j^{-1})^q_m \cO_{[q]}(0)\,,
\ee
where, we define
\be
(w_j)^m_q\equiv \binom{2j}{j+m}^{-\half}\binom{2j}{j+q}^\half d_{m,q}^j(\tfrac{\pi}{2})\,.
\ee
The contribution of a primary in radial quantization to some correlation function $\<AB\>$, where $A$ and $B$ are products of local operators is then given by
\be
\label{eq:AB-expanded-primary}
\<AB\>_\text{primary}=\sum_{p=-j}^j c_{\cO}^{-1} \,i^{2j}\,\binom{2j}{j+m}^{-1}\<A \cO_{\{m\}}(\oo\hat e_1)\>\<\cO_{\{-m\}}(0)B\>\,.
\ee
where $c_\cO$ is the normalization is the normalization of the two-point function given  by \eqref{eq:standard2pt}.\footnote{The normalization factor in \eqref{eq:AB-expanded-primary} can be obtained by setting $A=\cO_{\{-m\}}(0)$.} 

Using~\eqref{eq:AB-expanded-primary} twice in \eqref{eq:4-pt-function-rewriting}, we have
\be
\label{eq:expanding-4-pt-function-as-two-3-pt}
&\<\cO_{1,[q_1]}(0)\cO_{2,[q_2]}(\hat e_1)s^{-D} e^{-\phi M_{01}^E}\cO_{3,[q_3]}(\hat e_1)\cO_{4,[q_4]}(\oo\hat e_1)\>=\nn\\
&\sum_{m,m' = -j}^j c_{\cO}^{-2}i^{4j}\binom{2j}{j+m}^{-1}\binom{2j}{j+m'}^{-1} s^\De\<\cO_{1,[q_1]}(0)\cO_{2,[q_2]}(\hat e_1) \cO_{\{m\}}(\oo \hat e_1)\>\nn\\
&\quad \times\<\cO_{\{-m\}}(0)e^{-\phi M_{01}^E}\cO_{\{m'\}}(\oo\hat e_1)\>\<\cO_{\{-m'\}}(0)\cO_{3,[q_3]}(\hat e_1)\cO_{4,[q_4]}(\oo\hat e_1)\>
\ee 
We now  relate the 3-point functions appearing in the sum above to structures in the confromal frame basis and  find the $\phi$-dependence in of $\<\cO_{\{-m\}}(0)e^{-\phi M_{01}^E}\cO_{\{m'\}}(\oo\hat e_1)\>$.  

The 3-point coefficients in the $q$-basis are defined in section \ref{sec:tensor-struct} as
\be
		\<\cO_{1,[q_1]}(0)\cO_{2,[q_2]}(\hat e_2)\cO_{3,[q_3]}(\oo \hat e_2)\>=\l_{[q_1q_2q_3]}\,.
\ee
Performing a rotation we find, 
\be
\label{eq:lambda-q1-q2-q3-definition-review}
	\l_{[q_1q_2q_3]} &=\<e^{\frac{\pi}{2}M^{12}}\cO_{1,[q_1]}(0)\cO_{2,[q_2]}(\hat e_2)\cO_{3,[q_3]}(\oo \hat e_2)e^{-\frac{\pi}{2}M^{12}}\>\nn\\
	&=  \< \cO_{1,\{q_1\}}(0)\cO_{2,\{q_2\}}(\hat e_1)\cO_{3,\{q_3\}}(\oo \hat e_1)\>\,.
\ee

We then have that the first three-point function appearing in \eqref{eq:expanding-4-pt-function-as-two-3-pt} is given by,
\be
\label{eq:1st-3-pt-function}
\<\cO_{1,[q_1]}(0)\cO_{2,[q_2]}(\hat e_1) \cO_{\{m\}}(\oo\hat e_1)\>&=\sum_{m_1,m_2}\< \cO_{1,\{m_1\}}(0)\cO_{2,\{m_2\}}(\hat e_1)\cO_{\{m\}}(\oo \hat e_1)\>(w_{j_1})_{q_1}^{m_1}(w_{j_2})_{q_2}^{m_2}\nn\\
&=\sum_{m_1,m_2} (w_{j_1})_{q_1}^{m_1}(w_{j_2})_{q_2}^{m_2}\l_{[m_1m_2m]}^L\,,
\ee
For the second three-point function in \eqref{eq:expanding-4-pt-function-as-two-3-pt}  we have
\be
\label{eq:R-three-pt-function-OPE-coeff}
\< \cO_{\{-m'\}}(0)\cO_{3,[q_3]}(\hat e_1)\cO_{4,[q_4]}(\oo\hat e_1)\>&=
\sum_{m_3,m_4} (w_{j_3})_{q_3}^{m_3}(w_{j_4})_{q_4}^{m_2}\< \cO_{\{-m'\}}(0) \cO_{3,\{m_3\}}(\hat e_1) \cO_{4,\{m_4\}}(\oo\hat e_1)\>\nn\\
&=\sum_{m_3,m_4}(w_{j_3})_{q_3}^{m_3}(w_{j_4})_{q_4}^{m_2}\l^{034}_{[-m'\,m_3\,m_4]}\,.
\ee
We would like to instead provide structures in the order $\<\cO_4\cO_3\cO\>$. This requires exchanging operators at the 1st and 3rd positions. Under such interchange we have~\cite{Kravchuk:2016qvl}
\be
\l^{034}_{[-m'\,m_3\,m_4]}=i^{F_{34\cO}}(-1)^{j_3+j_4+j}\l^R_{[-m_4\,-m_3\,m']}
\,,
\ee
where $F_{34\cO}$ is the number of fermionic operators in the three-point function \eqref{eq:R-three-pt-function-OPE-coeff}. In total we have
\be
\label{eq:2nd-3-pt-function}
\< \cO_{\{-m'\}}(0)\cO_{3,[q_3]}(\hat e_1)\cO_{4,[q_4]}(\oo\hat e_1)\>=\sum_{m_3,m_4}i^{F_{340}}(-1)^{j_3+j_4+j}(w_{j_3})_{q_3}^{m_3}(w_{j_4})_{q_4}^{m_4}\l^R_{[-m_4\,-m_3\,m']}\,.
\ee
Finally, we want to rewrite $\<\cO_{\{-m\}}(0)e^{-\phi M_{01}^E}\cO_{\{m'\}}(\oo\hat e_1)\>$ in terms of a known function of $\phi$.  Using the conversion \eqref{eq:conversion-q-basis-m-basis-ops} between $\cO_{[q]}$ and $ \cO_{\{m\}}$ twice we find\footnote{Following a similar reasoning to that outlined in this appendix one should also be able to determine the thermal spinning conformal blocks, generalizing the initial work on the topic from \cite{Iliesiu:2018fao,Iliesiu:2018zlz}. }
\be
\label{eq:phi-angle-dependence}
\<\cO_{\{-m\}}(0)e^{-\phi M_{01}^E} \cO_{\{m'\}}(\oo\hat e_1)\>=&\sum_q (w_j^{-1})_{-m}^q\<\cO_{[q]}(0)e^{-\phi M_{01}^E}  \cO_{\{m'\}}(\oo\hat e_1)\>\nn\\
=&\sum_q (w_j^{-1})_{-m}^qe^{-iq\f}\<\cO_{[q]}(0)  \cO_{\{m'\}}(\oo\hat e_1)\>\nn\\
=&\sum_{q,m''} (w_j^{-1})_{-m}^qe^{-iq\f}(w_j)^{-m''}_q \< \cO_{\{-m''\}}(0) \cO_{\{m'\}}(\oo\hat e_1)\>\nn\\
=&\sum_{q} (w_j^{-1})_{-m}^qe^{-iq\f}(w_j)^{-m'}_q c_\cO(-i)^{2j}\binom{2j}{j+m'}\nn\\
=&c_\cO\,(-i)^{2j}\binom{2j}{j-m}^\half\binom{2j}{j-m'}^{\half}\sum_q  d_{q,-m}^j(-\tfrac{\pi}{2})e^{-iq\f}d_{-m',q}^j(\tfrac{\pi}{2})\nn \\
=& c_\cO \,i^{-2j+m'-m}\binom{2j}{j-m}^\half\binom{2j}{j-m'}^{\half}d^j_{-m',-m}(-\f)\,,
\ee
where in the last line we have used the identity, 
\be
	\sum_q  d_{q,-m}^j(-\tfrac{\pi}{2})e^{-iq\f}d_{-m',q}^j(\tfrac{\pi}{2})=d^j_{-m',-m}(-\f)i^{m'-m}\,.
\ee
Putting \eqref{eq:1st-3-pt-function}, \eqref{eq:2nd-3-pt-function} and \eqref{eq:phi-angle-dependence} together we find that the four-point function is given by,
\be
\label{eq:leading-eq-in-q-basis}
&\<\cO_{1,[q_1]}(0)\cO_{2,[q_2]}(\hat e_1)s^{-D} e^{-\phi M_{01}^E}\cO_{3,[q_3]}(\hat e_1)\cO_{4,[q_4]}(\oo\hat e_1)\>=\nn\\
&= s^\De \sum_{m_i,m,m'}(w_{j_1})_{q_1}^{m_1}(w_{j_2})_{q_2}^{m_2}(w_{j_3})_{q_3}^{m_3}(w_{j_4})_{q_4}^{m_4}\l_{m_1,m_2,m}^L\l^R_{-m_4,-m_3,m'}\nn\\
&\qquad\times c_{\cO}^{-1}i^{F_{340}}i^{-2(j_3+j_4)+m'-m}\binom{2j}{j-m}^{-\half}\binom{2j}{j-m'}^{-\half}d^j_{-m',-m}(-\f)\,.
\ee
Thus, from this,  we can extract the function $f_{[q_1 q_2 q_3 q_4], j}^{[m_1 m_2 m],\,[m_4  m_3 m']}(\phi) $ to be
\be
\label{eq:f-q-basis-derived-appendix}
f_{[q_1 q_2 q_3 q_4], j}^{[m_1 m_2 m],\,[m_4 m_3  m']}(\phi) & = b_{j}^{-1} (w_{j_1})_{q_1}^{m_1}(w_{j_2})_{q_2}^{m_2}(w_{j_3})_{q_3}^{-m_3}(w_{j_4})_{q_4}^{-m_4} i^{F_{340}-2(j_3+j_4)+m'-m}\nn \\ &\times\binom{2j}{j-m}^{-\half}\binom{2j}{j-m'}^{-\half}d^j_{-m',-m}(-\f)\,.
\ee

While \eqref{eq:f-q-basis-derived-appendix} is in the standard \(q\)-basis (the \textit{q}-basis), the residue matrices are expressed in the $\SO(3)$ basis defined in section \ref{sec:tensor-struct}. Therefore, it will be convenient to in re-express the leading term in the $\SO(3)$ basis. To do this we can use the conventions from section \ref{sec:tensor-struct} to relate the OPE coefficients between different bases, 
\be
\label{eq:q-basis-to-SO(2)-basis-OPE-coeff}
	\l_{[q_1q_2q]} &= \,\l_{(q_1q_2q)} (-1)^{j_1-j+q_2}  \prod_{i=1, 2, \cO} \binom{2j_i}{j_i+q_i}^{1/2}\,,\nn\\
	\l_{(m_1m_2m)} &=\,\sum_{j_{12}} \l_{(j_{12}, m)} \< j_1, m_1,j_2, m_2|j_{12}, m\>  \,,\nn \\
	\l_{(j_{12}, m)} &=\,\sum_{j_{12\cO}}\l_{( j_{12}, j_{12\cO})} \< j_{12}, -m, j, m|j_{12\cO}, 0\>  \,, 
\ee
 to re-express \eqref{eq:leading-eq-in-q-basis} in terms of OPE coefficients in the four-point function in the $\SO(3)$ basis. We thus find that, 
 \be
 f_{[q_1q_2q_3q_4], j}^{( j_{12}, j_{12\cO}), \,( j_{43}, j_{43\cO})}(\phi) =\sum_{\substack{m_1,m_2, m\\ m_4, m_3, m'}} \cW^{( j_{12}, j_{12\cO}), \,( j_{43}, j_{43\cO})}_{[m_1 m_2 m],\,[m_4 m_3 m']} f_{[q_1 q_2 q_3 q_4], j}^{[m_1 m_2 m],\,[m_4 m_3 m']}(\phi)\,,
 \ee
 where, 
 \be 
 \cW^{(j_{12\cO}, j_{12}), \,(j_{43\cO}, j_{43})}_{[m_1 m_2 m],\,[m_3 m_4 m']} &= (-1)^{j_1+j_4-2j+m_2-m_3} \<j_1, m_1; j_2, m_2|j_{12}, -m\>\<j_{12}, -m; j, m|j_{12\cO}, 0\>\nn \\ &\times \<j_4, m_4; j_3, m_3|j_{43}, m'\>\<j_{43}, m'; j, -m'|j_{43\cO}, 0\> \nn \\ &\times\left[\binom{2j}{j+m} \binom{2j}{j+m'}\prod_{i=1}^4 \binom{2j_i}{j_i+m_i}\right]^{1/2}\,.
 \ee

\bibliographystyle{JHEP}
\bibliography{refs}

\end{document}